\documentclass[12pt]{article}
\usepackage{amsfonts,amssymb,amsmath,fullpage,pst-node,rotating,graphics,epsfig, subfigure}
\usepackage{multirow}
\usepackage{authblk}

\newcommand{\R}{{\mathbb R}}
\newcommand{\Z}{{\mathbb Z}}
\newcommand{\C}{{\mathbb C}}
\newcommand{\mH}{{\cal Q}}

\newcommand{\T}{{\mathbb T}}
\newcommand{\I}{{\mathbb I}}
\newcommand{\V}{{\mathbb V}}
\newcommand{\mO}{{\mathbb O}}
\newcommand{\D}{{\mathbb D}}

\newcommand{\Fix}{{\rm Fix}\,}
\newcommand{\gc}{{\rm gcd}}
\newcommand{\ri}{{\rm i}}

\newcommand{\re}{{\rm e}}
\newcommand{\rd}{{\rm d}}

\newcommand{\br}{\bf r}

\newcommand{\bl}{\bf l}
\newcommand{\rl}{{\,|\,}}
\newcommand{\bx}{{\bf x}}
\newcommand{\by}{{\bf y}}

\newtheorem{definition}{Definition}
\newtheorem{theorem}{Theorem}
\newtheorem{lemma}{Lemma}
\newtheorem{proposition}{Proposition}

\newtheorem{remark}{Remark}
\newcommand{\proof}{\noindent{\bf Proof: }}

\newcommand{\qed}{\hfill{\bf QED}\vspace{5mm}}

\begin{document}
\title{Pseudo-simple heteroclinic cycles in $\R^4$}

\author[1]{Pascal Chossat}
\author[2]{Alexander Lohse}
\author[3]{Olga Podvigina}
\affil[1]{\small Universit\'e C\^ote d'Azur - CNRS, Parc Valrose, 06108 Nice cedex, France}
\affil[2]{\small University of Hamburg, Bundesstra\ss e 55, 20146 Hamburg, Germany}
\affil[3]{\small Institute of Earthquake Prediction Theory
and Mathematical Geophysics, 84/32 Profsoyuznaya St, 117997 Moscow, Russian Federation}

\maketitle

\begin{abstract}
We study \emph{pseudo-simple} heteroclinic cycles for a $\Gamma$-equivariant
system in $\R^4$ with finite $\Gamma \subset O(4)$, and their nearby dynamics.
In particular, in a first step towards a full classification -- analogous to
that which exists already for the
class of \emph{simple} cycles -- we identify all finite subgroups of $O(4)$
admitting pseudo-simple cycles. To this end we introduce a constructive method
to build equivariant dynamical systems possessing a robust heteroclinic cycle.
Extending a previous study we also investigate the existence of periodic orbits
close to a pseudo-simple cycle, which depends on the symmetry
groups of equilibria in the cycle.
Moreover, we identify subgroups $\Gamma\subset O(4)$, $\Gamma \not\subset SO(4)$,
admitting fragmentarily asymptotically stable
pseudo-simple heteroclinic cycles. (It has been previously shown that for $\Gamma\subset SO(4)$
pseudo-simple cycles generically are completely unstable.)
Finally, we study a generalized heteroclinic cycle, which involves
a pseudo-simple cycle as a subset.
\end{abstract}

\noindent {\em Keywords:} equivariant dynamics, quaternions, heteroclinic cycle, periodic orbit, stability

\noindent {\em Mathematics Subject Classification:} 34C14, 34C37, 37C29, 37C75, 37C80, 37G15, 37G40%, 70K44

\section{Introduction}\label{sec1}
A heteroclinic cycle is an invariant set of a dynamical system comprised of
equilibria $\xi_1, \ldots ,\xi_M$ and heteroclinic orbits $\kappa_i$ from
$\xi_i$ to $\xi_{i+1}$, $i=1\dots M$ with the convention $M+1=1$. For several
decades these objects have been of keen interest to the
nonlinear science community. A heteroclinic cycle is associated with
intermittent dynamics, where the system alternates between states of almost
stationary behaviour and phases of quick change.
It is well-known that a heteroclinic cycle can exist robustly in equivariant
dynamical systems, i.e. persist under generic equivariant perturbations,
namely when all heteroclinic orbits are saddle-sink connections in
(flow-invariant) fixed-point subspaces. Robust heteroclinic cycles, their nearby
dynamics and attraction properties have been thorougly studied, especially in
low dimensions. See \cite{cl2000, Kru97} for a general overview. In $\R^3$,
there are comparatively few possibilities for heteroclinic dynamics and these
are rather well-understood. In $\R^4$, the situation is significantly more
involved. We therefore consider systems
\begin{align}\label{sys1}
\dot{x}=f(x),
\end{align}
where $f: \R^4 \to \R^4$ is a smooth map that is equivariant with respect to
the action of a finite group $\Gamma \subset O(4)$, i.e.\
\begin{align}\label{equivariance}
 f(\gamma x)=\gamma f(x) \quad \text{for all} \ x \in \R^4,\ \gamma \in \Gamma.
\end{align}
In this setting, much attention has been paid to so-called \emph{simple} cycles,
see e.g. \cite{cl14,cl16,km95a,km04}, for which (i) all connections lie in
two-dimensional fixed-point spaces $P_j=\Fix(\Sigma_j)$ with
$\Sigma_j \subset \Gamma$, and (ii) the cycle intersects each connected
component of $P_{j-1} \cap P_j \setminus \{0\}$ at most once. This definition was introduced by
\cite{km04}, who also
suggested several examples of subgroups of $O(4)$ that admit such a cycle (in
the sense that there is an open set of $\Gamma$-equivariant vector fields
possessing such an invariant set). The classification of simple cycles was
completed in \cite{sot03,sot05} (for homoclinic cycles) and finally in
\cite{pc15} by finding all groups $\Gamma\subset O(4)$ admitting such a cycle.
In \cite{pc15} it was also discovered that the original definition of simple
cycles from \cite{km04} implicitly assumed a condition on the isotypic
decomposition of $\R^4$ with respect to the isotropy subgroup of an equilibrium,
see subsection \ref{sec21} for details. This prompted them to define
\emph{pseudo-simple} heteroclinic cycles as those satisfying (i) and (ii) above,
but not this implicit condition.

It is the primary aim of the present paper to
carry out a systematic study of pseudo-simple cycles in $\R^4$, by establishing
a complete list of all groups $\Gamma \subset O(4)$ that admit such a cycle.
This is done in a similar fashion to the classification of simple cycles in
\cite{pc15}, by using a quaternionic approach to describe finite subgroups of
$O(4)$. First examples for pseudo-simple cycles were investigated in
\cite{pc15,pc16}. The latter of those also addressed stability issues: it was
shown that a pseudo-simple cycle with $\Gamma\subset SO(4)$ is generically
completely unstable, while for the case $\Gamma\not\subset SO(4)$ a cycle
displaying a weak form of stability, called \emph{fragmentary asymptotic
stability}, was found. A fragmentarily asymptotically stable (f.a.s.)\ cycle has a positive measure basin
of attraction that does not necessarily include a full neighbourhood of the
cycle. We extend this stability study by showing
an example of group $\Gamma \not\subset SO(4)$ which admits an asymptotically
stable generalized heteroclinic cycle and pseudo-simple subcycles that are
f.a.s.. Moreover, we look at the dynamics near a pseudo-simple cycle and
discover that asymptotically stable periodic orbits may bifurcate from it.
Whether or not this happens depends on the isotropy subgroup
$\D_k$, $k \geq 3$ of equilibria comprising the cycle. The case $k=3$ was already considered in
\cite{pc16}. We illustrate our more general results by numerical simulations
for an example with $\Gamma=(\D_4\rl\D_2;\D_4\rl\D_2)$ in the case $k=4$.

This paper is organized as follows. Section \ref{sec2} recalls background
information on (pseudo-simple) heteroclinic cycles and useful properties of
quaternions as a means to describe finite subgroups of $O(4)$. Then, in section
\ref{sec3} we give conditions that allow us to decide whether or not such
a group $\Gamma\subset O(4)$ admits pseudo-simple heteroclinic cycles.
Section \ref{secth1} contains the statement and proofs of theorems \ref{th1}
and \ref{th2}, which use the previous results to list all subgroups of $O(4)$
admitting pseudo-simple heteroclinic cycles. The proof of theorem \ref{th1}
relies on properties of finite subgroups of $SO(4)$ that are given in
appendices A-C. In section \ref{sec6n} we
investigate the existence of asymptotically stable periodic orbits close to a
pseudo-simple cycle, depending on the symmetry groups $\D_k$, of equilibria.
The cases $k=3,4$ and $k\geq 5$ are covered by theorems \ref{thperorb} and
\ref{noorbit}, respectively.
In section \ref{sec8n} we employ the ideas of the previous sections
to provide a numerical example of a pseudo-simple cycle with a nearby
attracting periodic orbit.
Finally, in section \ref{sec6} for a family of subgroups $\Gamma\not\subset SO(4)$
we construct a generalized heteroclinic cycle (i.e., a cycle with
multidimensional connection(s)) and prove conditions for
its asymptotic stability in theorem \ref{as}. This cycle involves as
a subset a pseudo-simple heteroclinic cycle, that can be fragmentarily
asymptotically stable.
Section \ref{sec8} concludes and identifies possible continuations of this study.
The appendices contain additional information on subgroups of $SO(4)$ that is relevant for the proof of theorem \ref{th1}.

\section{Background}\label{sec2}
Here we briefly review basic concepts and terminology for pseudo-simple heteroclinic cycles and the quaternionic approach to describing subgroups of $SO(4)$ as needed in this paper.

\subsection{Pseudo-simple heteroclinic cycles}\label{sec21}
In this subsection we give the precise framework in which we investigate robust heteroclinic cycles and the associated dynamics. Given an equivariant system \eqref{sys1} with finite $\Gamma \subset O(4)$ recall that for $x \in \R^4$ the \emph{isotropy subgroup of $x$} is the subgroup of all elements in $\Gamma$ that fix $x$. On the other hand, given a subgroup $\Sigma \subset \Gamma$ we denote by $\Fix(\Sigma)$ its \emph{fixed point space}, i.e.\ the space of points in $\R^4$ that are fixed by all elements of $\Sigma$.

Let $\xi_1, \ldots ,\xi_M$ be hyperbolic equilibria of a system \eqref{sys1} with stable and unstable manifolds $W^s(\xi_j)$ and $W^u(\xi_j)$, respectively. Also, let $\kappa_j \subset W^u(\xi_j) \cap W^s(\xi_{j+1}) \neq \varnothing$ for $j=1,\ldots,M$ be connections between them, where we set $\xi_{M+1}=\xi_1$. Then the union of equilibria $\{\xi_1,\ldots ,\xi_M\}$ and connecting trajectories $\{ \kappa_1, \ldots ,\kappa_M\}$ is called a \emph{heteroclinic cycle}. Following \cite{km95a} we say it is \emph{structurally stable} or \emph{robust} if for all $j$ there are subgroups $\Sigma_j \subset \Gamma$ such that $\xi_{j+1}$ is a sink in $P_j:=\Fix(\Sigma_j)$ and $\kappa_j$ is contained in $P_j$. We also employ the established notation $L_j:=P_{j-1} \cap P_j=\Fix(\Delta_j)$, with a subgroup $\Delta_j \subset \Gamma$. As usual we divide the eigenvalues of the Jacobian $df(\xi_j)$ into \emph{radial} (eigenspace belonging to $L_j$), \emph{contracting} (belonging to $P_{j-1} \ominus L_j$), \emph{expanding} (belonging to $P_j \ominus L_j$) and \emph{transverse} (all others), where we write $X \ominus Y$ for a complementary subspace of $Y$ in $X$. In accordance with \cite{km95a} our interest lies in cycles where
\begin{itemize}
 \item[(H1)] $\dim P_j =2$ for all $j$,
 \item[(H2)] the heteroclinic cycle intersects each connected component of $L_j \setminus \{0\}$ at most once.
\end{itemize}
Then there is one eigenvalue of each type and we denote the corresponding contracting, expanding and transverse eigenspaces of $df(\xi_j)$ by $V_j$, $W_j$ and $T_j$, respectively. In \cite{pc15} it is shown that under these conditions there are three possibilities for the unique $\Delta_j$-isotypic decomposition of $\R^4$:
\begin{enumerate}
 \item[(1)] $\R^4=L_j \oplus V_j \oplus W_j \oplus T_j$
 \item[(2)] $\R^4=L_j \oplus V_j \oplus \widetilde{W}_j$, where $\widetilde{W}_j=W_j \oplus T_j$ is two-dimensional
 \item[(3)] $\R^4=L_j \oplus W_j \oplus \widetilde{V}_j$, where $\widetilde{V}_j=V_j \oplus T_j$ is two-dimensional
\end{enumerate}
Here $\oplus$ denotes the orthogonal direct sum. This prompts the following definition.

\begin{definition}[\cite{pc15}] \label{def:simcyc}
We call a heteroclinic cycle satisfying conditions (H1) and (H2) above \emph{simple} if case 1 holds true for all $j$, and \emph{pseudo-simple} otherwise.
\end{definition}

\begin{remark}\label{rem:grsig}
In case 1 the group $\Delta_j$ acts as $\Z_2$ on each one-dimensional component
other than $L_j$ and $\Delta_j\cong\D_2$ (which is always the case if
$\Gamma\subset SO(4)$) or $\Delta_j\cong(\Z_2)^3$.
In cases 2 and 3 the group acts on the two-dimensional isotypic
component as a dihedral group $\mathbb{D}_k$ in $\R^2$ for some $k \geq 3$ and
$\Delta_j\cong\D_k=<\rho_j,\sigma_j>$ (always for $\Gamma\subset SO(4)$) or
$\Delta_j\cong\D_k\times\Z_2$.
For $\Gamma \subset SO(4)$ in case 2 the
element $\rho_j$ acts as a $k$-fold rotation on $\widetilde{W}_j$ and trivially
on $P_{j-1}=L_j \oplus V_j$, while $\sigma_j$ acts as $-I$ on $V_j \oplus T_j$ and
trivially on $L_j \oplus W_j$. In case 3 the element $\rho_j$ acts as a $k$-fold rotation on
$\widetilde{V}_j$ and trivially on $P_j=L_j \oplus W_j$, while $\sigma_j$ acts as
$-I$ on $W_j \oplus T_j$ and trivially on $L_j\oplus V_j$.
\end{remark}
%In cases 2 and 3, the group $\Delta_j$ acts on the two-dimensional isotypic
%component as a dihedral group $\mathbb{D}_k$ in $\R^2$, for some $k \geq 3$.
%In contrast to that, in case 1 it acts as $\Z_2$ on each one-dimensional
%component other than $L_j$.
\begin{remark}\label{rem:eigv}
The existence of a two-dimensional isotypic component implies that in case 2 the contracting and transverse eigenvalues are equal ($c_j=t_j$) and the associated eigenspace is two-dimensional, while in case 3 the expanding and transverse eigenvalues are equal ($e_j=t_j$) and the associated eigenspace is two-dimensional. Hence, we say that $df(\xi_j)$ has a multiple contracting or expanding eigenvalue in cases 2 or 3, respectively.
\end{remark}

We are interested in identifying all subgroups of $O(4)$ that admit pseudo-simple heteroclinic cycles in the following sense. For simple cycles this task has been achieved step by step in \cite{km04,pc15, sot03, sot05}.

\begin{definition}[\cite{pc15}] \label{def:admits}
We say that a subgroup $\Gamma$ of O($n$) {\em admits} (pseudo-)simple heteroclinic cycles if there exists an open subset of the set of smooth $\Gamma$-equivariant vector fields in $\R^n$, such that all vector fields in this subset possess a (pseudo-)simple heteroclinic cycle.
\end{definition}

In order to establish the existence of a heteroclinic cycle it is sufficient to find a sequence of connections $\xi_1 \to \ldots \to \xi_{m+1}=\gamma \xi_1$ with some finite order $\gamma \in \Gamma$, that is minimal in the sense that no $i,j \in \{1, \ldots ,m\}$ satisfy $\xi_i=\gamma'\xi_j$ for any $\gamma' \in \Gamma$.

\begin{definition}[\cite{pc16}]
 Such a sequence $\xi_1 \to \ldots \to \xi_m$ together with the element $\gamma \in \Gamma$ is called a \emph{building block} of the heteroclinic cycle.
\end{definition}
\begin{remark}
Heteroclinic cycle in an equivariant system can be decomposed as a union
of building blocks. Usually, it is tacitly assumed that all blocks
in such a decomposition can be obtained from just one building block
by applying the associated symmetry $\gamma$. We also make this assumption.
\end{remark}

%Note that for a pseudo-simple cycle with $\Gamma \subset SO(4)$ there is at least one $j$ such that $\Sigma_j \cong \Z_k$ with $k \geq 3$.

\subsection{Asymptotic stability}\label{sec22}
Given a heteroclinic cycle $X$ and writing the flow of \eqref{sys1} as
$\Phi_t(x)$, the $\delta$-basin of attraction of $X$ is the set
$$
{\cal B}_\delta(X) = \{x\in \R^4~;~d(\Phi_t(x),X)<\delta\hbox{~for all~}t>0\mbox{~and~} \lim_{t\rightarrow+\infty}d(\Phi_t(x),X)=0 \}.
$$
\begin{definition} \label{def:asstable}
A heteroclinic cycle $X$ is {\em asymptotically stable} if for any $\delta>0$
there exists $\varepsilon>0$ such that
$B_{\varepsilon}\subset{\cal B}_\delta(X)$, where
$B_{\varepsilon}(X)$ denotes $\varepsilon$-neighbourhood of $X$.
\end{definition}
\begin{definition} \label{def:completelyunstable}
A heteroclinic cycle $X$ is {\em completely unstable} if there exists $\delta>0$
such that $l({\cal B}_\delta(X))=0$, where $l(\cdot)$ denotes Lebesgue measure
on $\R^4$.
\end{definition}
\begin{definition} \label{def:fragmstable}
A heteroclinic cycle $X$ is {\em fragmentarily asymptotically stable} if
$l({\cal B}_\delta(X))>0$ for any $\delta>0$.
\end{definition}

\subsection{Quaternions and subgroups of $SO(4)$}
\label{sec:quaternions}
We briefly recall some information on quaternions and their relation to subgroups of $SO(4)$, mainly following the notation and exposition in \cite[Chapter 3]{pdv}. For a more detailed background on this in general, and in the context of heteroclinic cycles, we also refer the reader to \cite{conw} and \cite{pc15}, respectively.

A quaternion ${\bf q} \in \mathbb{H}$ may be described by four real numbers as ${\bf q}=(q_1,q_2,q_3,q_4)$. With the convention $1=(1,0,0,0)$, $i=(0,1,0,0)$, $j=(0,0,1,0)$ and $k=(0,0,0,1)$ any ${\bf q} \in \mathbb{H}$ can be written as ${\bf q}=q_1+q_2i+q_3j+q_4k$. We denote the conjugate of ${\bf q}$ as ${\bf \tilde{q}}:=q_1-q_2i-q_3j-q_4k$. Multiplication is defined in the standard way through the rules $i^2=j^2=k^2=-1$, $ij=-ji=k$, $jk=-kj=i$, $ki=-ik=j$, such that for ${\bf p}, {\bf q} \in \mathbb{H}$ we have
$$
\renewcommand{\arraystretch}{1.2}
\begin{array}{ll}
{\bf p}{\bf q}=&(p_1q_1-p_2q_2-p_3q_3-p_4q_4,\ p_1q_2+p_2q_1+p_3q_4-p_4q_3,\\
&~p_1q_3-p_2q_4+p_3q_1+p_4q_2,\ p_1q_4+p_2q_3-p_3q_2+p_4q_1).
\end{array}
$$
By $\mH \subset \mathbb{H}$ we denote the multiplicative group of unit quaternions, with identity element $(1,0,0,0)$. %An element ${\bf q}=(q_1,q_2,q_3,q_4) \in \mH$ may alternatively be written as ${\bf q}=(\cos \theta, {\bf u} \sin \theta)$ with ${\bf u}=(q_2,q_3,q_4) \in \R^3$ and $\theta \in [0,2 \pi)$.
There is a 2-to-1 homomorphism from $\mH$ to $SO(3)$, relating ${\bf q} \in \mH$ to the map ${\bf v} \mapsto {\bf q}{\bf v}{\bf q}^{-1}$, which is a rotation in the three-dimensional space of points ${\bf v} = (0,v_2,v_3,v_4) \in \mathbb{H}$. Any finite subgroup of $\mH$ then falls into one of the following cases, which are pre-images of the respective subgroups of $SO(3)$ under this homomorphism:
\begin{equation}\label{finsg}
\renewcommand{\arraystretch}{1.5}
\begin{array}{ccl}
\Z_n&=&\displaystyle{\oplus_{r=0}^{n-1}}(\cos2r\pi/n,0,0,\sin2r\pi/n)\\
\D_n&=&\Z_{2n}\oplus\displaystyle{\oplus_{r=0}^{2n-1}}(0,\cos r\pi/n,\sin r\pi/n,0)\\
\V&=&((\pm1,0,0,0))\\
\T&=&\V\oplus\left(\pm{1\over2},\pm{1\over2},\pm{1\over2},\pm{1\over2}\right)\\
\mO&=&\T\oplus\sqrt{1\over2}((\pm1,\pm1,0,0))\\
\I&=&\T\oplus{1\over2}((\pm\tau,\pm1,\pm\tau^{-1},0)),
\end{array}\end{equation}
where $\tau=(\sqrt{5}+1)/2$. Double parenthesis denote all even  permutations of quantities within the parenthesis.

The four numbers $(q_1,q_2,q_3,q_4)$ can be regarded as Euclidean coordinates
of a point in $\R^4$. For any pair of unit quaternions $({\bf l};{\bf r})$,
the transformation ${\bf q}\to{\bf lqr}^{-1}$ is a rotation in $\R^4$, i.e.
an element of the group $SO(4)$. The mapping
$\Phi:\mH\times\mH\to SO(4)$ that relates the pair $({\bf l};{\bf r})$
with the rotation ${\bf q}\to{\bf lqr}^{-1}$ is a homomorphism onto,
whose kernel consists of two elements, $(1;1)$ and $(-1;-1)$; thus
the homomorphism is two to one.
Therefore, a finite subgroup of $SO(4)$ is a subgroup of a product of two
finite subgroups of $\mH$.
Following \cite{pdv} we write
$({\bf L}\rl{\bf L}_K;{\bf R}\rl{\bf R}_K)$ for the group $\Gamma$.
The isomorphism between ${\bf L}/{\bf L}_K$ and ${\bf R}/{\bf R}_K$ may not be
unique and different isomorphisms give rise to different subgroups of $SO(4)$.
The complete list of finite subgroups of $SO(4)$ is given in table \ref{listSO4}, where the subscript $s$ distinguishes subgroups obtained by different isomorphisms for $s < r/2$ and prime to $r$. This is explained in more detail in \cite[Chapter 3]{pdv} and in section 2.2 of \cite{pc15}.

\begin{table}[htp]
\hskip -1cm\begin{tabular}{|c|c|c|c|c|c|c|c|c|c|c|}
\hline \# & group & order && \# & group & order && \# & group & order \\ \hline
1 & $(\Z_{2nr}\rl\Z_{2n};\Z_{2kr}\rl\Z_{2k})_s$ & $2nkr$ &&
15 & $(\D_n\rl\D_n;\mO\rl\mO)$ & $96n$ &&
29 & $(\mO\rl\mO;\I\rl\I)$ & 2880  \\ \hline

2 & $(\Z_{2n}\rl\Z_{2n};\D_{k}\rl\D_{k})_s$ & $4nk$ &&
16 & $(\D_n\rl\Z_{2n};\mO\rl\T)$ & $48n$ &&
30 & $(\I\rl\I;\I\rl\I)$ & 7200 \\ \hline

3 & $(\Z_{4n}\rl\Z_{2n};\D_{k}\rl\Z_{2k})$ & $4nk$ &&
17 & $(\D_{2n}\rl\D_n;\mO\rl\T)$ & $96n$ &&
31 & $(\I\rl\Z_2;\I\rl\Z_2)$ & 120 \\ \hline

4 & $(\Z_{4n}\rl\Z_{2n};\D_{2k}\rl\D_{k})$ & $8nk$ &&
18 & $(\D_{3n}\rl\Z_{2n};\mO\rl\V)$ & $48n$ &&
32 & $(\I^{\dagger}\rl\Z_2;\I\rl\Z_2)$ & 120 \\ \hline

5 & $(\Z_{2n}\rl\Z_{2n};\T\rl\T)$ & $24n$ &&
19 & $(\D_n\rl\D_n;\I\rl\I)$ & $240n$ &&
33 & $(\Z_{2nr}\rl\Z_{n};\Z_{2kr}\rl\Z_{k})_s$ & nkr \\ \cline{1-7}

6 & $(\Z_{6n}\rl\Z_{2n};\T\rl\V)$ & $24n$ &&
20 & $(\T\rl\T;\T\rl\T)$ & 288 &&
   & $n\equiv k\equiv 1 (\mod 2)$ & \\ \hline

7 & $(\Z_{2n}\rl\Z_{2n};\mO\rl\mO)$ & $48n$ &&
21 & $(\T\rl\Z_2;\T\rl\Z_2)$ & 24 &&
34 & $(\D_{nr}\rl\Z_{n};\D_{kr}\rl\Z_{k})_s$ & 2nkr \\ \cline{1-7}

8 & $(\Z_{2n}\rl\Z_{2n};\mO\rl\T)$ & $48n$ &&
22 & $(\T\rl\V;\T\rl\V)$ & 96 &&
   & $n\equiv k\equiv 1$ (mod 2) & \\ \hline

9 & $(\Z_{2n}\rl\Z_{2n};\I\rl\I)$ & $120n$ &&
23 & $(\T\rl\T;\mO\rl\mO)$ & 576 &&
35 & $(\T\rl\Z_1;\T\rl\Z_1)$ & 12 \\ \hline

10 & $(\D_n\rl\D_n;\D_k\rl\D_k)$ & $8nk$ &&
24 & $(\T\rl\T;\I\rl\I)$ & 1440 &&
36 & $(\mO\rl\Z_1;\mO\rl\Z_1)$ & 24 \\ \hline

11 & $(\D_{nr}\rl\Z_{2n};\D_{kr}\rl\Z_{2k})_s$ & $4nkr$ &&
25 & $(\mO\rl\mO;\mO\rl\mO)$ & 1152 &&
37 & $(\mO\rl\Z_1;\mO\rl\Z_1)^{\dagger}$ & 24 \\ \hline

12 & $(\D_{2n}\rl\D_n;\D_{2k}\rl\D_k)$ & $16nk$ &&
26 & $(\mO\rl\Z_2;\mO\rl\Z_2)$ & 48 &&
38 & $(\I\rl\Z_1;\I\rl\Z_1)$ & 60 \\ \hline

13 & $(\D_{2n}\rl\D_n;\D_k\rl\Z_{2k})$ & $8nk$ &&
27 & $(\mO\rl\V;\mO\rl\V)$ & 192 &&
39 & $(\I^{\dagger}\rl\Z_1;\I\rl\Z_1)$ & 60 \\ \hline

14 & $(\D_n\rl\D_n;\T\rl\T)$ & $48n$ &&
28 & $(\mO\rl\T;\mO\rl\T)$ & 576 && &&\\ \hline

\end{tabular}
\caption{Finite subgroups of $SO(4)$}\label{listSO4}
\end{table}
The superscript $\dagger$ is employed to denote subgroups of $SO(4)$ where the
isomorphism between the quotient groups ${\bf L}/{\bf L}_K$ and
${\bf R}/{\bf R}_K\cong{\bf L}/{\bf L}_K$ is not the identity. The group
$\I^{\dagger}$, isomorphic to $\I$, involves the elements
$((\pm\tau^*,\pm1,\pm(\tau^*)^{-1},0))$, where $\tau^*=(-\sqrt{5}+1)/2$.
The groups 1-32 contain the central rotation $-I$, and the groups 33-39 do not.

A reflection in $\R^4$ can be expressed in the quaternionic presentation as
${\bf q}\to{\bf a\tilde qb}$, where ${\bf a}$ and ${\bf b}$ is a pair of
unit quaternions. We write this reflection as $({\bf a};{\bf b})^*$.
The transformations ${\bf q}\mapsto {\bf a\tilde qa}$ and
%${\bf q}\mapsto -{\bf a\tilde qa}$ are respectively the reflections about the
%axis $\bf a$ and through the hyperplane orthogonal to the vector $\bf a$.
%Here we call ${\cal R}$ a reflection about the axis $\bf a$ if $-{\cal R}$ is
%a reflection (in the usual sense) through the hyperplane orthogonal to $\bf a$. Therefore if $\bf a\perp b$ are two orthogonal unit quaternions, the rotation of angle $\pi$ about the plane $<{\bf a},{\bf b}>$ is ${\bf q}\to-{\bf a\tilde bq}({\bf\tilde ba)}$. We call this transformation the {\em plane reflection} about $<{\bf a},{\bf b}>$.% (see Remark \ref{rem:planereflection})
${\bf q}\mapsto -{\bf a\tilde qa}$ are respectively the axial reflection
in the $\bf a$-axis (leaving unchanged all vectors parallel to the axis $\bf a$
and reversing all those perpendicular to it) and reflection through the
hyperplane orthogonal to the vector $\bf a$.

\subsection{Lemmas}
In this subsection we recall lemmas \ref{lem2}-\ref{lem51} from \cite{pdv, op13,pc15} and prove
%state without proofs (due to their simplicity)
lemmas \ref{lem6} and \ref{lem8}. They provide basic geometric information that is used to prove theorem \ref{th1} in section \ref{secth1}.

\begin{lemma}[see proof in \cite{op13}]\label{lem2}
Let $N_1$ and $N_2$ be two planes in $\R^4$ and $p_j$, $j=1,2$, be
the elements of $SO(4)$ which act on $N_j$ as identity, and on $N_j^{\perp}$ as $-I$,
and $\Phi^{-1}p_j=({\bf l}_j,{\bf r}_j)$, where $\Phi$ is the homomorphism
defined in the previous subsection. Denote by $({\bf l}_1{\bf l}_2)_1$
and $({\bf r}_1{\bf r}_2)_1$ the first components of the respective quaternion
products. The planes $N_1$ and $N_2$ intersect if and only if
$({\bf l}_1{\bf l}_2)_1=({\bf r}_1{\bf r}_2)_1=\cos\alpha$ and $\alpha$ is
the angle between the planes.
\end{lemma}

\begin{lemma}[see proof in \cite{pc15}]\label{lem3}
Let $P_1$ and $P_2$ be two planes in $\R^n$,
$\dim(P_1\cap P_2)=1$, $\rho\in$\,O($n$) is a plane reflection about
$P_1$ and $\sigma\in$\,O($n$) maps $P_1$ into $P_2$. Suppose
that $\rho$ and $\sigma$ are elements of a finite subgroup
$\Delta\subset$\,O($n$). Then $\Delta\supset\D_m$, where $m\ge 3$.
\end{lemma}

\begin{lemma}[see proof in \cite{pc15}]\label{lem4}
Consider $g\in SO(4)$, $\Phi^{-1}g=((\cos\alpha,\sin\alpha{\bf v});(\cos\beta,\sin\beta{\bf w}))$.\\
Then $\dim\Fix<g>=2$ if and only if $\cos\alpha=\cos\beta$.
\end{lemma}

\begin{lemma}[see proof in \cite{pc15}]\label{lem5}
Consider $g,s\in SO(4)$, where
$\Phi^{-1}g=((\cos\alpha,\sin\alpha{\bf v});(\cos\alpha,\sin\alpha{\bf w}))$ and
$\Phi^{-1}s=((0,{\bf v});(0,{\bf w}))$. Then $\Fix<g>=\Fix<s>$.
\end{lemma}

\begin{lemma}[see proof in \cite{pdv}]\label{lem51}
If ${\bf l}=(\cos\omega,{\bf v}\sin\omega)$ and
${\bf r}=(\cos\omega',{\bf v}'\sin\omega')$, then the transformation
${\bf q}\to{\bf lqr}^{-1}$ is a rotation of angles $\omega\pm\omega'$ in a
pair of absolutely perpendicular planes.
\end{lemma}

\begin{lemma}\label{lem6}
If $\Gamma\subset SO(4)$,
$\Phi^{-1}\Gamma=({\bf L}\rl{\bf L}_K;{\bf R}\rl{\bf R}_K)$,
admits pseudo-simple heteroclinic cycles then ${\bf L}\supset\D_k$ and
${\bf R}\supset\D_k$, where $k\ge3$.
\end{lemma}

\proof
Let $\Phi^{-1}\rho_j=(\bl^{(1)};\br^{(1)})$ and
$\Phi^{-1}\sigma_j=(\bl^{(2)};\br^{(2)})$, where
$\Delta_j=<\rho_j,\sigma_j>\subset\Gamma$ is the group discussed in remark
\ref{rem:grsig}. Existence of at least one such $\Delta_j$ follows from
definitions \ref{def:simcyc} and \ref{def:admits}. Lemma \ref{lem51} implies
that the order of the elements $\bl^{(1)}$ and $\br^{(1)}$ is $k$, while
the order of $\bl^{(2)}$ and $\br^{(2)}$ is 2. Since $\Phi$ is a homomorphism,
${\bf L}\supset<\bl^{(1)},\bl^{(2)}>\cong\D_k$ and
${\bf R}\supset<\br^{(1)},\br^{(2)}>\cong\D_k$.
\qed

\begin{lemma}\label{lem8}
If $\Gamma\subset SO(4)$ admits pseudo-simple heteroclinic cycles,
then it has a symmetry axis $L=\Fix\Sigma$, where $\Sigma\subset\Gamma$ is
a maximal isotropy subgroup such that $\Sigma_L\cong\D_k$ with $k\ge3$.
\end{lemma}

The proof follows from definitions \ref{def:simcyc} and \ref{def:admits} and
remark \ref{rem:grsig}.

\section{Construction of a $\Gamma$-equivariant system, possessing
a heteroclinic cycle}\label{sec3}

In this section we prove the following lemma:

\begin{lemma}\label{lem1}
\begin{itemize}
\item[(i)]
If for a given finite subgroup $\Gamma\subset O(4)$ there exist two
sequences of isotropy subgroups $\Sigma_j$, $\Delta_j$, $j=1,\dots, m$, and an
element $\gamma\in\Gamma$ satisfying the following conditions:
\begin{itemize}
\item[\bf{C1}.] Denote $P_j=\Fix(\Sigma_j)$ and
$L_j=\Fix(\Delta_j)$. Then $\dim P_j=2$ and $\dim L_j=1$ for all $j$.
\item[\bf{C2}.] For $i\neq j$, $\Sigma_i$ and $\Sigma_j$ are not conjugate.
\item[\bf{C3}.] For $j=2,\dots,m$, $L_j=P_{j-1}\cap P_j$, and
$L_1=\gamma^{-1}P_m\gamma\cap P_1$. We set $\Delta_{m+1}=\gamma\Delta_1\gamma^{-1}$.
\item[\bf{C4}.] For all $j$, the subspaces
$L_j$, $P_{j-1}\ominus L_j$ and $P_j\ominus L_j$ belong to different
isotypic components in the isotypic decomposition of $\Delta_j$ in $\R^4$.
\item[\bf{C5}.] $\Sigma_j\cong\Z_{k_j}$ with $k_j\ge3$ for at least one $j$.
\end{itemize}
Consider $G_j=N_{\Gamma}(\Sigma_{j})/\Sigma_{j}\cong \D_{k_{j}}$,
the dihedral group of order $2k_j$, where we write $k_j=0$ for a trivial
$G_j$ or $k_j=1$ for $G_j\cong\Z_2$.
Let $n_j$ be the number of isotropy types of axes $\widetilde L_{sj}\subset P_j$
that are not fixed by an element of $G_j$ and
$\widetilde L_{sj}=\Fix\widetilde\Delta_{sj}$, $1\le s\le n_j$.
\begin{itemize}
\item[\bf{C6}.] Depending on $n_j$ one of the following takes place:\\
(a) if $n_j=0$ then either $k_j$ is even and the groups $\Delta_{j-1}$,
$\Delta_j$ are not conjugate, or $k_j$ is odd;\\
(b) if $n_j=1$ then the groups $\Delta_{j-1}$ and $\Delta_j$ are not conjugate
and one of $\Delta_{j-1}$ or $\Delta_j$ is conjugate to $\widetilde\Delta_{1j}$;\\
(c) if $n_j=2$ then $\Delta_{j-1}$ and $\Delta_j$ are conjugate to
$\widetilde\Delta_{1j}$ and $\widetilde\Delta_{2j}$.
\end{itemize}
then $\Gamma$ {\em admits} pseudo-simple heteroclinic cycles.
\item[(ii)]
%If $\Gamma\subset O(4)$ does not have two sequences of isotropy subgroups $\Sigma_j$, $\Delta_j$, $j=1,\dots,m$, where $m\ge2$, and an element $\gamma$, satisfying conditions {\bf C1},{\bf C3},{\bf C4},{\bf C5} and {\bf C6$^*$}, then the group does not admit pseudo-simple heteroclinic cycles.
If $\Gamma\subset O(4)$ admits pseudo-simple heteroclinic cycles, then there are
two sequences of isotropy subgroups $\Sigma_j$, $\Delta_j$, $j=1,\dots,m$, where
$m\ge2$, and an element $\gamma$, satisfying conditions {\bf C1},{\bf C3},{\bf C4},
{\bf C5} and {\bf C6$^*$}.
\begin{itemize}
\item[\bf{C6$^*$}.] If $-I\in\Gamma$, then $\Delta_j$ and $\Delta_i$ are
not conjugate for any $i\ne j$, $1\le i,j\le m$.
\end{itemize}
\end{itemize}
\end{lemma}

In \cite{pc15} we proved a similar lemma, stating necessary and sufficient
conditions for a group $\Gamma\subset\, O(n$) to admit simple heteroclinic cycles.
As noted in \cite{pc16}, with minor modifications the proof can
be used to prove sufficient conditions for a group $\Gamma\subset O(4)$ to
admit pseudo-simple heteroclinic cycles. Here, our proof of lemma \ref{lem1}
employs a different idea. We explicitly build a $\Gamma$-equivariant
dynamical system $\dot {\bf x}={\bf f}({\bf x})$ possessing a pseudo-simple
heteroclinic cycle and prove that the cycle persists under small
$\Gamma$-equivariant perturbations.
\bigskip

\begin{proof}
Starting with the proof of (i), we show that for any group $\Gamma\subset O(4)$
satisfying conditions {\bf C1}-{\bf C6} there is a vector field {\bf f} such
that the associated dynamics possess a heteroclinic cycle between equilibria in
$\Gamma L_j$ with connections in the fixed-point planes $\Gamma P_j$.
\medbreak
As a first step, for each plane $P_j$, $j=1,\ldots,m$, that contains
the axes $L_j$ and $L_{j+1}$ (in agreement with {\bf C3}, $L_{m+1}=\gamma L_1$),
we define a two-dimensional vector field ${\bf h}_j$, which in the polar coordinates $(r,\theta)$ is:
\begin{equation}\label{fieh}
{\bf h}_j(r,\theta)=\left(r(1-r), \ \sin(\theta)\prod_{i=1}^n \sin(\theta_{ij}-\theta)\right),
\end{equation}
where $0\le\theta_{ij}<\pi$ are the angles of all fixed-point axes in $P_j$
other than $L_j$, and the angle of $L_j$ is $\theta=0$.
For the flow of $(\dot r,\dot\theta)={\bf h}_j(r,\theta)$ each of these axes is
invariant and has an equilibrium $r=1$ which is attracting along the direction
of $r$. Moreover, there are heteroclinic connections between equilibria on
neighbouring axes, since the sign of $\dot{\theta}$ changes when an axis is
crossed.

We extend the vector fields ${\bf h}_j$ to ${\bf g}_j: \R^4 \to \R^4$
as follows: Denote by $\pi_j$ and $\pi^{\perp}_j$ the projections onto the plane $P_j$ and its orthogonal complement in $\R^4$, respectively. We set
\begin{equation}\label{gj}
\pi_j{\bf g}_j({\bf x})=
{{\bf h}_j(\pi_j{\bf x})\over 1+A|\pi^{\perp}_j{\bf x}|^2},\quad
\pi^{\perp}_j{\bf g}_j({\bf x})=0,
\end{equation}
with a positive constant $A$ (to be chosen sufficiently large later). The vector field ${\bf f}:\R^4\to\R^4$ is then defined as
\begin{equation}\label{f-equation}
{\bf f}({\bf x})=\sum\limits_{j=1}^{m}
\sum\limits_{\gamma_{ij}\in \mathcal{G}_j} \gamma_{ij}{\bf g}_j(\gamma_{ij}^{-1}{\bf x}),
\end{equation}
where $\mathcal{G}_j=\Gamma/N_{\Gamma}(\Sigma_j)$ and $N_{\Gamma}(\Sigma_j)$ is the normalizer
of $\Sigma_j$ in $\Gamma$.

As the second step, we show that the system
\begin{equation}\label{f-system}
\dot{\bf x}={\bf f}({\bf x})
\end{equation}
possesses steady states $\xi_j\in L_j$, $j=1,\ldots,m$, and heteroclinic
connections $\xi_j\to\xi_{j+1}'\subset P_j$,
where $\xi_{j+1}'=\gamma_j'\xi_{j+1}$ and $\gamma_j'\in N_{\Gamma}(\Sigma_j)$.
Note, that by construction the system (\ref{f-system}) is $\Gamma$-equivariant,
which implies invariance of the axes $L_j$ and planes $P_j$.

The system (\ref{f-system}) restricted to $L_j$ is
\begin{equation}\label{sysLj}
\dot x_j=\sum\limits_{k=1}^{m}\sum\limits_{\gamma_{ik}\in \mathcal{G}_k}
{x_j\cos\beta_{ik}(1-x_j\cos\beta_{ik})\over 1+Ax_j^2\sin^2\beta_{ik}},
\end{equation}
where $x_j$ is the coordinate along $L_j$ and $\beta_{ik}$ is the angle between
$L_j$ and $\gamma_{ik}P_k$. We split the sum in (\ref{sysLj}) into two,
for $L_j\subset\gamma_{ik}P_k$ and $L_j\not\subset\gamma_{ik}P_k$, and write
$$\dot x_j=s_jx_j(1-x_j)+\sum\limits_{i,k:\,L_j\not\subset\gamma_{ik}P_k}
{x_j\cos\beta_{ik}(1-x_j\cos\beta_{ik})\over 1+Ax_j^2\sin^2\beta_{ik}},$$
where $s_j$ is the number of planes $\gamma_{ik}P_k$ that contain $L_j$.
Hence, for sufficiently large positive $A$ there exists an equilibrium
$\xi_j\in L_j$ with $x_j=c_j\approx 1$, attracting in $L_j$.

To prove existence of a heteroclinic connection $\xi_j\to\xi_{j+1}'$,
we consider a sector in $P_j$ between $L_j$ and $L'_{j+1}$, where
$L'_{j+1}=\gamma_j'L_{j+1}$ with $\gamma_j'\in N_{\Gamma}(\Sigma_j)$ such that
there are no invariant axes between $L_j$ and $L'_{j+1}$.
Existence of such a sector follows from {\bf C6}.
(In case (a) the axes $L_j$ and $L'_{j+1}$ are invariant axes of $G_j$,
they are the only invariant axes in $P_j$.
In case (b) invariant axes of $G_j$ alternate with (symmetric copies of)
$\tilde L_{1j}$. In case (c) there is one $\tilde L_{1j}$ and one
$\tilde L_{2j}$ between any two neighbouring invariant axes of $G_j$.)
We choose a small number $a>0$ and divide this sector into three subsets, as
sketched in Figure \ref{V1V2V3}:
\begin{itemize}
 \item $V_1$: a strip of width $a$ near $L_j$
 \item $V_3$: a strip of width $a$ near $L_{j+1}$
 \item $V_2$: the rest of the sector
\end{itemize}
We now consider the dynamics of system (\ref{f-system}) in each of these regions.
\begin{itemize}
\item For $V_1$ we distinguish three cases: (a) $\xi_j$ is a simple equilibrium,
i.e. the isotypic decomposition of $\R^4$ w.r.t.\ $\Delta_j$ has only 1D
components, (b) $\xi_j$ is a pseudo-simple equilibrium, i.e. the isotypic
decomposition of $\R^4$ w.r.t.\ $\Delta_j$ has a 2D component, and the
component is the contracting eigenspace, (c) $\xi_j$ is a pseudo-simple
equilibrium with 2D expanding eigenspace.

In $V_1$ we employ the coordinates $(x_j,x_{j+1})=(r\cos(\theta),r\sin(\theta))$.
Choosing $a>0$ sufficiently small and $A>0$ sufficiently large, to approximate
the dynamics near $\xi_j$, we take into account only leading terms
in ${\bf h}_{j-1}$ and ${\bf h}_j$ and in (\ref{f-system}) we omit the terms
corresponding to the planes $\gamma_{ik}P_k$ that do not contain $L_j$.
The condition {\bf C4} implies that in case (a) the axis $L_j$ belongs
to planes $P_{j-1}$ and $P_j$ only; in case (b) it also belongs to several
symmetric copies of $P_{j-1}$; in case (c) to $P_{j-1}$, $P_j$ and several
symmetric copies of $P_j$.
In case (a) we have
$$\dot x_j=x_j(c_j-x_j)\biggl(1+{1\over 1+Ax_{j+1}^2}\biggr),\quad
\dot x_{j+1}=C_jx_{j+1},\quad\hbox{where } C_j=\prod_{i=1}^{n_j}\sin\theta_{ij},$$
and $C_j>0$ since $0<\theta_{ij}<\pi$. In case (b), $P_j$ is orthogonal to
$P_{j-1}$ and its $q_j$ symmetric copies. Hence, near $\xi_j$ we have
$$\dot x_j=x_j(c_j-x_j)\biggl(1+{q_j\over 1+Ax_{j+1}^2}\biggr),\quad
\dot x_{j+1}=C_jx_{j+1}.$$
In case (c), assuming that there are $q_j$ symmetric copies of $P_j$
containing $L_j$, hence the angles between neighbouring planes are $\pi/q_j$, we approximate
the dynamics in $P_j$ as
\begin{align*}
\dot{x}_j&=x_j(c_j-x_j)\biggl({1\over1+Ax_{j+1}^2}+
\sum\limits_{k=0}^{q_j-1}{1\over 1+A\sin^2(k\pi/q_j)x_{j+1}^2}\biggr)\\
\dot{x}_{j+1}&=C_jx_{j+1}
\sum\limits_{k=0}^{q_j-1}{\cos^2(k\pi/q_j)\over 1+A\sin^2(k\pi/q_j)x_{j+1}^2}.
\end{align*}
Thus, in all cases (a)-(c) the equilibrium $\xi_j$ is attracting in $L_j$ and we have
$\dot{x}_{j+1}>0$ in $V_1$, so trajectories leave $V_1$ and enter $V_2$.
At the point of entrance, $x_j\approx 1$.

\begin{figure}[ht]
\centerline{
\includegraphics[width=11cm]{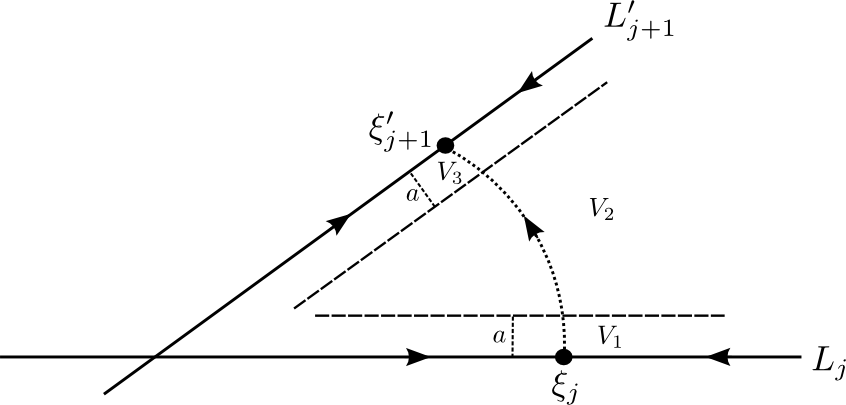}}
\caption{Division of the sector in $P_j$ into $V_1$, $V_2$, $V_3$.\label{V1V2V3}}
\end{figure}

\item The region $V_2$ is bounded away from the axes $L_j$ and
$L'_{j+1}$. Then, for any given $a>0$ there exists $A_0>0$ such that for all
$A>A_0$, the dynamics away from the fixed-point axes are essentially those of
${\bf h}_j$. Namely, the trajectories through $(x_j,x_{j+1})\approx(1,a)$
are attracted by $V_3$ and at the entrance point $r\approx 1$.
\item In $V_3$, for sufficiently small $a>0$ and sufficiently large $A>0$
all trajectories with $r\approx 1$ are attracted by $\xi'_{j+1}$ by arguments
that are analogous to those for $V_1$.
\end{itemize}

So we have shown that for the dynamics of (\ref{f-system}) in each $P_j$ there is a connection $\xi_j \to \xi'_{j+1}=\gamma_j'\xi_{j+1}$. Taking into account their symmetric copies we obtain a sequence of connections $\xi_1 \to \gamma_1'\xi_2 \to \gamma_1'\gamma_2'\xi_3 \to \ldots \to \gamma_1'\gamma_2'\ldots \gamma_m'\xi_{m+1}=:\gamma'\xi_1$ forming a building block of a heteroclinic cycle. The cycle is pseudo-simple because of {\bf C5}, and robust since by construction all connections lie in fixed-point subspaces that persist under equivariant perturbations.

For (ii) we note that necessity of conditions {\bf C1}, {\bf C3}, {\bf C4}, {\bf C5} and {\bf C6$^*$} follows directly from the definition of pseudo-simple heteroclinic cycles. The two sequences of subgroups are found by choosing $\Sigma_j$ as the isotropy subgroups of the planes $P_j$ and $\Delta_j$ as the isotropy subgroups of the equilibria $\xi_j$.
\qed
\end{proof}

\medbreak
In the following lemma we state sufficient conditions for a group
$\Gamma\subset O(4)$ to admit pseudo-simple cycles,
that are slightly different from the ones proven in lemma \ref{lem1}(i).
The difference is that in lemma \ref{lem11}
the subgroups $\Sigma_j$ can be conjugate in $\Gamma$. Since the proof
of lemma \ref{lem11} is similar to the one of lemma \ref{lem1}(i), it
is omitted.

\begin{lemma}\label{lem11}
If for a given finite subgroup $\Gamma\subset O(4)$, with $-I\in\Gamma$, there
exist two sequences of isotropy subgroups $\Sigma_j$, $\Delta_j$, $j=1,\dots,m$,
and an element $\gamma\in\Gamma$ satisfying conditions {\bf C1}, {\bf C2'},
{\bf C3}, {\bf C4}, {\bf C5} and {\bf C6'}, where
\begin{itemize}
\item[\bf{C2'}.] $\Delta_i$ and $\Delta_j$ are not conjugate for any $i\neq j$.
\item[\bf{C6'}.] For any $j$, there exists a sector in $P_j$, bounded by
$L_j$ and $L_{j+1}$, that does not contain any other isotropy axes of $\Gamma$.
\end{itemize}
then $\Gamma$ {\em admits} pseudo-simple heteroclinic cycles.
\end{lemma}

\begin{remark}\label{rem11}
Note that lemma \ref{lem11} can be generalised to $\R^n$ as follows:\\
If for a given finite subgroup $\Gamma\subset O(n)$, with $-I\in\Gamma$, there
exist two sequences of isotropy subgroups $\Sigma_j$, $\Delta_j$, $j=1,\dots,m$,
and an element $\gamma\in\Gamma$ satisfying conditions {\bf C1}, {\bf C2'},
{\bf C3}, {\bf C4} and {\bf C6'},
then $\Gamma$ {\em admits} heteroclinic cycles.
\end{remark}

\section{List of groups}\label{secth1}

\subsection{The groups $\Gamma$ in $SO(4)$}

In this subsection we prove theorem \ref{th1} that exhibits all finite subgroups of $SO(4)$, admitting robust pseudo-simple heteroclinic cycles. The proof employs lemmas \ref{lem1}(i) and \ref{lem11}, that give sufficient conditions for $\Gamma\subset SO(4)$ to admit pseudo-simple cycles, and lemma \ref{lem1}(ii) that gives necessary conditions. The lemmas allow us to split subgroups of $SO(4)$ into two classes, those admitting and those not admitting pseudo-simple heteroclinic cycles. Similarly to \cite{pc15}, we use the quaternionic presentation for subgroups of $SO(4)$, see subsection \ref{sec:quaternions}.
Appendices A-C contain detailed information on the geometry of various subgroups of $SO(4)$ which are used for proving the theorem.

\begin{theorem}\label{th1}
A group $\Gamma\subset SO(4)$ admits pseudo-simple heteroclinic cycles,
if and only if it is one of those listed in table \ref{table-th1}.

\pagebreak
\hskip -3cm\begin{table}[ht]
\begin{equation*}
\renewcommand{\arraystretch}{1.5}
\begin{array}{l}
%\hline
(\D_{2K_1}\rl\D_{2K_1};\D_{2K_2}\rl\D_{2K_2}),\ \gc(K_1,K_2)\ge2\\
(\D_{K_1r}\rl\Z_{2K_1};\D_{K_2r}\rl\Z_{2K_2})_s,\
\gc(K_1,K_2)\gc(r,K_1-sK_2)\ge3\\
(\D_{2K_1}\rl\D_{K_1};\D_{2K_2}\rl\D_{K_2}), \gc(K_1,K_2)\ge2\\
(\D_{2K_1}\rl\D_{K_1};\D_{K_2}\rl\Z_{2K_2}),\ \gc(K_1,K_2)\ge3\\
(\D_K\rl\D_K;\mO\rl\mO),\ K=3m_1\hbox{ and/or }K=4m_2\\
(\D_K\rl\Z_{2K};\mO\rl\T),\ K=3m\\
(\D_{2K}\rl\D_K;\mO\rl\T),\ K=3m_1\hbox{ and/or }K=2(2m_2+1)\\
(\D_{3K}\rl\Z_{2K};\mO\rl\V)\\
(\D_K\rl\D_K;\I\rl\I),\ K=3m_1\hbox{ and/or }K=5m_2\\
(\D_{K_1r}\rl\Z_{K_1};\D_{K_2r}\rl\Z_{K_2})_s,\ K_1\equiv K_2\equiv1(\hbox{mod } 2),\
\gc(K_1,K_2)\gc(r,K_1-sK_2)\ge3\\
%\hline
\end{array}
\end{equation*}
\caption{Groups $\Gamma\subset SO(4)$ admitting pseudo-simple heteroclinic cycles}\label{table-th1}
\end{table}
\end{theorem}

%\pagebreak
\bigskip\noindent
To prove the theorem, we proceed in four steps:

\medskip
In step [i], using lemmas \ref{lem6} and \ref{lem8} we identify subgroups of $SO(4)$,
that do not satisfy necessary conditions for existence of pseudo-simple
heteroclinic cycles stated in the lemmas. The groups 1-9 and 33 (see table
\ref{listSO4}) do not satisfy conditions of lemma \ref{lem6}. The groups 14,
20-32 and 35-39 do not satisfy conditions of lemma \ref{lem8}.
The groups 10-13, 15-19 and 34 should satisfy extra
conditions on $k_1$, $k_2$, $n$, $r$ and $s$.

\medskip
In step [ii], using lemmas \ref{lem2}-\ref{lem51} and the correspondence between
${\bf L}$ and ${\bf R}$ (see section \ref{sec:quaternions}\,), we identify
all subgroups $\Sigma$ such that $\dim\Fix\Sigma=2$, which are elements of
groups found in step [i]. The results are listed in appendix A.

\medskip
In step [iii], using the results obtained at step [ii], we determine the
(maximal) conjugacy classes of subgroups of $\Gamma$, isomorphic to $\Z_k$,
which have two-dimensional fixed-point subspaces
and (maximal) conjugacy classes of $\Delta\cong\D_k$ such that
$\dim\Fix(\Delta)=1$. The results are listed in appendix B.

\medskip
Finally, in step [iv], using the list in appendix B, we identify all groups
that possess sequences of subgroups $\Sigma_j$ and $\Delta_j$ satisfying
conditions {\bf C1}-{\bf C6} of lemmas \ref{lem1}(i) or \ref{lem11} (they are presented in appendix C). All the other groups do not have sequences satisfying conditions {\bf C1},{\bf C3}, {\bf C4},{\bf C5} and {\bf C6$^*$}. In fact, the only groups satisfying the conditions of lemma \ref{lem11}, but not those of lemma \ref{lem1}(i), are $(\D_{15K}\rl\D_{15K};\I\rl\I)$
and $(\D_{3K}\rl\Z_{6K};\mO\rl\T)$ with odd $K$.

\bigskip\noindent
{\bf Proof of the theorem}\\
Step [i]\\
The groups that satisfy conditions of lemmas \ref{lem6} and \ref{lem8} are:

\vskip 3mm
\begin{tabular}{c|c}
 \# & group \\ \hline
10 & $(\D_n\rl\D_n;\D_k\rl\D_k)$, $\gc(n,k)\ge3$\\
11 & $(\D_{nr}\rl\Z_{2n};\D_{kr}\rl\Z_{2k})_s$ $\gc(n,k)\gc(r,k-sn)\ge3$\\
12 & $(\D_{2n}\rl\D_n;\D_{2k}\rl\D_k)$, $\gc(n,k)\ge2$\\
13 & $(\D_{2n}\rl\D_n;\D_k\rl\Z_{2k})$, $\gc(n,k)\ge2$\\
15 & $(\D_n\rl\D_n;\mO\rl\mO)$, $n=3m_1$ and/or $n=4m_2$\\
16 & $(\D_n\rl\Z_{2n};\mO\rl\T)$, $n=3m$\\
17 & $(\D_{2n}\rl\D_n;\mO\rl\T)$, $n=3m_1$ and/or $n=2(2m_2+1)$\\
18 & $(\D_{3n}\rl\Z_n;\mO\rl\V)$\\
19 & $(\D_n\rl\D_n;\I\rl\I)$, $n=3m_1$ and/or $n=5m_2$\\
34 & $(\D_{nr}\rl\Z_{n};\D_{kr}\rl\Z_{k})_s$, $n\equiv k\equiv1(\mod 1)$,
$\gc(n,k)\gc(r,k-sn)\ge 3$\\
\end{tabular}

\bigskip
Below we show that the groups $(\D_{3K}\rl\Z_{6K};\mO\rl\T)$ and $(\D_{15K}\rl\D_{15K};\I\rl\I)$ admit heteroclinic cycles, while the groups $(\D_{K_1}\rl\D_{K_1};\D_{K_2}\rl\D_{K_2})$,
where at least one of $K_1$ or $K_2$ is odd, do not. For other groups the proofs are similar and we omit them.

\bigskip
{\bf The group $\Gamma=(\D_{K_1}\rl\D_{K_1};\D_{K_2}\rl\D_{K_2})$.}
\begin{itemize}
\item [[ii]] The group $\D_n$ (see (\ref{finsg})\,) is comprised of the elements
\begin{equation}\label{dn}
\rho_n(t)=(\cos t\pi/n,0,0,\sin t\pi/n),\
\sigma_n(t)=(0,\cos t\pi/n,\sin t\pi/n,0),\ 0\le t<2n.
\end{equation}
The pairs $({\bf l};{\bf r})\in(\D_{K_1}\rl\D_{K_1};\D_{K_2}\rl\D_{K_2})$
satisfy ${\bf l}\in\D_{K_1}$, ${\bf r}\in\D_{K_2}$, where all possible
combinations are elements of the group. If both $K_1$ and $K_2$ are odd,
then the elements $\gamma\in\Gamma$ satisfying $\dim\Fix\gamma=2$ are
\begin{equation}\label{pref1}
\begin{array}{l}
\kappa_1(\pm,n)=((\cos(n\theta),0,0,\pm\sin(n\theta));
(\cos(n\theta),0,0,\sin(n\theta)))\\
\kappa_2(n_1,n_2)=((0,\cos(n_1\theta_1),\sin(n_1\theta_1),0);
(0,\cos(n_2\theta_2),\sin(n_2\theta_2),0)),
\end{array}\end{equation}
where $\theta_1=\pi/K_1$, $\theta_2=\pi/K_2$, $\theta=\pi/m$,
$m=\gc(K_1,K_2)\ge3$, $0\le n_1<2K_1$, $0\le n_2<K_2$ and $0\le n<m$.
The elements $\kappa_2(n_1,n_2)$ are plane reflections, while
$\kappa_1(\pm,n)$ is a rotation by $2n\theta$ in the plane orthogonal to
$\Fix\kappa_1(\pm,n)$. For even $K_2$ the group possesses an additional
set of plane reflections
$$\kappa_3(n_1)=((0,\cos(n_1\theta_1),\sin(n_1\theta_1),0);(0,0,0,1)).$$

\item [[iii]] In the group $\D_n$ the elements
$(0,\cos(t\pi/n),\sin(t\pi/n),0)$ split into two conjugacy classes,
corresponding to odd and even $t$. Since
$\kappa_2(n_1,n_2)=\kappa_2(n_1+K_1,n_2+K_2)$,
in the case when both $K_1$ and $K_2$ are odd the group $\Gamma$ has three
maximal isotropy types of subgroups satisfying $\dim\Fix\Sigma=2$.
The subgroups are
$$\Sigma^{(1)}(\pm)=<\kappa_1(\pm,1)>,\
\Sigma^{(2)}(n_1,n_2)=<\kappa_2(n_1,n_2)>,\ n_1+n_2\hbox{ even or odd}.$$
The subgroups $\Sigma^{(1)}(+)$ and $\Sigma^{(1)}(-)$ are conjugate, e.g. by\\
$\sigma(n_1)=((0,\cos(n_1\theta_1),\sin(n_1\theta_1),0);(1,0,0,0))$.

For any plane $P=\Fix\Sigma^{(2)}(n_1,n_2)$ the only symmetry axes $L\subset P$
are the intersections with $\Sigma^{(1)}(\pm)$. The axes are conjugate by
$\sigma(n_1)\in N_{\Gamma}(\Sigma^{(2)}(n_1,n_2))$. Therefore, the group has two
maximal isotropy types of subgroups satisfying $\dim\Fix(\Delta)=1$:
$$\Delta(\pm,n_1,n_2)=
<\kappa_1(\pm,1),\kappa_2(n_1,n_2)>,\ n_1+n_2\hbox{ even or odd}.$$
Since planes $\Fix\Sigma^{(2)}(n_1,n_2)$ do not satisfy the condition {\bf C4$^*$}
and the remaining planes $\Fix\Sigma^{(1)}(+)$ and $\Fix\Sigma^{(1)}(-)$ do not
intersect, the group $(\D_{K_1}\rl\D_{K_1};\D_{K_2}\rl\D_{K_2})$ with
odd $K_1K_2$ does not admit heteroclinic cycles.

In the case when $K_1$ is odd and $K_2$ is even, a plane fixed by the
reflection $\kappa_3(n_1)$ does not intersect with any of
$\Fix\Sigma_1(\pm)$ and $\Fix\Sigma_2(n_1,n_2)$ (see lemma \ref{lem2}).
Moreover, $\Fix\kappa_3(n_1)$ does dot intersect with $\Fix\kappa_3(n_1')$ for
any $n_1\ne n_1'$. Similar arguments apply when $K_1$ is even and $K_2$ is odd.
Therefore, the group $(\D_{K_1}\rl\D_{K_1};\D_{K_2}\rl\D_{K_2})$
does not admit heteroclinic cycles when at least one of $K_1$ or $K_2$ is odd.
\end{itemize}

\bigskip
{\bf The group $\Gamma=(\D_{3K}\rl\Z_{6K};\mO\rl\T)$.}
\begin{itemize}
\item [[ii]]  The group $\mO$ can be decomposed as
$\mO=\T\oplus\sqrt{1\over2}((\pm1,\pm1,0,0))$, see (\ref{finsg}). Therefore, the
group $(\D_{3K}\rl\Z_{6K};\mO\rl\T)$ is comprised of the following elements:
\begin{equation}\label{elex2}
\begin{array}{l}
((\cos(n\theta),0,0,\sin(n\theta));\T)\\
((0,\cos(n\theta),\sin(n\theta),0);\sqrt{1\over2}((\pm1,\pm1,0,0))
\end{array}\end{equation}
where $\theta=\pi/3K$ and $0\le n<3K$. For odd $K$
the elements $\gamma\in\Gamma$ satisfying $\dim\Fix\gamma=2$ are
\begin{equation}\label{pref2}
\begin{array}{l}
\kappa_1(\pm,\pm,\pm,\pm)=((1,0,0,\pm\sqrt{3})/2);(1,\pm1,\pm1,\pm1)/2)\\
\kappa_2(n,r,\pm)=((0,\cos(n\theta),\sin(n\theta),0);\rho^r(0,1,\pm1,0)),
\end{array}\end{equation}
where $\rho(a,b,c,d)=(a,c,d,b)$. Here $\kappa_2$ are plane reflections
and $\kappa_1$ are rotations by $2\pi/3$ in the planes orthogonal to
$\Fix\kappa_1$. For even $K$ the group possesses an additional
set of plane reflections
$$\kappa_3(r,\pm)=((0,0,0,1);\rho^r(0,0,0,\pm1)).$$

\item [[iii]]
Since $\kappa_2(n,r,\pm)=-\kappa_2(n+3K,r,\pm)$ and in $\T$ the
elements $(0,1,\pm1,0)$ and $-(0,1,\pm1,0)$ are conjugate, for odd $K$ all
$\kappa_2(n,r,\pm)$ are conjugate in $\Gamma$. The elements\\
$\kappa_1(+,(-1)^{s_1},(-1)^{s_2},(-1)^{s_3})$ split into two
conjugacy classes, depending on whether $s_1+s_2+s_3$ is even or odd.
Hence, for odd $K$ the group has three maximal isotropy types of subgroups satisfying
$\dim\Fix\Sigma=2$:
\begin{equation}\label{iii2is}
\begin{array}{l}
\Sigma^{(1)}((-1)^{s_1},(-1)^{s_2},(-1)^{s_3}))=
<\kappa_1(+,(-1)^{s_1},(-1)^{s_2},(-1)^{s_3})>,\ s_1+s_2+s_3\hbox{ even or odd}\\
\Sigma^{(2)}(n,r,\pm)=<\kappa_2(n,r,\pm)>.
\end{array}\end{equation}
Each $\Fix\Sigma^{(1)}$ contains $3K$ isotropy axes, each of them are intersections
with three $\Fix<\kappa_2(n,r,\pm)>$, where $r=0,1,2$. Hence, the isotropy
groups of symmetry axes can be written as
\begin{equation}\label{iii1is}
\Delta(n,(-1)^{s_1},(-1)^{s_2},(-1)^{s_3})=
<\kappa_1(+,(-1)^{s_1},(-1)^{s_2},(-1)^{s_3}),\kappa_2(n,0,(-1)^{s_1+s_2+1})>.
\end{equation}
They split into two isotropy types, depending on whether $s_1+n$ is even or
odd. Any plane $\Fix\Sigma^{(2)}$ contains four isotropy axes which are
intersections with $\Fix\Sigma^{(1)}$. Since $N_{\Gamma}(\Sigma_2)=<\Sigma_2,-\Sigma_2>$
(this can be checked directly using the list (\ref{elex2})\,), all
four isotropy axes are of different types. Therefore, the group has four types of
isotropy subgroups (\ref{iii1is}) safisfying $\dim\Fix\Delta=2$, corresponding to
odd and even $s_1+n$ and $s_1+s_2+s_3$.

In the case when $K$ is even there exist five
isotropy types of subgroups satisfying $\dim\Fix\Sigma=2$:
\begin{equation}\label{pref22}
\begin{array}{l}
\Sigma^{(1)}((-1)^{s_1},(-1)^{s_2},(-1)^{s_3}))=
<\kappa_1(+,(-1)^{s_1},(-1)^{s_2},(-1)^{s_3})>,\ s_1+s_2+s_3\hbox{ even or odd}\\
\Sigma^{(2)}(n,r,\pm)=<\kappa_2(n,r,\pm)>,\ n\hbox{ even or odd}\\
\Sigma^{(3)}(r,\pm)=<\kappa_3(r,\pm)>.
\end{array}\end{equation}
A plane $\Fix\Sigma^{(2)}(n,r,\pm)$ orthogonally intersects with the ones
$\Fix\Sigma^{(2)}(n-(-1)^s3K/2,r,\mp)$ (and also with
$\Fix\Sigma^{(3)}(r,(-1)^s)$), hence for odd and even $K/2$ the isotropy axes are
different. Namely, for odd $K/2$ they are
\begin{equation}\label{pref22o}
\begin{array}{l}
\Delta^{(1)}(n,(-1)^{s_1},(-1)^{s_2},(-1)^{s_3})=\\
~~~<\kappa_1(+,(-1)^{s_1},(-1)^{s_2},(-1)^{s_3}),\kappa_2(n,0,(-1)^{s_1+s_2+1})>,\
s_1+s_2+s_3,\ n\hbox{ even or odd}\\
\Delta^{(2)}(n,r,\pm,(-1)^s)=
<\kappa_2(n,r,\pm),\kappa_3(r,(-1)^s)>,\ n+s\hbox{ even or odd},
\end{array}\end{equation}
while for even $K/2$ the second set of isotropy axes is
\begin{equation}\label{pref22e}
\Delta^{(2)}(n,r,\pm,\pm)=
<\kappa_2(n,r,\pm),\kappa_3(r,\pm)>,\ n\hbox{ even or odd}.
\end{equation}

\item [[iv]] According to (iii), for odd $K$ the
group $(\D_{3K}\rl\Z_{6K};\mO\rl\T)$ does not have isotropy subgroups satisfying
conditions {\bf C1}-{\bf C6} of lemma \ref{lem1}(i). Let us show that we
can find subgroups satisfying conditions of lemma \ref{lem11}. Set
\begin{equation}%\label{pref22o}
\begin{array}{l}
\Sigma_1=<\kappa_1(+,+,+,+)>,\ \Sigma_2=<\kappa_2(0,0,-)>,\
\Sigma_3=<\kappa_1(+,+,+,-)>,\\
\Sigma_4=<\kappa_2(1,0,-)>,\
\Delta_j=<\Sigma_{j-1},\Sigma_j>,\ j=2,3,4,\
\Delta_1=<\Sigma_4,\Sigma_1>\hbox{ and }\gamma=e.
\end{array}\end{equation}
By construction and due to (\ref{iii1is}) and (\ref{iii2is}), the subgroups
satisfy conditions {\bf C1},{\bf C2'},{\bf C3},{\bf C4} and {\bf C5}.

To show that $\Fix<\kappa_2(0,0,-)>$ satisfies condition {\bf C6'}, we
recall (see (iii)) that the plane involves four isotropy axes, all non-conjugate,
that are intersections with $\Fix\kappa_1(+,+,+,+)$, $\Fix\kappa_1(+,+,+,-)$,
$\Fix\kappa_1(+,-,-,+)$ and $\Fix\kappa_1(+,-,-,-)$. To determine the
angles between axes, we use lemmas \ref{lem5} and \ref{lem51}.
By lemma \ref{lem5},\\
$\Fix\kappa_1(+,\pm,\pm,\pm)=\Fix\kappa'$, where
$\kappa'(\pm,\pm,\pm)=((0,0,0,1);(0,\pm1,\pm1,\pm1)/\sqrt{3})$ is a plane
reflection about a plane that intersects with $\Fix<\kappa_2(0,0,-)>$
orthogonally. Therefore, $\Fix<\kappa_2(0,0,-)>$ is $\kappa'$-invariant.
Composition of two reflections about axes,
intersecting with the angle $\alpha$, is a rotation by $2\alpha$.
Since $\kappa'(+,+,+)\kappa'(+,+,-)=((1,0,0,0);(1,2,-2,0)/3)$, by
lemma \ref{lem51} the angle in $\Fix<\kappa_2(0,0,-)>$ between the lines of
intersections with $\kappa'(+,+,+)$ and
$\kappa'(+,+,-)$ is $\arccos(1/3)/2$, while the lines of intersections with
$\kappa'(+,+,+)$ and $\kappa'(-,-,-)$ are orthogonal. Hence, in
$\Fix<\kappa_2(0,0,-)>$ no other isotropy axes belong to the smaller sector
bounded by $\Fix\kappa_1(+,+,+,+)$ and $\Fix\kappa_1(+,+,+,-)$.
Similarly, it can be shown that the condition {\bf C6'} holds true
for $j=1,3,4$ as well.

For even $K$ we apply lemma \ref{lem1}. We choose
$$\Sigma_1=<\kappa_1(+,+,+,+)>,\ \Sigma_2=<\kappa_2(0,0,-)>,\
\Sigma_3=<\kappa_3(+,0)>,\ \Sigma_4=<\kappa_2(1,0,-)>,$$
$$\Delta_1=<\kappa_2(1,0,-),\kappa_1(+,+,+,+)>,\
\Delta_2=<\kappa_1(+,+,+,+)>,\kappa_2(0,0,-)>,$$
$$\Delta_3=<\kappa_2(0,0,-),\kappa_3(+,0)>,\
\Delta_4=<\kappa_3(+,0),\kappa_2(1,0,-)>,$$
which together with $\gamma=e$ satisfy conditions {\bf C1}-{\bf C6},
as follows from (\ref{pref22}), (\ref{pref22o}) and (\ref{pref22e}).
\end{itemize}

\bigskip
{\bf The group $\Gamma=(\D_{15K}\rl\D_{15K};\I\rl\I)$.}
\begin{itemize}
\item [[ii]] The group is comprised of the pairs $({\bf l};{\bf r})$, where
${\bf l}\in\D_{15K}$ and ${\bf r}\in\I$. Since for odd $K$ all elements
$((0,\cos(n\theta),\sin(n\theta),0);{\bf r})$ are conjugate, the
group has the following the elements satisfying $\dim\Fix\gamma=2$:
\begin{equation}\label{pref31}
\begin{array}{l}
\kappa_1(\pm,\pm,\pm,\pm)=((1,0,0,\pm\sqrt{3})/2;(1,\pm1,\pm1,\pm1)/2)\\
\kappa_1'(\pm,r,\pm,\pm)=
((1,0,0,\pm\sqrt{3})/2);\rho^r(1,\pm\tau^{-1},\pm\tau,0)/2)\\
\kappa_2(\pm,r,\pm,\pm)=
((\tau,0,0,\pm\tau^*)/2;\rho^r(\tau,\pm1,\pm\tau^{-1},0)/2)\\
\kappa_2'(\pm,r,\pm,\pm)=
((\tau^{-1},0,0,\pm\tau^{**})/2;\rho^r(\tau^{-1},\pm\tau,\pm1,0)/2)\\
\kappa_3(n,r,\pm)=((0,\cos(n\theta),\sin(n\theta),0);\rho^r(0,0,0,\pm1))\\
\kappa_3'(n,r,\pm,\pm)=
((0,\cos(n\theta),\sin(n\theta),0);\rho^r(0,1,\pm\tau,\pm\tau^{-1})),
\end{array}\end{equation}
where $\theta=\pi/15K$, $0\le n<30K$,
$\tau^*=2\sin(\pi/5)=\sqrt{5}(\tau)^{-1}$ and $\tau^{**}=2\sin(2\pi/5)=\sqrt{5}/2$.
By $\kappa_i$ and $\kappa_i'$ we denote elements that are conjugate in $\Gamma$.
Here $\kappa_3$ and $\kappa_3'$ are plane reflections,
$\kappa_1$ and $\kappa_1'$ are rotations by $2\pi/3$,
$\kappa_2$ is a rotation by $2\pi/5$ and $\kappa_2'$ is a rotation by $4\pi/5$.
For even $K$ the group possesses additional set of plane reflections:
$$\kappa_4(r,\pm)=((0,0,0,1);\rho^r(0,0,0,\pm1)),\
\kappa_4'(r,\pm,\pm)=((0,0,0,1);\rho^r(0,1,\pm\tau^{-1},\pm\tau)).$$

\item [[iii]] For odd $K$ all plane reflections are conjugate in $\Gamma$. The
rotations by $2\pi/3$ are conjugate, the rotations by $2\pi/5$ and
$4\pi/5$ are conjugate as well.
Hence, the group has three maximal isotropy types of subgroups satisfying
$\dim\Fix\Sigma=2$:
\begin{equation}\label{15K2is}
\begin{array}{l}
\Sigma^{(1)}(\pm,\pm,\pm)=<\kappa_1(+,\pm,\pm,\pm)>,\\
\Sigma^{(1')}(r,\pm,\pm)=<\kappa_1'(+,r,\pm,\pm)>,\\
\Sigma^{(2)}(r,\pm,\pm)=<\kappa_2(+,r,\pm,\pm)>,\\
\Sigma^{(3)}(n,r,\pm)=<\kappa_3(n,r,\pm)>,\\
\Sigma^{(3')}(n,r,\pm,\pm)=<\kappa_3'(n,r,\pm,\pm)>.
\end{array}\end{equation}
Each $\Fix\Sigma^{(1)}$ contains $30K$ isotropy axes, each of them is an (orthogonal)
intersection with three $\Fix<\kappa_3>$. Each $\Fix\Sigma^{(2)}$ contains $30K$
isotropy axes, each of them in an (orthogonal) intersection with five
$\Fix<\kappa_2>$. Hence, the isotropy
groups of symmetry axes can be written as
\begin{equation}\label{15K1is}
\begin{array}{l}
\Delta^{(1)}(n,(-1)^{s_1},(-1)^{s_2},(-1)^{s_3})=\\
~~~~<\kappa_1(+,(-1)^{s_1},(-1)^{s_2},(-1)^{s_3}),
\kappa_3'(n,0,(-1)^{s_1+s_2+1}(-1)^{s_1+s_3+1},))>,\ s_1+n\hbox{ even or odd},\\
\Delta^{(1')}(n,(-1)^{s_1},(-1)^{s_2})=
<\kappa_1'(+,r,(-1)^{s_1},(-1)^{s_2}),\kappa_3(n,0,\pm)>,\
s_1+s_2+n\hbox{ even or odd},\\
\Delta^{(2)}(n,(-1)^{s_1},(-1)^{s_2})=
<\kappa_2(+,(-1)^{s_1},(-1)^{s_2}),\kappa_3(n,0,\pm))>,\
s_1+s_2+n\hbox{ even or odd},\\
\Delta^{(2')}(n,(-1)^{s_1},(-1)^{s_2})=\\
~~~~<\kappa_2(+,(-1)^{s_1},(-1)^{s_2}),\kappa_3'(n,0,(-1)^{s_1+s_2+1},\pm))>,\
s_1+n\hbox{ even or odd}.
\end{array}\end{equation}

In the case when $K$ is even there exist five
isotropy types of subgroups satisfying $\dim\Fix\Sigma=2$:
\begin{equation}\label{15Kev}
\begin{array}{l}
\Sigma^{(1)},\ \Sigma^{(1')},\ \Sigma^{(2)},\\
\Sigma^{(3)}(n,r,\pm)=<\kappa_3(n,r,\pm)>,\ n\hbox{ even or odd},\\
\Sigma^{(3')}(n,r,\pm,\pm)=<\kappa_3'(n,r,\pm,\pm)>,\ n\hbox{ even or odd},\\
\Sigma^{(4)}(r,\pm)=<\kappa_4(\pm,r)>,\\
\Sigma^{(4')}(r,\pm,\pm)=<\kappa_4'(r,\pm,\pm)>.
\end{array}\end{equation}
Each of the planes $\Fix\Sigma^{(3)}$ has twelve isotropy axes. Four of them (of two isotropy
types) are orthogonal intersections with $\Fix\Sigma^{(4)}$, therefore
$N_{\Gamma}(\Sigma^{(3)})/\Sigma^{(3)}\cong\D_4$. The other eight
axes (of two isotropy types) are intersections with
$\Fix\Sigma^{(1)}$ and $\Fix\Sigma^{(2)}$. The respective isotropy subgroups
are different for odd or even $K/2$, as stated in appendix B.
A plane $\Fix\Sigma^{(1)}$ or $\Fix\Sigma^{(2)}$ involves two isotropy types
(with odd or even $n$) of symmetry axes, which are intersections with
$\Fix\Sigma^{(3)}$.

\item [[iv]] For odd $K$ we show that the groups
\begin{equation}\label{15Kgrps}
\begin{array}{l}
\Sigma_1=<\kappa_1'(+,0,+,+)>,\ \Sigma_2=<\kappa_3(0,0,+)>,\
\Sigma_3=<\kappa_2(+,0,+,+)>,\\
\Sigma_4=<\kappa_3(1,0,+)>,\
\Delta_j=<\Sigma_{j-1},\Sigma_j>,\ j=2,3,4,\
\Delta_1=<\Sigma_4,\Sigma_1>
\end{array}\end{equation}
and $\gamma=e$ satisfy conditions of lemma \ref{lem11}.
By construction and due to (\ref{15K2is}) and (\ref{15K1is}), the subgroups
satisfy conditions {\bf C1},{\bf C2'},{\bf C3},{\bf C4} and {\bf C5}.

Consider $\Fix\kappa_3(0,0,+)$. Denote by $\alpha_1$, $\alpha_2$ and
$\alpha_3$ the angles between the intersection
with $\Fix\kappa_1'(+,0,+,+)$ and the following three axes:
intersections with $\Fix\kappa_2(+,0,+,+)$, $\Fix\kappa_1'(+,0,+,-)$ and
$\Fix\kappa_2(+,0,+,-)$, respectively. By lemmas \ref{lem5} and \ref{lem51},
$$\cos2\alpha_1=(3+\sqrt{5})/(2\sqrt{15\tau}),\
\cos2\alpha_2=\sqrt{5}/3,\ \cos2\alpha_3=(\sqrt{5}-1)/(2\sqrt{15\tau}),$$
which implies that $\alpha_1<\alpha_2<\alpha_3$.

Since $N_{\Gamma}(\Sigma^{(3)})/\Sigma^{(3)}\cong\Z_4$ for old $K$ and
due to (\ref{15K1is}), in $\Fix\kappa_3(0,0,+)$ the angle between
the intersection with $\Fix\kappa_1'(+,0,+,+)$ and any
other isotropy axes is not smaller that $\alpha_1$. Therefore,
$j=2$ satisfies the condition {\bf C6'}. Similar arguments imply that
this condition is satisfied for $j=1,3,4$ as well.

Since for even $K$ the elements $\kappa_3(0,0,+)$ and $\kappa_3(1,0,+)$
are not conjugate, the set (\ref{15Kgrps}) satisfies conditions
{\bf C1}-{\bf C6} of lemma \ref{lem1}.
\end{itemize}
\qed

\subsection{The groups $\Gamma$ in $O(4)$ but not in $SO(4)$}\label{secth2}

In this subsection we prove theorem \ref{th2} that completes the list
of finite subgroups of $O(4)$, admitting pseudo-simple heteroclinic cycles.

A reflection in $\R^4$ can be expressed in the quaternionic presentation as
${\bf q}\to{\bf a\tilde qb}$, where ${\bf a}$ and ${\bf b}$ is a pair of
unit quaternions (see \cite{pdv,pc15}). We write this reflection as
$({\bf a};{\bf b})^*$. The transformations ${\bf q}\mapsto {\bf a\tilde qa}$
and ${\bf q}\mapsto -{\bf a\tilde qa}$ are respectively the reflections about
the axis $\bf a$ and through the hyperplane orthogonal to the vector $\bf a$.

A group $\Gamma^*\subset O(4)$, $\Gamma^*\not\subset SO(4)$, can be decomposed as
$\Gamma^*=\Gamma\oplus\sigma\Gamma$, where $\Gamma\subset SO(4)$
and $\sigma=({\bf a};{\bf b})^*\notin SO(4)$.
If $\Gamma^*$ is finite, then in the quaternionic form of $\Gamma$
\begin{equation}\label{gO4}
\Gamma=({\bf L}\rl{\bf L}_K;{\bf R}\rl{\bf R}_K),\hbox{ where }
{\bf L}\cong{\bf R}\hbox{ and }{\bf L}_K\cong{\bf R}_K.
\end{equation}

\begin{theorem}\label{th2}
A group $\Gamma^*\subset O(4)$,
\begin{equation}\label{decomp}
\Gamma^*=\Gamma\oplus\sigma\Gamma,
\hbox{ where $\Gamma\subset SO(4)$ and $\sigma\notin SO(4)$},
\end{equation}
admits pseudo-simple heteroclinic cycles, if and only if
${\Gamma}$ and $\sigma$ are listed in table \ref{table-th2}.
\begin{table}[h]
\begin{equation*}
\renewcommand{\arraystretch}{1.5}
\begin{array}{l|l}
\hline
{\Gamma} & \sigma \\
\hline
(\D_{Kr}\rl\Z_K;\D_{Kr}\rl\Z_K)_s,\ K\gc(r,K(1-s))\ge3
& -((0,1,0,0);(0,1,0,0))^* \\
\hline
(\D_{Kr}\rl\Z_{2K};\D_{Kr}\rl\Z_{2K})_s,\ K\gc(r,K(1-s))\ge3
& ((\cos\theta_0,0,0,\sin\theta_0);(1,0,0,0))^*,\\
& \theta_0=\pi/(2K) \\
\hline
\end{array}
\end{equation*}
\caption{Groups $\Gamma\oplus\sigma\Gamma \subset O(4)$ admitting pseudo-simple heteroclinic cycles}\label{table-th2}
\end{table}
\end{theorem}

\proof
Lemma 8 in \cite{pc15} states that if a group $\Gamma^*$ admits simple
heteroclinic cycles, then so does $\Gamma$. By similar arguments the same
holds true for pseudo-simple heteroclinic cycles.
Therefore (see lemma \ref{lem1}), the group $\Gamma$ has two sequences of isotropy
subgroups $\Sigma_j$, $\Delta_j$, $j=1,\dots,m$,
satisfying conditions {\bf C1},{\bf C3},{\bf C4},{\bf C5} and
{\bf C6$^*$}. Let $\Sigma_1$ be the subgroup satisfying {\bf C5}, i.e.
$\Sigma_1\cong\Z_{k_1}$ with $k_1\ge3$.
An element $\sigma'\in\Gamma^*$, $\sigma'\notin SO(4)$, maps
$P_1=\Fix\Sigma_1$ either to itself, or to another $P'=\Fix\Sigma'$ with
$\Sigma'\cong\Z_{k_1}$.

First, we assume the existence of $\sigma'$, such
that $\sigma'P_1=P_1$. Hence, there exists $\sigma\in\Gamma^*$ which
is a reflection through a hyperplane that contains $P_1$. Let
the hyperplane be spanned by ${\bf e}_1$, ${\bf e}_3$ and ${\bf e}_4$ and
$P_1=<{\bf e}_1,{\bf e}_4>$. The hyperplane is mapped by
elements of $\Sigma_1$ to
\begin{equation}\label{hplanes}
<{\bf e}_1,{\bf e}_4,\cos\theta_n{\bf e}_2+\sin\theta_n{\bf e}_3>,\
0\le n<k_1/2,\ \theta_n=2\pi n/k_1.
\end{equation}
Any isotropy plane of $\Gamma$, that intersects with $P_1$, is
$P(\theta',\theta_n)=
<\cos\theta'{\bf e}_1+\sin\theta'{\bf e}_4,\cos\theta_n{\bf e}_2+\sin\theta_n{\bf e}_3>$.
An isotropy plane $P'=\Fix\Sigma'\ne P_1$, such that $\Sigma'\cong\Z_{k'}$
with $k'\ge3$, is orthogonal to all hyperplanes (\ref{hplanes}).
Therefore (if such an isotropy plane exists), it is $<{\bf e}_2,{\bf e}_3>$.
Any other isotropy plane of $\Gamma$ (different from $P_1$, $P'$ and $P(\theta',\theta_n)$) either intersects all hyperplanes (\ref{hplanes})
orthogonally, or the line of intersection belongs to $P_1$ or $P'$.
Since there is no isotropy plane that satisfies these conditions, we conclude that the only isotropy planes of $\Gamma$ are
$P_1$, $P'$ and $P(\theta',\theta_n)$.
The groups listed in table \ref{table-th1} satisfying these conditions
and (\ref{gO4}) are
$(\D_{Kr}\rl\Z_K;\D_{Kr}\rl\Z_K)_s$, $K\gc(r,K(1-s))\ge3$.
The element $\sigma$ acting as a reflection through
$<{\bf e}_1,{\bf e}_3,{\bf e}_4>$ is $-((0,1,0,0),(0,1,0,0))^*$.

Second, we assume that there is no $\sigma\in\Gamma^*$, $\sigma\notin SO(4)$,
such that $\sigma P_1=P_1$. Therefore, $\sigma\notin SO(4)$ satisfies
$\sigma P_1=P'$, where $P'=\Fix\Sigma'$ with $\Sigma'\cong\Z_{k_1}$, and
the subgroups $\Sigma_1$ and $\Sigma'$ are not conjugate in $\Gamma$. The only
groups in table \ref{table-th1} that contain such $\Sigma_1$ and $\Sigma'$ are
$(\D_{Kr}\rl\Z_{2K};\D_{Kr}\rl\Z_{2K})_s$, $K\gc(r,K(1-s))\ge3$. Moreover,
$\Sigma'=\Sigma_3$ (see appendix C). The element $\sigma$ maps a symmetry axis in
$P_1$ to a symmetry axis in $\Fix\Sigma_3$. For definiteness, we assume that
$\sigma$ maps $\Fix\Delta_1$ to $\Fix\Delta_3$, where according to
the appendices\\
$\Delta_1=<\kappa_2(1,0,0),\kappa_1(+,1)>$,
$\Delta_3=<\kappa_2(0,0,0),\kappa_1(-,1)>$,\\
$\kappa_2(n,0,0)=((0,\cos n\theta_1,\sin n\theta_1,0);(0,1,0,0))$,
$\kappa_1(\pm,1)=((\cos\theta,0,0,\pm\sin\theta);(\cos\theta,0,0,\sin\theta))$,
$\theta_1=\pi/K$, $\theta=\pi/m$ and $m=K\gc(r,K(1-s))$. Such
$\sigma$ is $((\cos\theta_0,0,0,\sin\theta_0);(1,0,0,0))^*$.
\qed

\begin{remark}\label{O4_1}
A heteroclinic cycle in a $\Gamma^*$-equivariant system, where in
the decomposition (\ref{decomp})
$\Gamma=(\D_{Kr}\rl\Z_{2K};\D_{Kr}\rl\Z_{2K})_s,\ K\gc(r,K(1-s))\ge3$
and $\sigma=((\cos\theta_0,0,0,\sin\theta_0);(1,0,0,0))^*$, in general
is completely unstable. The proof follows the same arguments as the
proof of theorem 1 in \cite{pc16}. Similarly, the conditions for existence of
a nearby periodic orbit are the ones given in theorems \ref{thperorb} and
\ref{noorbit} in section \ref{sec6n} below.
\end{remark}

\begin{remark}\label{O4_2}
A heteroclinic cycle in a $\Gamma^*$-equivariant system, where in
the decomposition (\ref{decomp})
$\Gamma=(\D_{Kr}\rl\Z_K;\D_{Kr}\rl\Z_K)_s,\ K\gc(r,K(1-s))\ge3$
and $\sigma=-((0,1,0,0),(0,1,0,0))^*$, can be fragmentarily asymptotically
stable. The conditions for stability can be obtained by algebra, which is
standard (see, e.g., theorem 3 in \cite{pc16}) but tedious; we do not
present it here.
\end{remark}

\section{Existence of nearby periodic orbits when $\Gamma\subset SO(4)$}
\label{sec6n}

As shown in \cite{pc16}, despite complete instability of a pseudo-simple
heteroclinic cycle in a $\Gamma$-equivariant system for
$\Gamma\subset SO(4)$, trajectories staying in a small neighbourhood of
a pseudo-simple cycle for all $t>0$ can possibly exist.
Namely, it was proven {\it ibid} that
in a one-parameter dynamical system
an asymptotically stable periodic orbit can bifurcate from a cycle. More specifically, in their example such an asymptotically stable periodic orbit exists as long as a double positive eigenvalue is sufficiently small.
Building blocks of the considered cycles were comprised of two equilibria,
whose isotropy groups were isomorphic to $\D_3$. One of these equilibria had
a multiple expanding eigenvalue, while the other equilibrium had a multiple
contracting one. In this section we prove that similar periodic orbits can bifurcate
in a more general setup -- we do not restrict the number of equilibria in
a building block (note that building block of a pseudo-simple cycle
in $\R^4$ is comprised of at least two equilibria) and assume that
their isotropy groups are isomorphic to $\D_k$ with $k\le4$.
However, we assume that building block of a heteroclinic cycle
involves only one equilibrium with a multiple expanding eigenvalue.
In the case of several such equilibria,
the bifurcation of a periodic orbit has codimension two or higher,
which is beyond the scope of this paper.
By contrast, no such periodic orbits bifurcate in a codimension one
bifurcation if a building block
involves an equilibrium with the isotropy group $\D_k$ with $k\ge5$.

\subsection{The case $\D_3$ and $\D_4$}\label{d3d4}

Consider the $\Gamma$-equivariant system
\begin{equation}\label{eqal_ode}
\dot{\bf x}=f({\bf x},\mu),\hbox{ where }
f(\gamma{\bf x},\mu)=\gamma f({\bf x},\mu)
\mbox{ for all }\gamma\in\Gamma\subset SO(4),
\end{equation}
and $f:\R^4\times\R\to\R^4$ is a smooth map. We assume that the system
possesses a pseudo-simple heteroclinic cycle
with a building block $\{\xi_1\to\ldots\xi_m;\ \gamma\}$. By $-c_j$, $e_j$ and $t_j$ we denote the non-radial eigenvalues of $df(\xi_j)$, $1\le j\le m$.
Let $\xi_2$ be an equilibrium with a two-dimensional expanding eigenspace
(hence, $e_2=t_2$) and a symmetry group $\Delta_2=\D_k$, $k=3$ or 4,
acting naturally on the expanding eigenspace, and all other equilibria have
one-dimensional expanding eigenspaces.
A general $\D_k$-equivariant dynamical system in $\C$ in the leading order is
$\dot z=\alpha z+\beta \bar{z}^2$ (for $k=3$) and
$\dot z=\alpha z+\beta_1z^2\bar z+\beta_2\bar z^3$ ($k=4$). A necessary
condition for existence of a heteroclinic trajectory $\xi_2\to\xi_3$
along the direction of real $z$ is that $e_2=\alpha>0$ and $\beta>0$ or
$\beta_1+\beta_2>0$ (for $k=3$ or $k=4$, respectively).
Suppose that there exists $\mu_0>0$ such that
\begin{itemize}
\item[(i)] $e_2<0$ for $-\mu_0<\mu<0$ and
$e_2>0$ for $0<\mu<\mu_0$;
\item[(ii)] for any $0<\mu<\mu_0$ there exist heteroclinic
connections $\kappa_j=(W_u(\xi_j)\cap P_j)\cap W_s(\xi_{j+1})\ne\varnothing$,
for all $1\le j\le m$, where $\xi_{m+1}=\gamma\xi_1$.
\end{itemize}
Denote by $X$ the group orbit of heteroclinic connections $\kappa_j$:
$$X=\cup_{\sigma\in\Gamma}\sigma\biggl(\bigcup_{1\le j\le m}\kappa_j\biggr),$$
$\eta$ is the product $\eta=\prod_{3\le j\le m}\min(c_j/e_j,1-t_j/e_j)$, where
we set $\eta=1$ if $m=2$, and $\zeta=3$ (for $k=3$) or
$\zeta=2\beta_2/(\beta_1+\beta_2)$ (for $k=4$).
\begin{theorem}\label{thperorb}
\item[(a)] If $\eta\zeta c_1<e_1$ then there exist $\mu'>0$ and $\delta>0$, such that
for any $0<\mu<\mu'$ almost all trajectories escape from
$B_{\delta}(X)$ as $t\to\infty$.
\item[(b)] If $\eta\zeta c_1>e_1$ then generically there exists a periodic orbit
bifurcating from $X$ at $\mu=0$.
To be more precise, for any $\delta>0$ we can find $\mu(\delta)>0$
such that for all $0<\mu<\mu(\delta)$ the system
(\ref{eqal_ode}) possesses an asymptotically stable periodic
orbit that belongs to $B_{\delta}(X)$.
\end{theorem}

We give the proof only for $k=4$, for $k=3$ it can be obtained by a simple
modification combined with results of \cite{pc16}. Since it follows
closely the proof of theorem 2 {\it ibid}, some details are omitted and
the reader is referred to that paper. We first formulate lemma \ref{d3_30} below, describing properties of trajectories of a generic $\D_4$-equivariant systems in $\C$, which in the leading order is
\begin{equation}\label{sysA}
\dot z=\alpha z+\beta_1z^2\bar z+\beta_2\bar z^3.
\end{equation}
In polar coordinates, $z=r\re^{\ri\theta}$, it takes the form
\begin{equation}\label{pcor1}
\begin{array}{rcl}
\dot r&=&\alpha r+r^3(\beta_1+\beta_2\cos4\theta),\\
\dot\theta&=&-\beta_2 r^2\sin4\theta.
\end{array}
\end{equation}
We assume that
\begin{equation}\label{pcoraux}
\alpha>0,\ \beta_2>0\hbox{ and }\beta_1+\beta_2>0.
\end{equation}
The system has four invariant axes with $\theta=K\pi/4$, $K=0,1,2,3$.
The two axes with even $K$ are symmetric images of one another, as are the two axes with odd $K$. In case
$\beta_1-\beta_2<0$ there are four equilibria that are not at the origin with
$r^2=\alpha/(\beta_2-\beta_1)$ and $\theta=(2k+1)\pi/4$, $k=0,1,2,3$.
We consider the system in the sector $0\le\theta<\pi/4$, the complement part
of $\C$ is related to this sector by symmetries of the group $\D_4$.

Trajectories of the system satisfy
\begin{equation}\label{pcor2}
{\rd\tilde r\over\rd\theta}=-{2\alpha+2\tilde r(\beta_1+\beta_2\cos4\theta)
\over \beta_2\sin4\theta},
\end{equation}
where we have denoted $\tilde r=r^2$.
Re-writing this equation as
$${\rd\tilde r\over\rd\theta}+\tilde r{2(\beta_1+\beta_2\cos4\theta)
\over\beta_2\sin4\theta}=-{2\alpha\over\beta_2\sin4\theta},$$
multiplying it by
$s(\theta)=(\sin4\theta)^{(\beta_1+\beta_2)/2\beta_2}(1+\cos4\theta)^{-\beta_1/2\beta_2}$
and integrating, we obtain that
\begin{equation}\label{trajr}
r^2s(\theta)=-{2\alpha\over\beta_2}S(\theta)+C,\hbox{ where }
S(\theta)=\int_0^{\theta}{s(\theta)\over\sin4\theta}\rd\theta,
\end{equation}
which implies that
\begin{equation}\label{trajr1}
r^2s(\theta)+{2\alpha\over\beta_2}S(\theta)=
r^2_0s(\theta_0)+{2\alpha\over\beta_2}S(\theta_0)
\end{equation}
for the trajectory through the point $(r_0,\theta_0)$.

\begin{lemma}\label{d3_30}
Let $\tau(r_0,\theta_0)$ denote the time it takes the trajectory of the
system (\ref{pcor1}),(\ref{pcoraux}) starting at $(r_0,\theta_0)$ to reach
$r=1$ and $\vartheta(r_0,\theta_0)$ denote the value of $\theta$ at $r=1$. Then
\begin{itemize}
\item[(i)] $\tau(r_0,0)$ satisfies
$${\rm e}^{\alpha\tau(r_0,0)}={r_0+\alpha/(\beta_1+\beta_2)\over
r_0(1+\alpha/(\beta_1+\beta_2))}.$$
\item[(ii)] $\tau(r_0,\theta_0)$ satisfies
\begin{equation}\label{estr}
\tau(r_0,\theta_0)>\tau(r_0,0)\hbox{ for any }0<\theta_0<\pi/4.
\end{equation}
\item[(iii)] $\vartheta(r_0,\theta_0)$ satisfies
\begin{equation}\label{estt}
s(\vartheta(r_0,\theta_0))+
{2\alpha\over\beta_2}S(\vartheta(r_0,\theta_0))=
r_0^2s(\theta_0)+{2\alpha\over\beta_2}S(\theta_0).
\end{equation}
\item[(iv)] Given $C>0$, $\beta_1+\beta_2>0$ and $0<\theta_0<\pi/4$, for
sufficiently small $\alpha$ and $r_0$
$${\rm e}^{-C\tau(r_0,\theta_0)}\ll\vartheta(r_0,\theta_0).$$
\end{itemize}
\end{lemma}

The proof is similar to the proof of lemma 3(i-iv) in \cite{pc16} and is omitted.

\bigskip\noindent
{\bf Proof of the theorem}\\
As usual, we approximate trajectories in the vicinity of the cycle by
superposition of local and global maps, $\phi_j:\ H^{(in)}_j\to H^{(out)}_j$
and $\psi_j:\ H^{(out)}_j\to H^{(in)}_{j+1}$, respectively, where $H^{(in)}_j$ and $H^{(out)}_j$ are cross sections transversal to the incoming and outgoing connections at an equilibrium $\xi_j$. We consider
$g=\gamma\phi_1\psi_2\phi_2...\psi_m\phi_m\psi_1:\ H^{(out)}_1\to H^{(out)}_1$,
where the $\gamma$ is the symmetry in the definition of a building block.
Since the expanding eigenspace of $\xi_2$ is two-dimensional, the contracting
eigenspace of $\xi_1$ is two-dimensional as well. By the assumption of the theorem,
other equilibria in the cycle have one-dimensional expanding and contracting
eigenspaces. We employ the coordinates $(w_j,q_j)$ in $H^{(in)}_j$ and
$(v_j,q_j)$ in $H^{(out)}_j$, similarly to \cite{pc16}.
We also employ the coordinates $(\rho_1,\theta_1)$ and
$(\rho_2,\theta_2)$, in $H^{(out)}_1$ and $H^{(in)}_2$,
respectively, such that $v_1=\rho_1\cos\theta_1$, $q_1=\rho_1\sin\theta_1$,
$w_2=\rho_2\cos\theta_2$ and $q_2=\rho_2\sin\theta_2$.
In the leading order the maps $\phi_1$ is
$$(v_1^{(out)},q_1^{(out)})=\phi_1(w_1^{(in)},q_1^{(in)})=
(v_{0,1}(w_1^{(in)})^{c_1/e_1},q_1^{(in)}(w_1^{(in)})^{c_1/e_1}),$$
which in polar coordinates takes the form
\begin{equation}\label{phmap1}
(\rho_1^{(out)},\theta_1^{(out)})=\phi_1(w_1^{(in)},q_1^{(in)})=
(v_{0,1}(w_1^{(in)})^{c_1/e_1},\arctan(q_1^{(in)}/v_{0,1})).
\end{equation}
The maps $\phi_j$, $j=3,\ldots,m$, are
\begin{equation}\label{phmapj}
(v_j^{(out)},q_j^{(out)})=\phi_j(w_j^{(in)},q_j^{(in)})=
(v_{0,j}(w_j^{(in)})^{c_j/e_j},q_j(w_j^{(in)})^{-t_j/e_j}).
\end{equation}
(Here superscripts indicate coordinates in $H^{(in)}_1$ or $H^{(out)}_1$.
Below, where it does not create ambiguity, we do not use superscripts.)
In the leading order the map $\psi_1$ is
\begin{equation}\label{glmap1}
(\rho_2,\theta_2)=\psi_1(\rho_1,\theta_1)=(A\rho_1,\theta_1+\Theta),
\end{equation}
where generically $\Theta\ne N\pi/4$ for $N=1,...,8$.
The maps $\psi_j$, $j=2,...,m$, are
\begin{equation}\label{psi2}
(w_1,q_1)=\psi_j(v_2,q_2)=(B_{j,11}v_2+B_{j,12}q_2,B_{j,21}v_2+B_{j,22}q_2).
\end{equation}
Because of (i), for small $\mu$ the expanding eigenvalue of $\xi_2$ depends
linearly on $\mu$, therefore without restriction of generality we can assume that
$e_2=\mu$.
Generically, all other eigenvalues and coefficients in the expressions for local
and global maps do not vanish for sufficiently small $\mu$ and are of the order
of one. We assume them to be constants independent of $\mu$.  From (ii),
the eigenvalues satisfy $e_1>0$, $-c_1<0$ and $-c_2<0$.

For small enough $\tilde\delta$, in the scaled neighbourhoods
$B_{\tilde\delta}(\xi_2)$ the restriction of the system to the unstable
manifold of $\xi_2$ in the leading order is
$\dot z=\mu z+\beta_1 z^3+\beta_2\bar z^3$,
where we have denoted $z=w_2+\ri q_2$.
We assume that the local bases near $\xi_1$ and $\xi_2$ are chosen in such
a way that the heteroclinic connections $\gamma^{-1}\xi_m\to\xi_1$ and $\xi_2\to\xi_3$
go along the directions $\arg(\theta_j)=0$ for both $j=1,2$.
In the complement subspace the system is
approximated by the contractions $\dot u=-r_2 u$ and $\dot v=-c_2 v$.
In terms of the functions $\tau(r,\theta)$ and $\vartheta(r,\theta)$ introduced
in lemma \ref{d3_30}, the map $\phi_2$ is
$$(v_2,q_2)=\phi_2(\rho_2,\theta_2)=
(v_{0,2}\re^{-c_2\tau(\rho_2,\theta_2)},\sin\vartheta(\rho_2,\theta_2)).$$

According to lemma \ref{d3_30}(iv), for small $\rho_2$ and $\mu$
$$\re^{-c_2\tau(\rho_2,\theta_2)}\ll\sin\vartheta(\rho_2,\theta_2),$$
which implies that the superposition $\psi^*=\psi_3...\psi_m\phi_m\psi_1$ can
be approximated as $\psi^*(v_2,q_2)\approx(B_{1,*}q_2^\eta,B_{2,*}q_2^\eta)$,
where $\eta=\prod_{3\le j\le m}\min(c_j/e_j,1-t_j/e_j)$ and the constants
$B_{1,*}$ and $B_{2,*}$ depend on $B_{j,kl}$, $2\le j\le m$, and eigenvalues
of $\rd f(\xi_j)$, $3\le j\le m$.
For small $\theta_1$ we have $\sin\theta_1\approx\tan\theta_1\approx\theta_1$.
Taking into account (\ref{phmap1}), (\ref{glmap1})
and lemma \ref{d3_30}(iii), we obtain that
\begin{equation}\label{mapg}
g(\rho_1,\theta_1)\approx
\biggl(C_1(\rho_1^2A\beta_2 s(\Theta)+\mu S(\Theta))^{\eta\zeta c_1/2e_1},
C_2(\rho_1^2A\beta_2 s(\Theta)+\mu S(\Theta))^{\eta\zeta/2}\biggr),
\end{equation}
where we have denoted $\zeta=2\beta_2/(\beta_1+\beta_2)$,
$C_1=v_{0,1}\beta_2^{-\eta\zeta c_1/2e_1}|B_{1,*}|^{c_1/e_1}$ and
$C_2=v_{0,1}^{-1}4^{-1}\beta_2^{-\eta\zeta/2}B_{2,*}$.

\medskip
(a) From (\ref{mapg}), the $\rho$-component of $g$ satisfies
$$g_{\rho}(\rho_1,\theta_1)> C_3\rho_1^{\eta\zeta c_1/e_1},
\hbox{ where }C_3=C_1(A\beta_2 s(\Theta))^{\eta\zeta c_1/2e_1},$$
hence if $\eta\zeta c_1<e_1$ then for any $0<\delta<C_3^{e_1/(e_1-c_1\eta\zeta)}$
the iterates $g^n(\rho_1,\theta_1)$ with initial $0<\rho_1<\delta$ satisfy
$g_{\rho}^n(\rho_1,\theta_1)>\delta$ for sufficiently large $n$.

\medskip
(b) Assume that $\eta\zeta c_1>e_1$. Existence and stability of a fixed
point of the map $g$ (\ref{mapg}) for small $\mu$ can be proven by
the same arguments as employed to prove theorem 2(b) in \cite{pc16}.
We omit the proof. The fixed point can be approximated by
$(\rho_p,\theta_p)=(C_1(\mu S(\Theta))^{\eta\zeta c_1/2e_1},
C_2\mu S(\Theta)^{\eta\zeta/2})$.
This fixed point is an intersection
of a periodic orbit with $H^{(out)}_1$. The distance from $(\rho_p,\theta_p)$
to $X$ depends on $\mu$ as $\mu^{c_1\eta\zeta/2e_1}$, therefore the trajectory approaches
$X$ as $\mu\to0$.
\qed

\subsection{The case $\D_k$, $k\ge5$}

In this subsection we prove that a bifurcation of a periodic orbit,
that was discussed in the previous subsection, does not take place for $k\ge5$:

\begin{theorem}\label{noorbit}
Suppose that for $0<\mu<\mu_0$ the system (\ref{eqal_ode}) possesses a
pseudo-simple heteroclinic cycle $X=\xi_1\to\ldots\to\xi_M$, where
$\xi_2$ has a two-dimensional expanding eigenspace with the associated
eigenvalue $e_2=\mu$ and the symmetry group $\Delta_2=\D_k$, $k\ge5$, acting
naturally on the expanding eigenspace.
There exist $\varepsilon>0$ and $\mu'>0$, such that for any $0<\mu<\mu'$ almost all
trajectories $\Phi(x,t)$ of the system (\ref{eqal_ode}), such that
$d(\Phi(x_0,0),X)<\varepsilon$, satisfy $d(\Phi(x_0,t_0),X)>\varepsilon$ for some
$t_0>0$. By $d(\cdot,\cdot)$ we denoted the distance between a point and a set.
\end{theorem}

\proof
Similarly to the proof of theorem 1 in \cite{pc16}, we consider the map
$\phi_2\psi_1\phi_1:\,H^{(in)}_1\to H^{(out)}_2$ and prove
existence of $\varepsilon>0$ such that
\begin{equation}\label{nonincl}
\phi_2\psi_1\phi_1(H^{(in)}_1(\varepsilon))\cap H^{(out)}_2(\varepsilon)=\varnothing,
\end{equation}
where
$$H^{(in)}_1(\varepsilon)=\{(w,q)\in H^{(in)}_1~:~|(w,q)|<\varepsilon\}
\hbox{ and }H^{(out)}_2(\varepsilon)=
\{(v,q)\in H^{(out)}_2~:~|(v,q)|<\varepsilon\}.$$
Equation (\ref{nonincl}) shows that all points in $H^{(in)}_1(\varepsilon)$
are mapped outside $H^{(out)}_2(\varepsilon)$, which implies the statement of
the theorem.

The maps $\phi_1$ and $\psi_1$ are the same as for the $\D_4$ system, they are
given by (\ref{phmap1}) and (\ref{glmap1}), respectively. In (\ref{glmap1})
generically $\Theta\ne N\pi/k$ for $N=1,2,...,2k$. Moreover,
there exist $\Theta'>0$ and $\mu'>0$, such that
$\min_{1\le N\le 2k} |\Theta- N\pi/k|>\Theta'$ for
all sufficiently small $\delta$  and $0<\mu<\mu'$ (recall that $\delta$ is the distance from
$H^{(in)}_2$ and $H^{(out)}_2$ to $\xi_2$).

For small enough $\delta$, in a $\delta$-neighbourhoods of $\xi_2$ the
restriction of the system to the unstable manifold of $\xi_2$ in the leading
order is $\dot z=\mu z+\beta_1 z^3+\beta_2\bar z^{k-1}$,
where $z=w_2+\ri q_2$. In polar coordinates the system takes the
form $\dot r=\alpha r+r^3\beta_1$, $\dot\theta=-\beta_2 r^{k-1}\sin k\theta$,
which implies that the map
$\phi_2(\rho_2^{in},\theta_2^{in})=(v_2^{out},q_2^{out})$ satisfies
$q_2^{out}=\delta\tan\theta_2^{out}$ and
\begin{equation}\label{phmap2n}
|\theta_2^{out}-\theta_2^{in}|<\int_{\rho_2^{in}}^{\delta}
{|\beta_2|\over|\beta_1|}r^{k-5}\rd r={|\beta_2|\over|\beta_1|(k-4)}
(\delta^{k-4}-(\rho_2^{in})^{k-4}).
\end{equation}

We choose $0<\delta<\Theta'/4$ and set
\begin{equation}\label{setde}
0<\varepsilon<\min\biggl(\delta\tan{\Theta'\over4},~v_{0,1}\tan{\Theta'\over4}\biggr).
\end{equation}
Any $(w_1,q_1)\in H^{(in)}_1(\varepsilon)$ satisfies $q_1<\varepsilon$,
therefore (\ref{phmap1}) and (\ref{setde}) imply that $\theta_1<\Theta'/4$.
Hence, due to (\ref{glmap1}),(\ref{phmap2n}) and (\ref{setde}), $|\theta_2- N\pi/k|>\Theta'/4$ for any
$N$. The steady state $\xi_2$ has $k$ symmetric copies (under the action of
symmetries $\sigma\in\Sigma_2$) of the heteroclinic connection
$\kappa_2:\xi_2\to\xi_3$ which belong to the hyperplanes
$\theta_2=N\pi/k$ with some integer $N$'s. Due to (\ref{phmap2n}) and (\ref{setde}),
the distance of $(v_2,q_2)$ to any of these hyperplanes is larger than
$\delta\tan(\Theta'/4)$, which implies (\ref{nonincl}).
\qed

\section{Example: periodic orbit near a heteroclinic cycle in
a $(\D_4\rl\D_2;\D_4\rl\D_2)$-equivariant system}\label{sec8n}

We proved (see section \ref{sec6n}) that an attracting periodic orbit can
exist near a pseudo-simple heteroclinic cycle if the isotropy subgroup of one
of its equilibria is $\D_3$ or $\D_4$. For the case when the isotropy subgroup
is $\D_3$, examples of $\Gamma$-equivariant systems possessing periodic orbit in
a neighbourhood of a heteroclinic cycle were given in \cite{pc16} for
$\Gamma=(\D_3\rl\Z_1;\D_3\rl\Z_1)$ and $\Gamma=(\D_3\rl\Z_2;\mO\rl\V)$.
The vector fields considered {\it ibid} were third order normal forms commuting
with the considered actions of $\Gamma$.
Here we present a numerical example of a heteroclinic cycle with a nearby
attracting periodic orbit, where the isotropy subgroup of an equilibrium is
$\D_4$ and a $\Gamma$-equivariant vector field is constructed using ideas
employed in the proofs of lemma \ref{lem1} and theorem \ref{thperorb}.

We consider a $\Gamma$-equivariant dynamical system where
$\Gamma=(\D_4\rl\D_2;\D_4\rl\D_2)$ (recall that the quaternionic group $\D_2$
is usually denoted by $\V$). The elements of $\Gamma$ are:
\begin{eqnarray*}\label{listex1}
(\V&;&\V),\\
((1,0,0,\pm1)/\sqrt{2}&;&(\pm1,0,0,\pm1)/\sqrt{2}),\\
((1,0,0,\pm1)/\sqrt{2}&;&(0,\pm1,\pm1,0)/\sqrt{2}),\\
((0,1,\pm1,0)/\sqrt{2}&;&(\pm1,0,0,\pm1)/\sqrt{2}),\\
((0,1,\pm1,0)/\sqrt{2}&;&(0,\pm1,\pm1,0)/\sqrt{2}).
\end{eqnarray*}
The group has five isotropy types of subgroups $\Sigma$ satisfying
$\dim\Fix\Sigma=2$ (see appendix A). In agreement with appendix C,
we take $\Sigma_1=<\kappa_1>\cong\Z_4$ and
$\Sigma_2=<\kappa_5>\cong\Z_2$. For convenience, we use different notation for
generating elements. Namely, we write
$\Sigma_1=<\gamma_1>$ and $\Sigma_2=<\gamma_2>$, where
$$\gamma_1(s)=((1,0,0,1)/\sqrt{2};(1,0,0,(-1)^s)/\sqrt{2})\hbox{ and }$$
$$\gamma_2(q,r,t)=((0,1,(-1)^q,0)/\sqrt{2};(0,(-1)^r,(-1)^t,0)/\sqrt{2}).$$
The action $(\bl;\br):\bx\to\bl\bx\br^{-1}$ on $\R^4$ of (some) elements of
$\Gamma$ is
\begin{equation}\label{action}
\begin{array}{r|l}
(\bl;\br)&\bx\to\bl\bx\br^{-1}\\
\hline\\
((0,0,0,1);(0,0,0,1)) & \bx\to(x_1,-x_2,-x_3,x_4)\\
((0,0,1,0);(0,0,1,0)) & \bx\to(x_3,x_4,-x_1,-x_2)\\
((1,0,0,1)/\sqrt{2};(1,0,0,1)/\sqrt{2}) & \bx\to(x_1,x_3,-x_2,x_4)\\
((0,1,1,0)/\sqrt{2};(0,1,1,0)/\sqrt{2}) & \bx\to(x_1,x_3,x_2,-x_4).
\end{array}\end{equation}

The isotropy planes can be labelled as follows:
\begin{eqnarray*}\label{listex12}
P_1(s)=\Fix\Sigma_1(s),&\hbox{ where }\Sigma_1(s)=<\gamma_1(s)>,\\
P_2(q,r,t)=\Fix\Sigma_2(q,r,t),&
\hbox{ where }\Sigma_2(q,r,t)=<\gamma_2(q,r,t)>,
\end{eqnarray*}
hence there exist two different planes $P_1$ with $s=0,1$ and eight different
planes $P_2$ corresponding to $q,r,t=0,1$.
A plane $P_1$ contain four symmetry axes of two isotropy types with
isotropy groups of the axes isomorphic to $\D_4$.
An axis is an intersection of $P_1$ with two planes $P_2$ (and also with two
other planes fixed by $\kappa_4$, that is irrelevant), namely
$P_1(s)$ intersects with $P_2(0,r,t)$ and $P_2(1,r+s,t+s+1)$.
The axes split into two isotropy classes, with odd or even
$s+r+t$. A plane $P_2$ contains two isotropy axes which are intersections
with $P_1(0)$ and $P_1(1)$.

\begin{figure}[h]
%[p]

\vspace*{-62mm}
\hspace*{-10mm}\includegraphics[width=10cm]{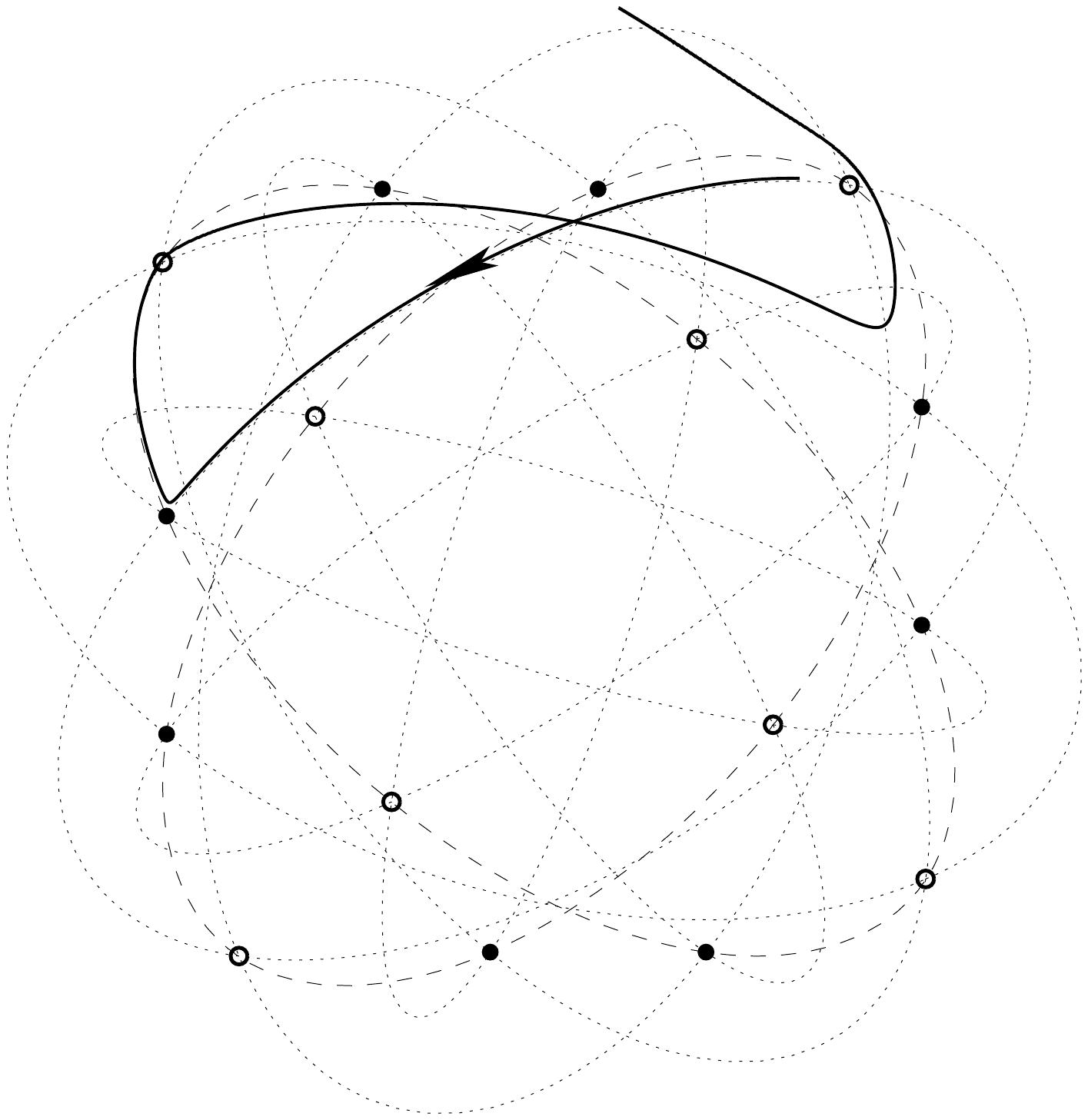}\hspace*{-20mm}\includegraphics[width=112mm]{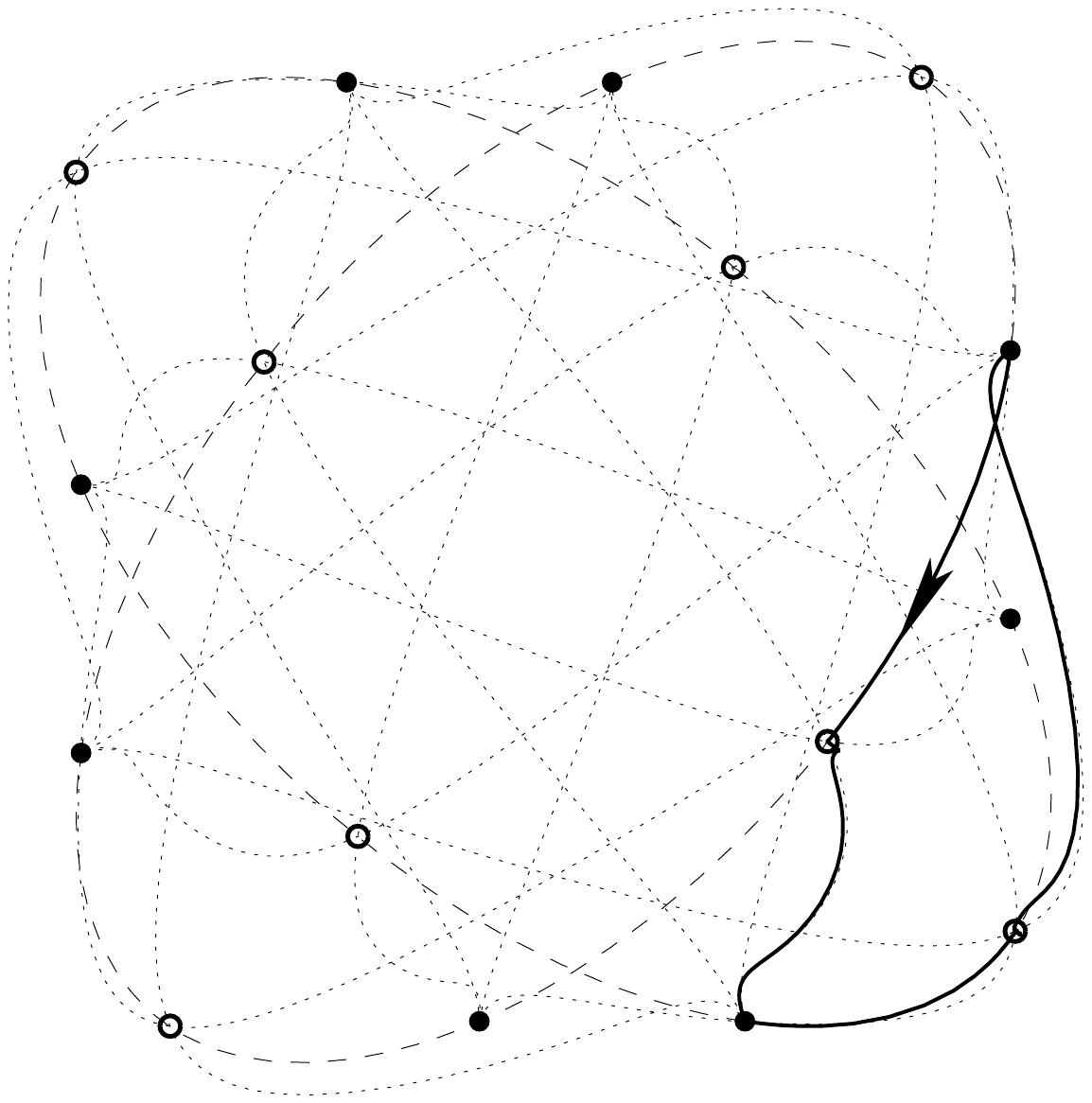}

\vspace*{-17mm}
\hspace*{60mm}{\large (a)}\hspace*{70mm}{\large (b)}

\vspace*{3mm}
\noindent
\caption{Projection of the heteroclinic connections $\xi_2\to \xi_1$
(dashed lines), $\xi_1\to \xi_2$ (dotted lines),
a trajectory of the system $\dot\bx=f(\bx))$ (a) and a periodic orbit
of the system $\dot\bx=f^*(\bx))$ (b) (solid lines) into the plane
$<{\bf v}_1,{\bf v}_2>$, where ${\bf v}_1=(4,2,4,1.5)$ and
${\bf v}_2=(2,4,-1.5,4)$, (a). The steady state $\xi_1$ is denoted by
a hollow circle and $\xi_2$ by filled one.}
\label{figs9}\end{figure}

We choose $P_1(0)=(x_1,0,0,x_4)$, $P_2(0,0,0)=(x_1,x_2,x_2,0)$ and,
in agreement with (\ref{fieh}), set
\begin{equation}\label{fieh12}
{\bf h}_1(r_1,\theta_1)=(r_1(1-r_1),\sin(4\theta_1),\
{\bf h}_2(r_2,\theta_2)=(r_2(1-r_2),-\sin(2\theta_1),
\end{equation}
where $x_1=r_1\cos\theta_1$ and $x_4=r_1\sin\theta_1$ in $P_1$ and
$x_1=r_2\cos\theta_2$ and $x_2=(r_2\sin\theta_2)/\sqrt{2}$ in $P_2$.
Hence, $\xi_1\approx(1/\sqrt{2},0,0,-1/\sqrt{2})\in P_1(0)\cap P_2(0,1,0)$ is unstable in
$P_1$ and stable in $P_2$;
$\xi_2\approx(1,0,0,0)\in P_1(0)\cap P_2(0,0,0)$ is stable in
$P_1$ and unstable in $P_2$. Following the proof of lemma \ref{lem1},
we construct the system (\ref{gj})-(\ref{f-system})
that possesses a heteroclinic cycle with a building block
$\xi_1\to\xi_2\to\gamma\xi_1$, where $\gamma=((1,0,0,0);(0,0,1,0))$. In agreement
with theorem 1 in \cite{pc16}, the cycle is not asymptotically stable, hence
trajectories starting near the cycle escape from it (see fig. \ref{figs9}(a)).

Theorem \ref{thperorb} states that a periodic orbit exists near a heteroclinic
cycle with $\Delta_2\cong\D_4$ if the multiple expanding eigenvalue $e_2$ is
sufficiently small and $2c_1\beta_2/(\beta_1+\beta_2)>e_1$ (recall that $\alpha$,
$\beta_1$ and $\beta_2$ are the coefficients of the system (\ref{sysA})\,).
To be more precise, in the proof we use the fact that the ratio
$\alpha/\beta_2$ is small. Therefore, we introduce a modified system
\begin{equation}\label{fm-system}
\dot{\bf x}={\bf f}^*({\bf x}),\hbox{ where }
{\bf f}^*({\bf x})={\bf f}({\bf x})+
\sum\limits_{\gamma^*\in G_2} \gamma^*{\bf g}^*((\gamma^*)^{-1}{\bf x}),
\end{equation}
\begin{equation}\label{g*}
{\bf g}^*({\bf y})=(0,0,{c y_3^3\over 1+B|\pi^{\perp}{\bf y}|^2},
{-b y_4y_3^2\over 1+B|\pi^{\perp}{\bf y}|^2}),
\end{equation}
$G_2=\Gamma/N_{\Gamma}(\Sigma_2(0,0,0,0))$ and
$\by=(x_1,x_2,(x_3+x_4)/\sqrt(2),(x_3-x_4)/\sqrt(2))$.

In a small neighbourhood of $\xi_2$ the projection of the local field
(\ref{fm-system}) into the plane $x_1=x_2=0$ is
$$\dot y_3=ay_3+cy_3^3-by_3y_4^2,\quad \dot y_4=ay_4+cy_4^3-by_4y_3^2.$$
Comparing the above expression with (\ref{sysA}), we obtain that
$$\beta_1=(c-3b)/2\hbox{ and }\beta_2=(c+b)/4.$$
If the coefficient $B$ in (\ref{g*}) is sufficiently large, then by the same
arguments as applied in the proof in lemma \ref{lem1}, the system
(\ref{fm-system}) possesses the heteroclinic cycle $\xi_1\to\xi_2\to\gamma\xi_1$.
Theorem \ref{thperorb} indicates that for sufficiently large $b\gg c>0$ there
exists a stable periodic orbit close to the cycle. Therefore, we set
\begin{equation}\label{parr}
B=100,\ b=1000,\ c=0.1\ .
\end{equation}
In agreement with our arguments, the system (\ref{fm-system})-(\ref{parr})
has an attracting periodic orbit near the heteroclinic cycle, as
shown on fig. \ref{figs9}(b).

\section{An example of stability when $\Gamma\not\subset SO(4)$}\label{sec6}

In this section we show that a family of subgroups $\Gamma\subset O(4)$,
$\Gamma\not\subset SO(4)$, admits heteroclinic cycles involving
multidimensional heteroclinic orbits. Following \cite{cgl1999}, we call such
heteroclinic cycles {\it generalized}. We derive conditions for
asymptotic stability of such generalized cycle and show that it involves
as a subset a pseudo-simple heteroclinic cycle, that can be fragmentarily
asymptotically stable. Numerical studies indicate that addition
of small perturbation that breaks an $O(4)$ symmetry can result on
emergence of asymptotically stable periodic orbit or on chaotic
dynamics in the vicinity of a pseudo-simple heteroclinic cycle.

We shall in fact consider a class of subgroups of $O(4)$ defined as follows. \\
Let $(x_1,y_1,x_2,y_2)\in\R^4$ and $z_j=x_j+iy_j$. Fix an integer $n\geq 3$ and let $\Gamma$ be the group generated by the transformations
\begin{equation} \label{gen:Gamma}
\rho:~(z_1,z_2)\mapsto (z_1,e^{\frac{2\pi i}{n}}z_2),~~\kappa(z_1,z_2)\mapsto (\bar z_1,z_2),~~\sigma:~(z_1,z_2)\mapsto (z_1,\bar z_2)
\end{equation}
(Choosing coordinates $z_1=q_1+iq_4$ and $z_2=q_2+iq_3$, we obtain that
in quaternionic presentention the $SO(4)$ subgroup of $\Gamma$ is
$(\D_n\rl\Z_1;\D_n\rl\Z_1)$, in agreement with theorem \ref{th2}.)
This group action decomposes $\R^4$ into the direct sum of three irreducible
representations of the dihedral group $\Gamma=\D_n\times\Z_2$: \\
(i) the trivial representation acting on the component $x_1$, \\
(ii) the one-dimensional representation acting on $y_1$ by $\kappa y_1=-y_1$, \\
(iii) the two-dimensional natural representation of $\D_n$ acting on $z_2=(x_2,y_2)$. \\
There are four types of fixed-point subspaces for this action:
\begin{itemize}
\item $L=P_1\cap P_2=Fix(\Gamma)$,
\item $P_1=\{(x_1,y_1,0,0)\}=Fix(\rho,\sigma)$,
\item $P_2=\{(x_1,0,x_2,0)\}=Fix(\kappa,\sigma)$,
\item $V=\{(x_1,y_1,x_2,0)\}=Fix(\sigma)$,
\item $W=\{(x_1,0,x_2,y_2)\}=Fix(\kappa)$.
\end{itemize}
When $n$ is even there are two more types of invariant subspaces:
\begin{itemize}
\item $P'_2=\{(x_1,0,x_2\cos(\pi/n),x_2\sin(\pi/n))\}=Fix(\kappa,\rho\sigma)$,
\item $V'=\{(x_1,y_1,x_2\cos(\pi/n),x_2\sin(\pi/n)))\}=Fix(\rho\sigma)$.
\end{itemize}
Note that $P_1$ is fixed by $\Gamma$. When $n$ is odd $P_2$ and $V$ have $n-1$ symmetric copies $\rho^j P_2$, $\rho^j V$, $j=1,\dots,n-1$. When $n$ is even each of $P_2$, $P'_2$, $V$, $V'$ has $n/2-1$ symmetric copies. \\
It can be shown that for an open set of $\Gamma$-equivariant vector fields, there exists an equilibrium $\xi_1$ on the negative semi-axis in $L$, an equilibrium $\xi_2$ on the positive semi-axis, and heteroclinic orbits lying in the planes $P_1$ and $P_2$ and realizing a cycle between $\xi_1$ and $\xi_2$. Moreover this cycle is pseudo-simple due to the action of the rotation $\rho$ on the plane $P_2$, which forces the eigenvalues along the $x_2$ direction in $P_2$ to be double. To fix ideas we assume the double eigenvalue is stable at $\xi_1$ and unstable at $\xi_2$. In order to study the stability of this pseudo-simple cycle we shall exploit a property that was observed in the case $n=3$ in \cite{pc16} and appears to also occur when $n>3$. First, the two dimensional unstable manifold at $\xi_2$ lies entirely in the invariant subspace $W$, which also contains the axis $L$. Second, for an open set of vector fields any orbit on this unstable manifold lies in the stable manifold of $\xi_1$, hence realizing a two dimensional manifold of saddle-sink connections in $W$. Therefore the pseudo-simple heteroclinic cycle is part of a cycle involving multidimensional heteroclinic orbits, which was called a generalized heteroclinic cycle in \cite{cgl1999}. Let us prove this claim.

\begin{proposition}\label{prop:existence gen cycle}
There exists an open set ${\cal V}$ of $\Gamma$-equivariant smooth vector fields which possess a generalized heteroclinic cycle. This cycle, which we denote by ${\cal X}$, connects two equilibria $\xi_1$ and $\xi_2$ which lie on the negative, resp. positive semi axis in $L$. It is composed of a single heteroclinic orbit in $P_1$ and a two dimensional manifold of heteroclinic orbits in the space $W$.
This manifold in $W$ contains heteroclinic orbits in $P_2$ and in $P'_2$ (when $n$ is even), which realize two isotropy types of pseudo-simple heteroclinic cycles.
\end{proposition}
\proof Let us consider the group $\Gamma_\infty$ defined by relations \eqref{gen:Gamma} where we replace the transformation $\rho$ by $\rho_\varphi~(z_1,z_2)\mapsto (z_1,e^{i\varphi}z_2)$, $\varphi\in S^1$. This group has the same invariant subspaces as $\Gamma$, but in addition any copy of the plane $P_2$ by $\rho_\varphi$ is also invariant, and moreover $W$ is spanned by letting $\rho_\varphi$ rotate $P_2$ with any $\varphi\in S^1$. Therefore if a saddle-sink connection between equilibria $\xi_1$, $\xi_2$ lying on $L$ exists in $P_2$, then a two dimensional manifold of connections exists in $W$. The fact that such equilibria and connections exist for an open set of smooth vector fields follows from a slight adaptation of lemma \ref{lem1}, which shows that the group $\Gamma_\infty$ {\em admits} robust heteroclinic cycles with connections in $P_1$ and $P_2$.
Since any $\Gamma$ equivariant perturbation of this vector field leaves $W$, $P_1$ and $P_2$ invariant, we conclude by structural stability that generalized heteroclinic cycles persist for an open set of $\Gamma$ equivariant smooth vector fields. The same argument applies if replacing $P_2$ by $P'_2$ when $n$ is even.
\qed

We denote by $e_j>0$ and $-c_j<0$ the non-radial eigenvalues at $\xi_j$, $j=1,2$, and further assume that $-c_1$ and $e_2$ are the double eigenvalues. Hence $\xi_2$ is a source while $\xi_1$ is a sink in $W$, along the eigendirections $(x_2,y_2)$.

\begin{figure}[ht!]
\centerline{\includegraphics[width=12cm]{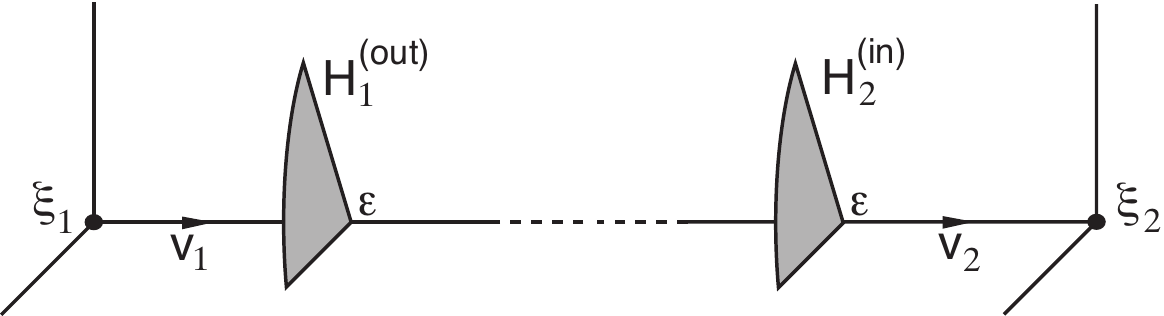}}
\caption{Cross-sections to the heteroclinic orbit $\xi_1\to\xi_2$.}
\label{fig:crosssec1}
\end{figure}

\begin{theorem}\label{as}
The generalized heteroclinic cycle ${\cal X}$ defined in Proposition \ref{prop:existence gen cycle} is asymptotically stable if $c_1c_2>e_1e_2$ and is completely unstable if $c_1c_2<e_1e_2$. Moreover there exists an open subset of ${\cal V}$ such that for any vector field in this subset, a pseudo-simple heteroclinic subcycle of ${\cal X}$ is fragmentarily asymptotically stable.
\end{theorem}
\proof
As usual we want to define a first return map in the vicinity of the heteroclinic cycle, and to do so we decompose the dynamics close to ${\cal X}$ into local maps and global transition maps between suitably chosen cross-sections to the heteroclinic orbits near the equilibria. Possibly after a smooth $\Gamma$-equivariant change of coordinates we can always assume that in a neighborhood of the equilibria their stable and unstable manifolds are linear. Let $v_j$, resp. $r_je^{i\theta_j}$ denote the local coordinates near $\xi_j$  along $y_1$, resp. $z_2$. The "radial" direction (along the axis $L$, coordinate $x_1$) can be neglected.
We define the cross-sections along the (single) heteroclinic orbit from $\xi_1$ to $\xi_2$ (Fig. \ref{fig:crosssec1}) by
\begin{equation}\label{cross-sections1}
\begin{array}{l}
H_1^{out}=\{(v_1=\varepsilon, r_1>0,\theta_1<\pi/n)\} \\
H_2^{in}=\{(v_2=\varepsilon, r_2>0,\theta_2<\pi/n)\}
\end{array}
\end{equation}
where $\varepsilon>0$ is a small constant value.

\begin{figure}[ht!]
\centerline{\includegraphics[width=8cm]{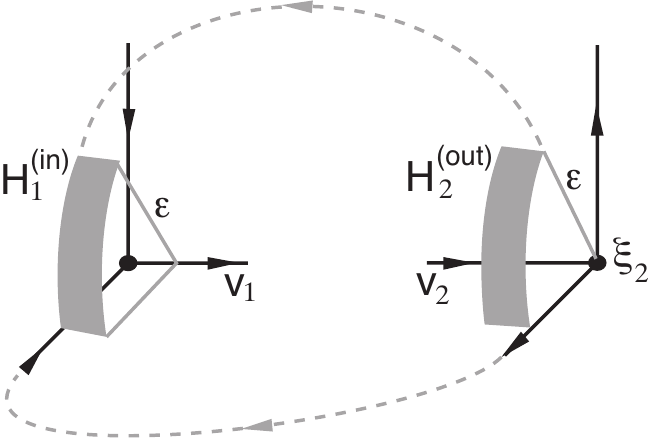}}
\caption{Cross-sections to the heteroclinic manifold $\xi_2\to\xi_1$.}
\label{fig:crosssec2}
\end{figure}

Similarly we define the cross-sections along the two-dimensional manifold of connections from $\xi_2$ to $\xi_1$ by (see Fig. \ref{fig:crosssec2}):
\begin{equation}\label{cross-sections2}
\begin{array}{l}
H_1^{in}=\{(v_1>0, r_1=\varepsilon,\theta_1<\pi/n)\} \\
H_2^{out}=\{(v_2>0, r_2=\varepsilon,\theta_2<\pi/n)\}
\end{array}
\end{equation}

The boundaries of the cross-sections at the limit values $\theta_j=0$ ($j=1,2$) lie in the space $V$ while at $\theta_j=\pi/n$ they lie in the space $\rho V$ (when $n$ is odd) or $V'$ (when $n$ is even). Since these spaces are flow-invariant, the sections defined above are mapped to each other by the flow in the order $H_1^{in}\to H_1^{out}\to H_2^{in}\to H_2^{out}\to H_1^{in}$. We can therefore define the local first hit maps $\Phi_j~:~H_j^{(in)}\rightarrow H_j^{(out)}$ and global maps $\Psi_j~:~H_j^{(out)}\rightarrow H_j^{(in)}$, $j=1,2$. \\
By choosing $\varepsilon$ small enough and if non resonance conditions are satisfied between the eigenvalues at each equilibrium, we can approximate the local vector fields by their linear parts. Therefore near $\xi_1$ the flow is defined by the equations
$$
\frac{d(r_1e^{i\theta_1})}{dt}=-c_1r_1e^{i\theta_1}~\mbox{ and }~\frac{dv_1}{dt}=e_1v_1~,
$$
which gives
\begin{equation}\label{eq:phi1}
(r'_1,\theta'_1)=\Phi_1(v_1,\theta_1)=(v_1^{c_1/e_1},\theta_1)
\end{equation}
and near $\xi_2$ the flow is defined by
$$
\frac{dv_2}{dt}=-c_2v_2~\mbox{ and }~\frac{d(r_2e^{i\theta_2})}{dt}=e_2r_2e^{i\theta_2}~,
$$
which gives
\begin{equation}\label{eq:phi2}
(v'_2,\theta'_2)=\Phi_2(r_2,\theta_2)=(r_2^{c_2/e_2},\theta_2)
\end{equation}
The far map $\Psi_1$ is a $\Gamma$-equivariant near identity diffeomorphism which can be linearized under generic conditions. We therefore have
\begin{equation}\label{eq:psi1}
\Psi_1(r_1,\theta_1)=(ar_1,\theta_1)
\end{equation}
where $a$ is a positive constant. \\
The far map $\Psi_2$ is also $\Gamma$-equivariant, however it is not near identity and it can't be expressed as simply as $\Psi_1$. Let us set
\begin{equation}\label{eq:psi2}
\Psi_2(v_2,\theta_2)=(v(v_2,\theta_2),\theta(v_2,\theta_2))
\end{equation}
The component $v(v_2,\theta_2)$ vanishes when $v_2=0$, hence there exists a smooth function $h$ such that $v'_1(v_2,\theta_2)=v_2h(v_2,\theta_2)$. Moreover because $\Psi_2$ is a diffeomorphism $h(0,\theta_2)\neq 0$, which allows by a smooth change of variables to set $v'_1(v_2,\theta_2)=v_2b(\theta_2)$ where $b$ is a bounded function. The map $k(\theta_2):=\theta(0,\theta_2)$ is differentiably defined in the interval $[0,\pi/n]$ and has fixed-points at $0$ and $\pi/n$.

Now we can define the first return map in $H_1^{in}$ by
$g=\Psi_2\circ\Phi_2\circ\Psi_1\circ\Phi_1$ and we write $(v'_1,\theta'_1)=g(v_1,\theta_1)$. \\
Applying the above expressions for $\Phi_j$ and $\Psi_j$ one obtains
\begin{equation}\label{eq:v'1}
v'_1 = b(\theta_1)a^{\frac{c_2}{e_2}}v_1^{\frac{c_1c_2}{e_1e_2}}
\end{equation}
Since $b$ is a bounded function the iterates of the first component of $g$ tend to 0 if and only if $c_1c_2>e_1e_2$. This proves the first part of the theorem. \\
The second component of $g$ has the form
\begin{equation}\label{eq:theta'1}
\theta'_1 = \theta(a^{\frac{c_2}{e_2}}v_1^{\frac{c_1c_2}{e_1e_2}},\theta_1)
\end{equation}
Assume $c_1c_2>e_1e_2$, then by iteration the first argument of the function $\theta$ tends to 0. Therefore the dynamics of $\theta$ converges to the dynamics of the map $k$. By an argument similar to Prop. 4.9 of \cite{km95a}, $k$ has generically hyperbolic fixed points at $0$ and $\pi/n$. Moreover there exists an open subset of ${\cal V}$ such that for vector fields in this subset, $k$ has no fixed point inside $(0,\pi/n)$. In this case we can conclude that the iterates of $g$ converge to a pseudo-simple heteroclinic cycle.
\qed

In order to illustrate this result we built a $\D_n$ equivariant polynomial system with $n>2$ satisfying the hypotheses of the theorem and performed numerical simulations. We use bifurcation method to find the equilibria and corresponding heteroclinic orbits.
Applying classical methods in computing equivariant bifurcation systems
\cite{GSS} we construct
\begin{equation}\label{eq:Dnequivariant}
\begin{array}{l}
\dot z_1= a_1z_1+a_2\bar z_1+a_3z_1^2+a_4\bar z_1^2+a_5z_1\bar z_1+
a_6z_2\bar z_2+a_7z_1^2\bar z_1+a_8z_1z_2\bar z_2+a_9(z_2^n +\bar z_2^n) \\
\dot z_2=z_2[b_1+b_2(z_1+\bar z_1)+b_3z_1\bar z_1+b_4z_2\bar z_2]+b_5\bar z_2^{n-1},
\end{array}
\end{equation}
where $a_1$, $a_2$ and $b_1$ are small parameters. Suitable coefficient values for the system to possess generalized heteroclinic cycles can be found as was done in \cite{pc16} in the $n=3$ case.

\begin{figure}[ht!]
	\centering
	\mbox{\subfigure[Case $n=3$]{\includegraphics[width=8cm]{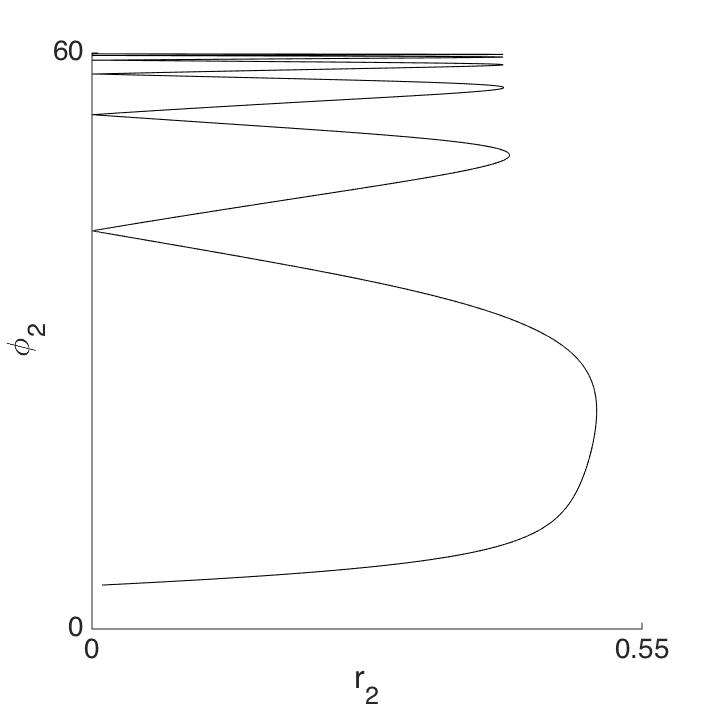}}\quad
      \subfigure[Case $n=5$]{\includegraphics[width=8cm]{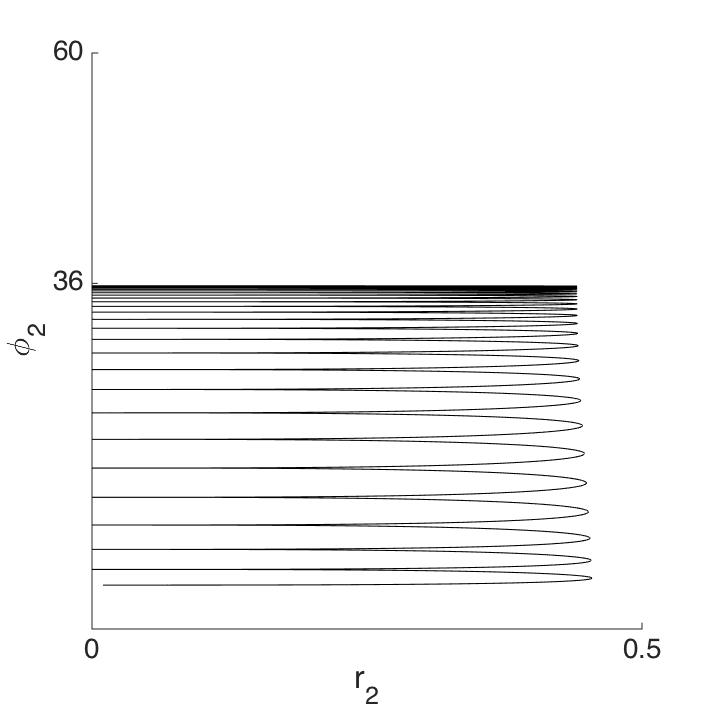}}}
	\caption{\small{Dynamics of the $z_2=r_2e^{i\phi_2}$ variable in polar coordinates. Horizontal axis: $r_2$, vertical axis: $\phi_2$ (in degrees).}}
	\label{fig:simus}
\end{figure}

We additionally assume that $a_3+a_4+a_5$ is close to 0 in order to ensure supercritical bifurcation of two equilibria on the $x_1$ axis. There is no loss of generality to take this sum equal to 0, so that the bifurcation is a pitchfork. Moreover in this bifurcation context it is suitable to take negative cubic coefficients in both equations, in order to keep the dynamics bounded. We normalize these coefficients to $-1$.
Then the bifurcated equilibria are $\xi_1=-\sqrt{a_1+a_2}$ and $\xi_2=+\sqrt{a_1+a_2}$. The non radial eigenvalues at $\xi_1$ and $\xi_2$ are
\begin{equation}\label{evs}
\begin{array}{l}
e_1=2(a_1-(a_3-a_4)\sqrt{a_1+a_2}),~ -c_1=b_1-2b_2\sqrt{a_1+a_2}-a_1-a_2 \\
e_2=b_1+2b_2\sqrt{a_1+a_2}-a_1-a_2,~ -c_2=2(a_1+(a_3-a_4)\sqrt{a_1+a_2})
\end{array}
\end{equation}
The heteroclinic cycles exist for a range of coefficient values which includes the following:
\begin{equation}\label{eq:coefficients}
\begin{array}{l}
a_1=0.2,~ a_2=0,~ a_3=-0.3,~ a_4=0.05,~ a_5=0.25,~ a_6=-0.6,~ a_7=a_8=-1 \\
b_1=0.05,~ b_2=0.4,~ b_3=b_4=-1,~ b_5=-0.1
\end{array}
\end{equation}
The eigenvalues are
\begin{equation*}
\begin{array}{l}
e_1 = 0.283406,~ c_1 = -0.428634 \\
e_2 = 0.228634,~ c_2 = -0.483406
\end{array}
\end{equation*}
so that the generalized heteroclinic cycle is an attractor. The numerical simulations (with Matlab) were done with $n=3$ and $n=5$. The two pictures in Figure \ref{fig:simus} show the dynamics of the $z_2$ variable in polar coordinates: $z_2=r_2e^{i\phi_2}$. The horizontal axis is the radial variable $r_2$ while the vertical axis is the angle $\phi_2$ (in degrees). Observe that in both cases, taking an initial condition close to $\xi_2$ even with a small angle $\phi_2$ (hence close to the plane $P_2$) the trajectory comes back to the vertical axis sequentially (as expected since it corresponds to going close to $\xi_2$), but with an increasing value of the angle. In the $n=3$ case the angle converges to $60^\circ$ while in the case $n=5$ it converges to $36^\circ$. In both cases this corresponds to convergence to a pseudo-simple cycle with a connection in $\rho P_2$.
It is clear from this figure that when $n=3$ the convergence to the pseudo-simple cycle is faster and in particular the trajectory near
the equilibria (near the vertical coordinate axis in the figure) is oblique while it is nearly horizontal in the $n=5$ case. This is consistent with the results of \cite{pc16} where the case $n=3$ was studied using a different approach in which the double unstable eigenvalue $e_2$ is small enough for nonlinear effects to be felt by the flow near $\xi_2$. This argument doesn't work however when $n>4$ because one essential property of the case $n=3$ is that on the center manifold which exists at $\xi_2$ when $e_2$ is small enough, an unstable equilibrium point always exists near $\xi_2$ in $P_2$, which obliges the flow to "bend" back to $P_2$ or to $\rho P_2$ in the vicinity of $\xi_2$. A similar idea holds when $n=4$. The advantage of the method of \cite{pc16} is that it does not require the existence of a generalized heteroclinic cycle, however only fragmentarily asymptotic stability can be proved in such case.

\begin{figure}[ht]
	\centering
	\mbox{\subfigure[Coordinates $x_1,~y_1$]{\includegraphics[width=8cm]{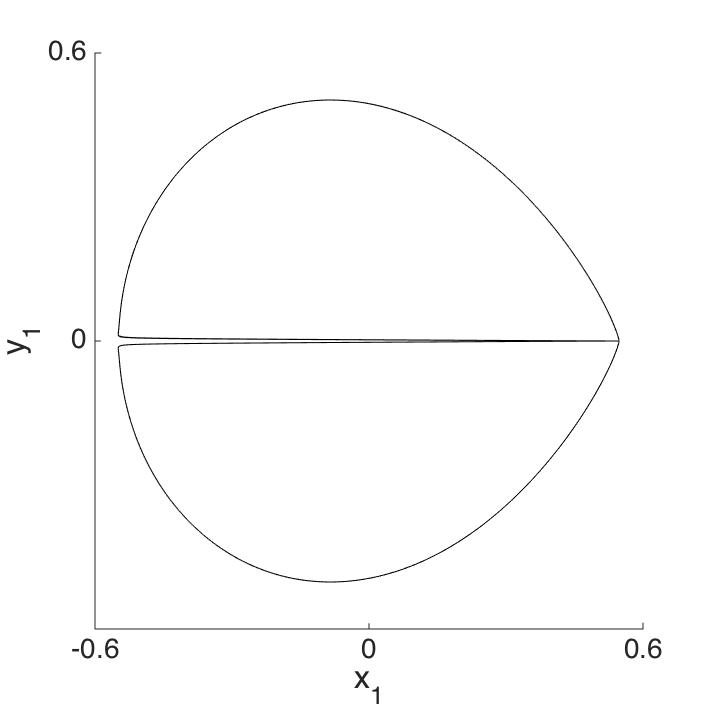}}\quad
       \subfigure[Coordinates $x_1,~x_2,~y_2$]{\includegraphics[width=8cm]{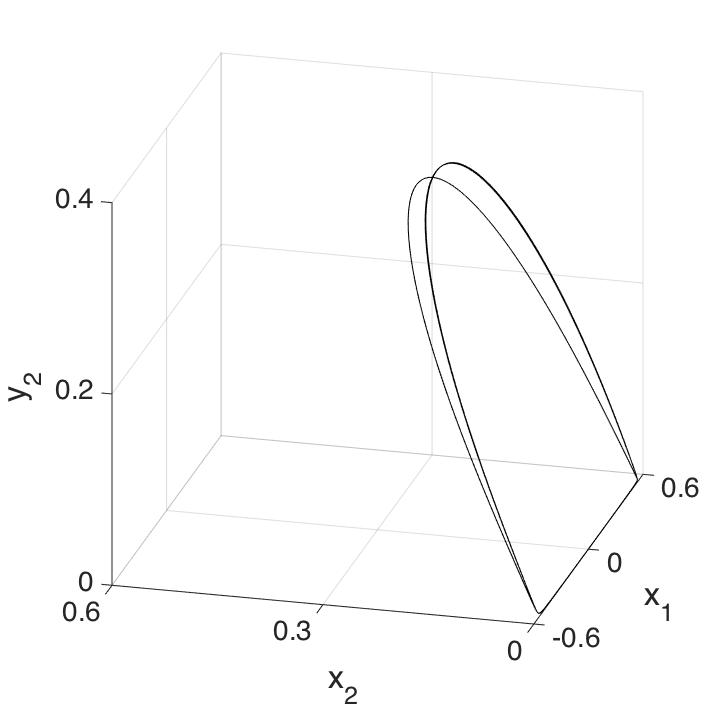}}}
	\caption{\small{Asymptotic dynamics with $a_9=0.9$, $a'_9=1.05$, $b_2=0.292$, $b'_2=0.31$.}}
	\label{fig:simus-D5-1}
\end{figure}

\begin{figure}[ht]
	\centering
	\mbox{\subfigure[Coordinates $x_1,~y_1$]{\includegraphics[width=8cm]{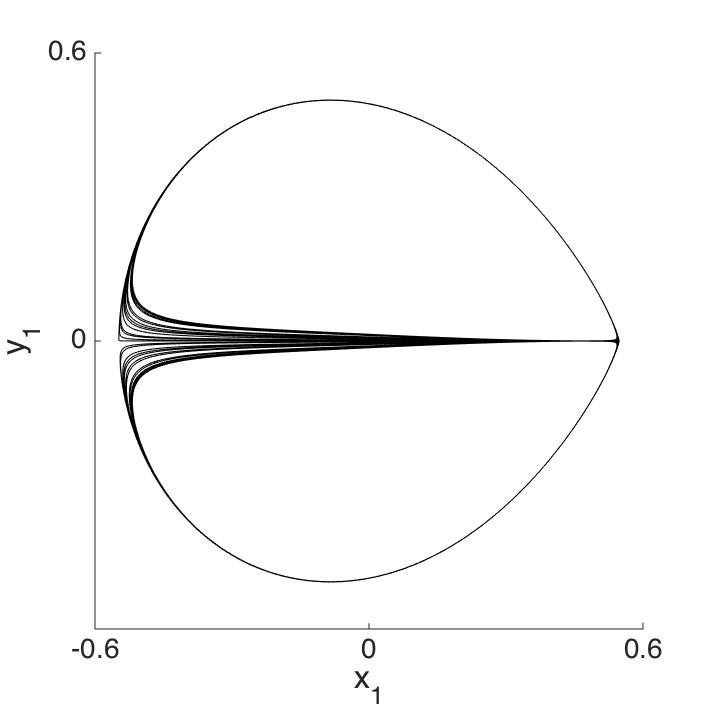}}\quad
         \subfigure[Coordinates $x_1,~x_2,~y_2$]{\includegraphics[width=8cm]{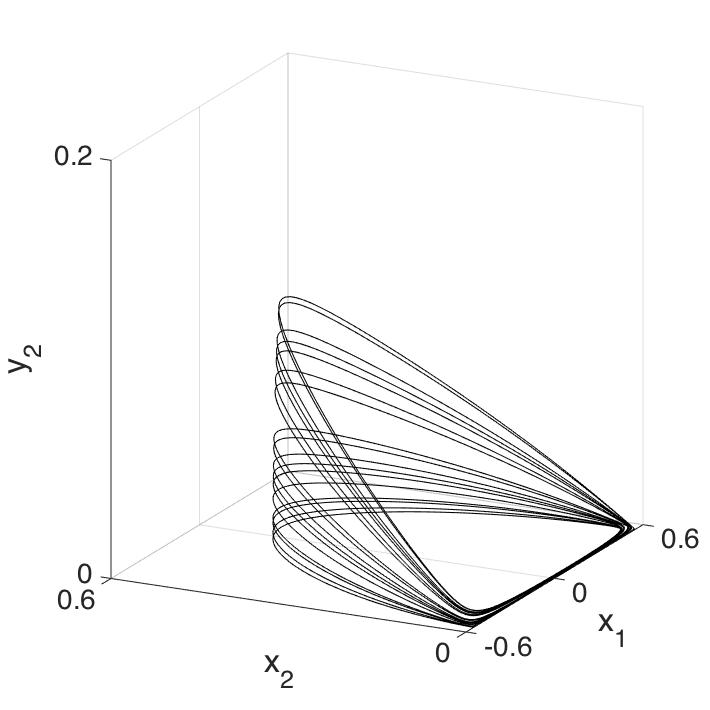}}}
	\caption{\small{Asymptotic dynamics with $a_9=1.1$, $a'_9=0.85$, $b_2=0.28$, $b'_2=0.32$.}}
	\label{fig:simus-D5-2}
\end{figure}

Let us assume now that a perturbation is added to the vector field, which breaks the symmetry $\kappa$. The symmetry group is therefore reduced to the action of $\D_n$ generated by the transformations $\rho$ and $\kappa\sigma$. The invariant planes $P_1$, $P_2$ (and its copies by $\rho^k$) and $V$ are preserved, but not the invariant space $W$. If the perturbation is not too large the equilibria in $L=P_1\cap P_2$ and their heteroclinic connections in the invariant planes persist, hence a pseudosimple heteroclinic cycle exists, however we know it is completely unstable. The question is what happens to the dynamics when this perturbation is switched on. Some preliminary numerical experiments have been performed on the system \eqref{eq:Dnequivariant}, where $n=5$ and the perturbation consists in replacing the terms $a_9(z_2^5 +\bar z_2^5)$ by $a_9z_2^5 + a'_9\bar z_2^5$ and $b_2(z_1 + \bar z_1)z_2$ by $(b_2z_1 + b'_2\bar z_1)z_2$, where $|a_9- a'_9|$ and $|b_2-b'_2|$ are small but non zero. Other coefficients are the same as in \eqref{eq:coefficients} except $a_1=0.25, a_2=0.05, b_1=0.2$. It has been observed that the dynamics remains in a neighborhood of the cycle and converges in certain cases to a periodic orbit (Fig. \ref{fig:simus-D5-1}) while in other cases it exhibits a clear a aperiodic, possibly chaotic behavior (Fig. \ref{fig:simus-D5-2}). The mathematical analysis of this behavior will be a subject for future study.

\section{Conclusion}\label{sec8}

In this paper we completed the study of pseudo-simple heteroclinic cycles in
$\R^4$, which have been discovered and distinguished from simple cycles
only recently \cite{pc15,pc16}. Our primary contribution is a complete list of
finite subgroups of $O(4)$ admitting pseudo-simple heteroclinic cycles. Similar
to the completion of the classification of simple cycles in \cite{pc15}, and
as projected ibid, this was achieved using the quaternionic presentation of
such groups.

Up to now stability of pseudo-simple cycles had only been addressed in
\cite{pc16}, where generic complete instability for the case
$\Gamma\subset SO(4)$ was shown, and an example of a \emph{fragmentarily
asymptotically stable} cycle, an intermediate weak form of stability, with
$\Gamma \not\subset SO(4)$ was given. We extended the stability analysis for
pseudo-simple cycles in subsection \ref{secth2} by identifying all subgroups
of $O(4)$ admitting f.a.s.\ pseudo-simple heteroclinic cycles.
A more comprehensive study, e.g.\ derivation of conditions for fragmentary asymptotic stability or calculation of stability indices along the heteroclinic
connections as defined in \cite{pa11}, is beyond the scope of this work.

We have also studied the behaviour of trajectories close to pseudo-simple
cycles. Namely, we proved that asymptotically stable periodic orbits can bifurcate from the cycle in
a codimension one bifurcation at a point where a multiple expanding
eigenvalue vanishes. Necessary and sufficient conditions for such a
bifurcation are given in theorems \ref{thperorb} and \ref{noorbit}. In section
\ref{sec8n} we illustrated this through a numerical example of a heteroclinic
cycle with a nearby attracting periodic orbit
with symmetry group $\Gamma =(\D_4\rl\D_2;\D_4\rl\D_2)$.

In contrast with \cite{pc15}, the proof of lemma \ref{lem1} to characterize
conditions for a group to be admissible relies upon an explicit construction of
corresponding equivariant systems. This allows us to build examples of
pseudo-simple heteroclinic cycles for any admissible group. As we noted
(see remark \ref{rem11}), lemma \ref{lem11} can be generalized to $\R^n$ with
$n>4$ to provide sufficient conditions for a subgroup of $O(n)$ to admit
heteroclinic cycles. Moreover, the explicit construction of an equivariant
system in $\R^n$ is applicable for this subgroup.

In addition to simple and pseudo-simple heteroclinic cycles other types
of structurally stable heteroclinic cycles can exist in $\R^4$.
One example is the generalized heteroclinic cycle that we studied in
section \ref{sec6}. Another example is the cycle
considered in \cite{mrwp}.
To describe all robust heteroclinic cycles existing in $\R^4$ is
an open problem which is beyond the scope of this paper.

Other possible continuations of our work include the full classification of
pseudo-simple cycles in $\R^5$, similar to the full classification of
homoclinic cycles in \cite{op13},
as well as the study of networks, which are connected unions of more than one
cycle. In principle we think this can be achieved by the same means as we used
here, even though a complete classification of networks has not even been done
for simple cycles yet, partial results to this end can be found in \cite{cl16}.

\newpage
\appendix
\section{Elements of for the groups listed in Table \ref{listSO4},
satisfying $\dim\Fix\gamma=2$\label{planereflections}}
We denote the quaternions ${\bf u}=(0,0,0,1)$,
${\bf v}_{\pm}=(0,1,\pm1,0)/\sqrt{2}$, ${\bf v}'_{\pm}=(1,0,0,\pm1)/\sqrt{2}$,
${\bf h}_{\pm,\pm,\pm}=(1,\pm1,\pm1,\pm1)/2$,
${\bf w}_{\pm\pm}=(0,1,\pm\tau\pm\tau^{-1})/2$,
${\bf w}'_{\pm\pm}=(1,\pm\tau^{-1},\pm\tau,0)/2$,
${\bf w}''_{\pm\pm}=(\tau,\pm1,\pm\tau^{-1},0)/2$,
where $\tau=2\cos(\pi/5)=(\sqrt{5}+1)/2$,
$\tau^*=2\sin(\pi/5)=\sqrt{5}(\tau)^{-1}$ and
the permutation $\rho:~(a,b,c,d)\mapsto~(a,c,d,b)$.
\begin{table}[htp]
{\footnotesize
\begin{center}
\begin{tabular}{|c|l|}
\hline
group ${\Gamma}$ & elements $\gamma$ \\
\hline
$(\D_{K_1}\rl\D_{K_1};\D_{K_2}\rl\D_{K_2})$ &
$\kappa_1(\pm,n)=((\cos(n\theta),0,0,\pm\sin(n\theta));(\cos(n\theta),0,0,\sin(n\theta)))$ \\
$K_1,K_2$ odd, $m=\gc(K_1,K_2)$&
$\kappa_2(n_1,n_2)=((0,\cos(n_1\theta_1),\sin(n_1\theta_1),0);
(0,\cos(n_2\theta_2),\sin(n_2\theta_2),0))$ \\
$\theta=\pi/m$,\ $\theta_1=\pi/K_1$, $\theta_2=\pi/K_2$ &\\
\hline
$(\D_{K_1}\rl\D_{K_1};\D_{K_2}\rl\D_{K_2})$ &
$\kappa_1(\pm,n)=((\cos(n\theta),0,0,\pm\sin(n\theta));(\cos(n\theta),0,0,\sin(n\theta)))$ \\
$K_1$ odd, $K_2$ even&
$\kappa_2(n_1,n_2)=((0,\cos(n_1\theta_1),\sin(n_1\theta_1),0);
(0,\cos(n_2\theta_2),\sin(n_2\theta_2),0))$ \\
&$\kappa_3(n_1)=((0,\cos(n_1\theta_1),\sin(n_1\theta_1),0);(0,0,0,1))$\\
\hline
$(\D_{2K_1}\rl\D_{2K_1};\D_{2K_2}\rl\D_{2K_2})$ &
$\kappa_1(\pm,n)=((\cos(n\theta),0,0,\pm\sin(n\theta));(\cos(n\theta),0,0,\sin(n\theta)))$ \\
$\theta=\pi/(2m)$, $\theta_1=\pi/(2K_1)$ &
$\kappa_2(n_1)=((0,\cos(n_1\theta_1),\sin(n_1\theta_1),0);(0,0,0,1))$ \\
$\theta_2=\pi/(2K_2)$ & $\kappa_3(n_2)=((0,0,0,1);(0,\cos(n_2\theta_2),\sin(n_2\theta_2),0))$ \\
& $\kappa_4(n_1,n_2)=((0,\cos(n_1\theta_1),\sin(n_1\theta_1),0);
(0,\cos(n_2\theta_2),\sin(n_2\theta_2),0))$ \\
\hline
$(\D_{2K_1r}\rl\Z_{4K_1};\D_{2K_2r}\rl\Z_{4K_2})_s$ &
$\kappa_1(\pm,n)=((\cos(n\theta),0,0,\sin(n\theta));(\cos(n\theta),0,0,\pm\sin(n\theta)))$ \\
$m=\gc(K_1,K_2)\gc(r,K_1-sK_2)$
&$\kappa_2(n_1,n_2,n_3)=((0,\cos(n_1\theta_1+n_3\theta_1^*),
\sin(n_1\theta_1+n_3\theta_1^*),0)$; \\
$\theta_1^*=\theta_1/r,\ \theta_2^*=\theta_2/r$
& ~~~~~~~~~~~~~~~~~~~~~~~~~$(0,\cos(n_2\theta_2+n_3s\theta_2^*),\cos(n_2\theta_2+n_3s\theta_2^*),0))$ \\
\hline
$(\D_{2K_1r}\rl\Z_{2K_1};\D_{2K_2r}\rl\Z_{2K_2})_s$ &
$\kappa_1(\pm,n)=((\cos(n\theta),0,0,\pm\sin(n\theta));(\cos(n\theta),0,0,\sin(n\theta)))$ \\
& $\kappa_2(n_1,n_2,n_3)=((0,\cos(2n_1\theta_1+n_3\theta_1^*),
\sin(2n_1\theta_1+n_3\theta_1^*),0);$ \\
& ~~~~~~~~~~~~~~~~~~~~~$(0,\cos(2n_2\theta_2+n_3s\theta_2^*),\sin(2n_2\theta_2+n_3s\theta_2^*),0)),$ \\
& $\kappa_3(n_1,n_2,n_3)=((0,\cos((2n_1+1)\theta_1+n_3\theta_1^*), \sin((2n_1+1)\theta_1+n_3\theta_1^*),0);$ \\
& ~~~~~~~~~~~~~~~~~~$(0,\cos((2n_2+1)\theta_2+n_3s\theta_2^*), \sin((2n_2+1)\theta_2+n_3s\theta_2^*),0))$ \\
\hline
$(\D_{2K_1}\rl\D_{K_1};\D_{2K_2}\rl\D_{K_2})$ &
$\kappa_1(\pm,n)=((\cos(n\theta),0,0,\pm\sin(n\theta));
(\cos(n\theta),0,0,\sin(n\theta)))$ \\
$K_1,K_2\hbox{ even}$ & $\kappa_2(n_1)=((0,\cos(2n_1\theta_1),\sin(2n_1\theta_1),0);(0,0,0,1))$ \\
$m=\gc(K_1,K_2)$& $\kappa_3(n_2)=((0,0,0,1);(0,\cos(2n_2\theta_2),\sin(2n_2\theta_2,0)))$ \\
& $\kappa_4(n_1,n_2)=((0,\cos(2n_1\theta_1),\sin(2n_1)\theta_1),0);
(0,\cos(2n_2\theta_2),\sin(2n_2\theta_2),0))$ \\
& $\kappa_5(n_1,n_2)=((0,\cos((2n_1+1)\theta_1),\sin((2n_1+1)\theta_1),0);$ \\
& ~~~~~~~~~~~~~~~~~~~~~$(0,\cos((2n_2+1)\theta_2),\sin((2n_2+1)\theta_2),0))$ \\
\hline
$(\D_{2K_1}\rl\D_{K_1};\D_{2K_2}\rl\D_{K_2})$ &
$\kappa_1(\pm,n)=((\cos(n\theta),0,0,\pm\sin(n\theta));
(\cos(n\theta),0,0,\sin(n\theta)))$ \\
$K_1,K_2,m\hbox{ odd}$ & $\kappa_2(n_1)=((0,\cos((2n_1+1)\theta_1),\sin((2n_1+1)\theta_1),0);(0,0,0,1))$ \\
& $\kappa_3(n_2)=((0,0,0,1);(0,\cos((2n_2+1)\theta_2),\sin((2n_2+1)\theta_2),0))$ \\
& $\kappa_4(n_1,n_2)=((0,\cos(2n_1\theta_1),\sin(2n_1\theta_1),0);
(0,\cos(2n_2\theta_2),\sin(2n_2\theta_2),0))$ \\
& $\kappa_5(n_1,n_2)=((0,\cos((2n_1+1)\theta_1),\sin((2n_1+1)\theta_1),0);$ \\
& ~~~~~~~~~~~~~~~~~~~~~$(0,\cos((2n_2+1)\theta_2),\sin((2n_2+1)\theta_2),0))$ \\
\hline
$(\D_{2K_1}\rl\D_{K_1};\D_{2K_2}\rl\D_{K_2})$ &
$\kappa_1(\pm,n)=((\cos(n\theta),0,0,\pm\sin(n\theta));
(\cos(n\theta),0,0,\sin(n\theta)))$ \\
$K_1\hbox{ even}$, $K_2\hbox{ odd}$ &
$\kappa_2(n_1)=((0,\cos((2n_1+1)\theta_1),\sin((2n_1+1)\theta_1),0);(0,0,0,1))$ \\
& $\kappa_3(n_2)=((0,0,0,1);(0,\cos(2n_2\theta_2),\sin(2n_2\theta_2),0))$ \\
& $\kappa_4(n_1,n_2)=((0,\cos(2n_1\theta_1),\sin(2n_1\theta_1),0);
(0,\cos(2n_2\theta_2),\sin(2n_2\theta_2),0))$ \\
& $\kappa_5(n_1,n_2)=((0,\cos((2n_1+1)\theta_1),\sin((2n_1+1)\theta_1),0);$ \\
&  ~~~~~~~~~~~~~~~~~~~~~$(0,\cos((2n_2+1)\theta_2),\sin((2n_2+1)\theta_2),0))$\\
\hline
\end{tabular}
\end{center}
}
\end{table}
\newpage

\noindent {\bf Annex A continued.} \\
\begin{table}[htp]
{\footnotesize
\begin{center}
\begin{tabular}{|c|l|}
\hline
group ${\Gamma}$ & elements $\gamma$ \\
\hline
$(\D_{2K_1}\rl\D_{K_1};\D_{2K_2}\rl\Z_{4K_2})$ &
$\kappa_1(\pm,n)=((\cos(n\theta),0,0,\pm\sin(n\theta));
(\cos(n\theta),0,0,\sin(n\theta)))$ \\
$K_1\hbox{ even}$ & $\kappa_2(n_1)=((0,\cos(2n_1\theta_1),\sin(2n_1\theta_1),0);(0,0,0,1))$ \\
& $\kappa_3(n_1,n_2)=((0,\cos((2n_1+1)\theta_1),\sin((2n_1+1)\theta_1),0);$ \\
& ~~~~~~~~~~~~~~~~~~~~~$(0,\cos(n_2\theta_2),\sin(n_2\theta_2),0))$ \\
\hline
$(\D_{2K_1}\rl\D_{K_1};\D_{2K_2}\rl\Z_{4K_2})$ &
$\kappa_1(\pm,n)=((\cos(n\theta),0,0,\pm\sin(n\theta));
(\cos(n\theta),0,0,\sin(n\theta)))$ \\
$K_1\hbox{ odd}$ &  $\kappa_2(n_1)=((0,\cos(2n_1\theta_1),\sin(2n_1\theta_1),0);(0,0,0,1))$ \\
 & $\kappa_3(n_2)=((0,0,0,1);(0,\cos(n_2\theta_2),\sin(n_2\theta_2),0))$ \\
& $\kappa_4(n_1,n_2)=((0,\cos((2n_1+1)\theta_1),\sin((2n_1+1)\theta_1),0);$ \\
& ~~~~~~~~~~~~~~~~~~~~~$(0,\cos(n_2\theta_2),\sin(n_2\theta_2),0))$ \\
\hline
$(\D_{2K_1}\rl\D_{K_1};\D_{K_2}\rl\Z_{2K_2})$ &
$\kappa_1(\pm,n)=((\cos(n\theta),0,0,\sin(n\theta));
(\cos(n\theta),0,0,\pm\sin(n\theta)))$ \\
$K_1,K_2\hbox{ odd}$ &
$\kappa_2(n_1,n_2)=((0,\cos((2n_1+1)\theta_1),\sin((2n_1+1)\theta_1),0);$ \\
$\theta=\pi/m$& ~~~~~~~~~~~~~~~~~~~~~$(0,\cos(n_2\theta_2),\sin(n_2\theta_2),0))$ \\
& $\kappa_3(n_2)=((0,0,0,1);(0,\cos(n_2\theta_2),\sin(n_2\theta_2),0))$ \\
\hline
$(\D_{2K_1}\rl\D_{K_1};\D_{K_2}\rl\Z_{2K_2})$ &
$\kappa_1(\pm,n)=((\cos(n\theta),0,0,\pm\sin(n\theta));
(\cos(n\theta),0,0,\sin(n\theta)))$ \\
$K_1\hbox{ even},\ K_2\hbox{ odd}$ &
$\kappa_2(n_1,n_2)=((0,\cos((2n_1+1)\theta_1),\sin((2n_1+1)\theta_1),0);$ \\
& ~~~~~~~~~~~~~~~~~~~~~$(0,\cos(n_2\theta_2),\sin(n_2\theta_2),0))$ \\
\hline
$(\D_{3K}\rl\D_{3K};\mO\rl\mO)$ &
$\kappa_1(\pm,\pm,\pm,\pm)=(1,0,0,\pm\sqrt{3})/2;h_{\pm\pm\pm})$\\
$K\hbox{ odd}$ &
$\kappa_2(n,r)=((0,\cos(n\theta),\sin(n\theta),0);\rho^r{\bf u})$ \\
$\theta=\pi/(3K)$& $\kappa_3(n,r,\pm)=((0,\cos(n\theta),\sin(n\theta),0);\rho^r{\bf v}_{\pm})$ \\
\hline
$(\D_{6K}\rl\D_{6K};\mO\rl\mO)$ &
$\kappa_1(\pm,\pm,\pm,\pm)=(1,0,0,\pm\sqrt{3})/2;h_{\pm\pm\pm})$,\
$\kappa_2(\pm,r)=((0,0,0,\pm1);\rho^r{\bf u})$ \\
$K\hbox{ odd}$ &
$\kappa_3(\pm,r,\pm)=((0,0,0,\pm1);\rho^r{\bf v}_{\pm})$,\
$\kappa_4(n,r)=((0,\cos(n\theta),\sin(n\theta),0);\rho^r{\bf u})$ \\
$\theta=\pi/(6K)$& $\kappa_5(n,r,\pm)=((0,\cos(n\theta),\sin(n\theta),0);\rho^r{\bf v}_{\pm})$ \\
\hline
$(\D_{4K}\rl\D_{4K};\mO\rl\mO)$ &
$\kappa_1(\pm,r,\pm)=(1,0,0,\pm1)/\sqrt{2};\rho^r{\bf v}'_{\pm})$\\
$K\ne 3m$&$\kappa_2(\pm,r,\pm)=((0,0,0,\pm1);\rho^r{\bf v}_{\pm})$,
$\kappa_3(n,r)=((0,\cos(n\theta),\sin(n\theta),0);\rho^r{\bf u})$ \\
$\theta=\pi/(4K)$&
$\kappa_4(n,r,\pm)=((0,\cos(n\theta),\sin(n\theta),0);\rho^r{\bf v}_{\pm})$,
$\kappa_5(\pm,r)=((0,0,0,\pm1);\rho^r{\bf u})$ \\
\hline
$(\D_{12K}\rl\D_{12K};\mO\rl\mO)$ &
$\kappa_1(\pm,\pm,\pm,\pm)=(1,0,0,\pm\sqrt{3})/2;h_{\pm\pm\pm})$,
$\kappa_2(\pm,r,\pm)=(1/\sqrt{2},0,0,\pm1/\sqrt{2});\rho^r{\bf v}'_{\pm}))$\\
$\theta=\pi/(12K)$&$\kappa_3(\pm,r,\pm)=((0,0,0,\pm1);\rho^r{\bf v}_{\pm})$,\
$\kappa_4(n,r)=((0,\cos(n\theta),\sin(n\theta),0);\rho^r{\bf u})$\\
& $\kappa_5(n,r,\pm)=((0,\cos(n\theta),\sin(n\theta),0);\rho^r{\bf v}_{\pm})$,
$\kappa_6(\pm,r)=((0,0,0,\pm1);\rho^r{\bf u})$\\
\hline
$(\D_{3K}\rl\Z_{6K};\mO\rl\T)$ &
$\kappa_1(\pm,\pm,\pm,\pm)=(1,0,0,\pm\sqrt{3})/2;h_{\pm\pm\pm})$\\
$K\hbox{ odd}$, $\theta=\pi/(3K)$&
$\kappa_2(n,r,\pm)=((0,\cos(n\theta),\sin(n\theta),0);\rho^r{\bf v}_{\pm})$ \\
\hline
$(\D_{6K}\rl\Z_{12K};\mO\rl\T)$ &
$\kappa_1(\pm,\pm,\pm,\pm)=(1/2,0,0,\pm\sqrt{3}/2);h_{\pm\pm\pm})$,\
$\kappa_2(\pm,r)=((0,0,0,\pm1);\rho^r{\bf u})$ \\
$\theta=\pi/(6K)$& $\kappa_3(n,r,\pm)=((0,\cos(n\theta),\sin(n\theta),0);\rho^r{\bf v}_{\pm})$ \\
\hline
$(\D_{4K}\rl\D_{2K};\mO\rl\T)$&
$\kappa_1(\pm,r,\pm)=(1,0,0,\pm1)/\sqrt{2};\rho^r{\bf v}'_{\pm}))$,
$\kappa_2(n,r)=((0,\cos(2n\theta),\sin(2n\theta),0);\rho^r{\bf u})$ \\
$K\hbox{ odd},\ K\ne3m$, $\theta=\pi/(4K)$&
$\kappa_3(n,r,\pm)=((0,\cos((2n+1)\theta),\sin((2n+1)\theta),0);\rho^r{\bf v}_{\pm})$\\
&$\kappa_4(\pm,r)=((0,0,0,\pm1);\rho^r{\bf u})$\\
\hline
\end{tabular}
\end{center}
}
\end{table}

\newpage
\noindent {\bf Annex A continued.} \\
\begin{table}[htp]
{\footnotesize
\begin{center}
\begin{tabular}{|c|l|}
\hline
group ${\Gamma}$ & elements $\gamma$ \\
\hline
$(\D_{6K}\rl\D_{3K};\mO\rl\T)$&
$\kappa_1(\pm,\pm,\pm,\pm)=(1,0,0,\pm\sqrt{3})/2;h_{\pm\pm\pm})$,\\
$K\hbox{ odd}$&$\kappa_2(\pm,r,\pm)=((0,0,0,\pm1);\rho^r{\bf v}_{\pm})$,
$\kappa_3(n,r)=((0,\cos(2n\theta),\sin(2n\theta),0);\rho^r{\bf u})$\\
$\theta=\pi/(6K)$&
$\kappa_4(n,r,\pm)=((0,\cos((2n+1)\theta),\sin((2n+1)\theta),0);\rho^r{\bf v}_{\pm})$ \\
\hline
$(\D_{12K}\rl\D_{6K};\mO\rl\T)$&
$\kappa_1(\pm,\pm,\pm,\pm)=(1,0,0,\pm\sqrt{3})/2;h_{\pm\pm\pm})$,\
$\kappa_2(\pm,r)=(0,0,0,\pm1);\rho^r{\bf u}))$\\
$K\hbox{ even}$&
$\kappa_3(n,r)=((0,\cos(2n\theta),\sin(2n\theta),0);\rho^r{\bf u})$ \\
$\theta=\pi/(12K)$& $\kappa_4(n,r,\pm)=((0,\cos((2n+1)\theta),\sin((2n+1)\theta),0);\rho^r{\bf v}_{\pm})$ \\
\hline
$(\D_{12K}\rl\D_{6K};\mO\rl\T)$&
$\kappa_1(\pm,\pm,\pm,\pm)=(1,0,0,\pm\sqrt{3})/2;h_{\pm\pm\pm})$,
$\kappa_2(\pm,r,\pm)=(1,0,0,\pm1)/\sqrt{2};\rho^r{\bf v}'_{\pm}))$\\
$K\hbox{ odd}$ &
$\kappa_3(n,r)=((0,\cos(2n\theta),\sin(2n\theta),0);\rho^r{\bf u})$ \\
& $\kappa_4(n,r,\pm)=((0,\cos((2n+1)\theta),\sin((2n+1)\theta),0);\rho^r{\bf v}_{\pm})$ \\
&$\kappa_5(\pm,r)=((0,0,0,\pm1);\rho^r{\bf u})$\\
\hline
$(\D_{3K}\rl\Z_{2K};\mO\rl\V)$ &
$\kappa_1(\pm,\pm,\pm,\pm)=(1,0,0,\pm\sqrt{3})/2;h_{\pm\pm\pm})$\\
$\theta=\pi/(3K)$ & $\kappa_2(n,\pm)=((0,\cos(3n\theta),\sin(3n\theta),0);{\bf v}_{\pm})$ \\
$K\hbox{ odd}$ & $\kappa_2'(n,\pm)=((0,\cos(3n+1)\theta),\sin((3n+1)\theta),0);\rho{\bf v}_{\pm})$ \\
& $\kappa_2''(n,\pm)=((0,\cos(3n+2)\theta),\sin((3n+2)\theta),0);\rho^2{\bf v}_{\pm})$ \\
\hline
$(\D_{6K}\rl\Z_{4K};\mO\rl\V)$ &
$\kappa_1(\pm,\pm,\pm,\pm)=(1,0,0,\pm\sqrt{3})/2;h_{\pm\pm\pm})$\\
& $\kappa_2(\pm,r)=((0,0,0,\pm1);\rho^r{\bf u}))$ \\
$\theta=\pi/(6K)$ & $\kappa_3(n,\pm)=((0,\cos(3n\theta),\sin(3n\theta),0);{\bf v}_{\pm})$ \\
& $\kappa_3'(n,\pm)=((0,\cos(3n+1)\theta),\sin((3n+1)\theta),0);\rho{\bf v}_{\pm})$ \\
& $\kappa_3''(n,\pm)=((0,\cos(3n+2)\theta),\sin((3n+2)\theta),0);\rho^2{\bf v}_{\pm})$ \\
\hline
$(\D_{3K}\rl\D_{3K};\I\rl\I)$ &
$\kappa_1(\pm,\pm,\pm,\pm)=(1,0,0,\pm\sqrt{3})/2;h_{\pm\pm\pm})$\\
$K\hbox{ odd}$, $K\ne5m$&
$\kappa_1'(\pm,r,\pm,\pm)=(1,0,0,\pm\sqrt{3})/2;\rho^r{\bf w}'_{\pm\pm})$\\
$\theta=\pi/(3K)$&
$\kappa_2(n,r)=((0,\cos(n\theta),\sin(n\theta),0);\rho^r{\bf u})$ \\
&$\kappa_2'(n,r,\pm,\pm)=((0,\cos(n\theta),\sin(n\theta),0);\rho^r{\bf w}_{\pm\pm}$)\\
\hline
$(\D_{6K}\rl\D_{6K};\I\rl\I)$ &
$\kappa_1(\pm,\pm,\pm,\pm)=(1,0,0,\pm\sqrt{3})/2;h_{\pm\pm\pm})$\\
$K\hbox{ odd}$, $K\ne5m$&
$\kappa_1'(\pm,r,\pm,\pm)=(1,0,0,\pm\sqrt{3})/2;\rho^r{\bf w}'_{\pm\pm})$\\
$\theta=\pi/(6K)$&
$\kappa_2(n,r)=((0,\cos(n\theta),\sin(n\theta),0);\rho^r{\bf u})$ \\
&$\kappa_2'(n,r,\pm,\pm)=((0,\cos(n\theta),\sin(n\theta),0);\rho^r{\bf w}_{\pm\pm}$)\\
& $\kappa_3(\pm,r)=((0,0,0,\pm1);\rho^r{\bf u})$,
$\kappa_3'(\pm,r,\pm,\pm)=((0,0,0,\pm1);\rho^r{\bf w}_{\pm\pm})$\\
\hline
$(\D_{5K}\rl\D_{5K};\I\rl\I)$ &
$\kappa_1(\pm,r,\pm,\pm)=(\tau,0,0,\pm\tau^*)/2;\rho^r{\bf w}''_{\pm\pm})$\\
$K$ odd, $K\ne3m$&
$\kappa_2(n,r)=((0,\cos(n\theta),\sin(n\theta),0);\rho^r{\bf u})$ \\
$\theta=\pi/(5K)$&$\kappa_2'(n,r,\pm,\pm)=
((0,\cos(n\theta),\sin(n\theta),0);\rho^r{\bf w}_{\pm\pm}$)\\
\hline
$(\D_{10K}\rl\D_{10K};\I\rl\I)$ &
$\kappa_1(\pm,r,\pm,\pm)=(\tau,0,0,\pm\tau^*)/2;\rho^r{\bf w}''_{\pm\pm})$,
$\kappa_2(n,r)=((0,\cos(n\theta),\sin(n\theta),0);\rho^r{\bf u})$ \\
$K\ne3m$ &
$\kappa_2'(n,r,\pm,\pm)=((0,\cos(n\theta),\sin(n\theta),0);\rho^r{\bf w}_{\pm\pm}$)\\
$\theta=\pi/(10K)$&
$\kappa_3(\pm,r)=((0,0,0,\pm1);\rho^r{\bf u})$,\\
& $\kappa_3'(\pm,r,\pm,\pm)=((0,0,0,\pm1);\rho^r{\bf w}_{\pm\pm})$ \\
\hline
\end{tabular}
\end{center}
}
\end{table}

\newpage
\noindent {\bf Annex A continued.} \\
\begin{table}[htp]
{\footnotesize
\begin{center}
\begin{tabular}{|c|l|}
\hline
group ${\Gamma}$ & elements $\gamma$ \\
\hline
$(\D_{15K}\rl\D_{15K};\I\rl\I)$ &
$\kappa_1(\pm,\pm,\pm,\pm)=(1,0,0,\pm\sqrt{3})/2;h_{\pm\pm\pm})$\\
$K\hbox{ odd}$&$\kappa_1'(\pm,r,\pm,\pm)=(1,0,0,\pm\sqrt{3})/2;\rho^r{\bf w}'_{\pm\pm})$\\
$\theta=\pi/(15K)$&$\kappa_2(\pm,r,\pm,\pm)=(\tau,0,0,\pm\tau^*)/2;\rho^r{\bf w}''_{\pm\pm})$\\
&$\kappa_3(n,r)=((0,\cos(n\theta),\sin(n\theta),0);\rho^r{\bf u})$\\
&$\kappa_3'(n,r,\pm,\pm)=((0,\cos(n\theta),\sin(n\theta),0);\rho^r{\bf w}_{\pm\pm}$)\\
\hline
$(\D_{30K}\rl\D_{30K};\I\rl\I)$ &
$\kappa_1(\pm,\pm,\pm,\pm)=(1,0,0,\pm\sqrt{3})/2;h_{\pm\pm\pm})$\\
$\theta=\pi/(6K)$&$\kappa_1'(\pm,r,\pm,\pm)=(1,0,0,\pm\sqrt{3})/2;\rho^r{\bf w}'_{\pm\pm})$\\
&$\kappa_2(\pm,r,\pm,\pm)=(\tau,0,0,\pm\tau^*)/2;\rho^r{\bf w}''_{\pm\pm})$\\
&$\kappa_3(n,r)=((0,\cos(n\theta),\sin(n\theta),0);\rho^r{\bf u})$\\
&$\kappa_3'(n,r,\pm,\pm)=((0,\cos(n\theta),\sin(n\theta),0);\rho^r{\bf w}_{\pm\pm}$)\\
& $\kappa_4(\pm,r)=((0,0,0,\pm1);\rho^r{\bf u})$,
$\kappa_4'(\pm,r,\pm,\pm)=((0,0,0,\pm1);\rho^r{\bf w}_{\pm\pm})$\\
\hline
$(\D_{2rK_1}\rl\Z_{K_1};\D_{2rK_2}\rl\Z_{K_2})_s$ &
$\kappa_1(n)=((\cos(n\theta),0,0,\sin(n\theta));(\cos(n\theta),0,0,\sin(n\theta)))$ \\
$K_1,K_2\hbox{ odd}$ , $m=\gc(K_1,K_2)(K_2-sK_1)$
& $\kappa_2(n_1,n_2,n_3)=((0,\cos(2n_1\theta_1+n_3\theta_1^*),\sin(2n_1\theta_1+n_3\theta_1^*),0)$ \\
$\theta_1=\pi/K_1,\theta_2=\pi/K_2$ & $(0,\cos(2n_2\theta_2+sn_3\theta_2^*),\cos(2n_2\theta_2+sn_3\theta_2^*),0))$ \\
$\theta_1^*=\theta_1/(2r),\theta_2^*=\theta_2/(2r)$ & $\kappa_3(n_1,n_2,n_3)=((0,\cos((2n_1+1)\theta_1+n_3\theta_1^*),\sin((2n_1+1)\theta_1+n_3\theta_1^*),0);$ \\
$\theta=\pi/m$& $(0,\cos((2n_2+1)\theta_2+sn_3\theta_2^*),\sin((2n_2+1)\theta_2+sn_3\theta_2^*),0))$ \\
\hline
\end{tabular}
\end{center}
}
\end{table}

\section{Conjugacy classes of isotropy subgroups of finite groups $\Gamma$ satisfying
$\dim\Fix\,(\Sigma)=2$ and $\dim\Fix\,(\Delta)=1$\label{conjugacyclasses}}

We list subgroups $\Sigma$ and $\Delta$ of $\Gamma$ that satisfy
$\dim\Fix\,(\Sigma)=2$ and $\dim\Fix\,(\Delta)=1$.
{\small
\begin{sidewaystable}
$$
\begin{array}{|l|l|l|}
\hline
{\Gamma} & \Sigma & \Delta \\
\hline
(\D_{K_1}\rl\D_{K_1};\D_{K_2}\rl\D_{K_2})&<\kappa_1(\pm,1)>;&
<\kappa_1(\pm,1),\kappa_2(n_1,n_2)>:\ n_1+n_2\hbox{ even or odd}\\
K_1,K_2\hbox{ odd}&<\kappa_2(n_1,n_2)>:\ n_1+n_2\hbox{ even or odd}&\\
\hline
(\D_{K_1}\rl\D_{K_1};\D_{K_2}\rl\D_{K_2})&<\kappa_1(\pm,1)>;&
<\kappa_1(\pm,1),\kappa_2(n_1,n_2)>:\ n_2\hbox{ even or odd}\\
K_1\hbox{ odd},\ K_2\hbox{ even}&
<\kappa_2(n_1,n_2)>:\ n_2\hbox{ even or odd},\ <\kappa_3(n_1)>&\\
\hline
(\D_{2K_1}\rl\D_{2K_1};\D_{2K_2}\rl\D_{2K_2})&<\kappa_1(\pm,1)>;\
<\kappa_2(n_1)>:\ n_1\hbox{ even or odd};&
<\kappa_2(n_1),\kappa_3(n_2)>:\ n_1,\ n_2\hbox{ even or odd};\\
K_1+K_2\hbox{ even}&<\kappa_3(n_2)>:\ n_2\hbox{ even or odd;}&
<\kappa_1(\pm,1),\kappa_4(n_1,n_2)>:\ n_1+n_2\hbox{ even or odd}\\
&<\kappa_4(n_1,n_2)>:\ n_1,\ n_2\hbox{ even or odd}&\\
\hline
(\D_{2K_1}\rl\D_{2K_1};\D_{2K_2}\rl\D_{2K_2})&<\kappa_1(\pm,1)>,\
<\kappa_2(n_1)>:\ n_1\hbox{ even or odd};&
<\kappa_2(n_1),\kappa_3(n_2)>:\ n_1,\ n_2\hbox{ even or odd};\\
K_1\hbox{ odd},\ K_2\hbox{ even}&<\kappa_3(n_2)>:\ n_2\hbox{ even or odd;}&
<\kappa_1(\pm,1),\kappa_4(n_1,n_2)>:\ n_2\hbox{ even or odd}\\
&<\kappa_4(n_1,n_2)>:\ n_1,\ n_2\hbox{ even or odd}&\\
\hline
(\D_{2K_1r}\rl\Z_{4K_1};\D_{2K_2r}\rl\Z_{4K_2})_s&
<\kappa_1(+,1)>;\ <\kappa_1(-,1)>;&
<\kappa_1(+,1),\kappa_2(n_1,n_2,n_3)>:\ n_1+n_2\hbox{ even or odd;}\\
K_1,K_2,r\hbox{ odd},&
<\kappa_2(n_1,n_2,n_3)>:\ n_1+n_3,&
<\kappa_1(-,1),\kappa_2(n_1,n_2,n_3)>:\ n_1+n_2\hbox{ even or odd}\\
&n_2+n_3\hbox{ even or odd}&\\
\hline
(\D_{2K_1r}\rl\Z_{4K_1};\D_{2K_2r}\rl\Z_{4K_2})_s&
<\kappa_1(+,1)>;\ <\kappa_1(-,1)>;&
<\kappa_1(+,1),\kappa_2(n_1,n_2,n_3)>:\ n_3\hbox{ even or odd;}\\
K_1,K_2\hbox{ odd},r\hbox{ even}&
<\kappa_2(n_1,n_2,n_3)>:n_1+n_2,\ n_3\hbox{ even or odd}&
<\kappa_1(-,1),\kappa_2(n_1,n_2,n_3)>:\ n_3\hbox{ even or odd}\\
\hline
(\D_{2K_1r}\rl\Z_{4K_1};\D_{2K_2r}\rl\Z_{4K_2})_s&
<\kappa_1(+,1)>;\ <\kappa_1(-,1)>;&
<\kappa_1(+,1),\kappa_2(n_1,n_2,n_3)>:\ n_1\hbox{ even or odd};\\
K_1\hbox{ even},\ K_2,r\hbox{ odd}&
<\kappa_2(n_1,n_2,n_3)>:\ n_1, n_2+n_3\hbox{ even or odd}&
<\kappa_1(-,1),\kappa_2(n_1,n_2,n_3)>:\ n_1\hbox{ even or odd}\\
\hline
(\D_{2K_1r}\rl\Z_{2K_1};\D_{2K_2r}\rl\Z_{2K_2})_s&
<\kappa_1(+,1)>;\ <\kappa_1(-,1)>;&
<\kappa_1(+,1),\kappa_2(n_1,n_2,n_3)>:\ n_1\hbox{ even or odd};\\
K_1\hbox{ even, }K_2\hbox{ odd}
&<\kappa_2(n_1,n_2,n_3)>:\ n_1\hbox{ even or odd}&
<\kappa_1(-,1),\kappa_2(n_1,n_2,n_3)>:\  n_1 \hbox{ even or odd}\\
\hline
(\D_{2K_1}\rl\D_{K_1};\D_{2K_2}\rl\D_{K_2})&<\kappa_1(\pm,1)>;\
<\kappa_2(n_1)>;\ <\kappa_3(n_2)>;&
<\kappa_2(n_1),\kappa_3(n_2)>:\ n_1+n_2\hbox{ even or odd};\\
&<\kappa_4(n_1,n_2)>:\ n_1+n_2\hbox{ even or odd};&
<\kappa_1((-1)^s,1),\kappa_5(n_1,n_2)>:\ s+n_1+n_2\hbox{ even or odd}\\
&<\kappa_5(n_1,n_2)>&\\
\hline
(\D_{2K_1}\rl\D_{K_1};\D_{2K_2}\rl\Z_{4K_2})&<\kappa_1(\pm,1)>;\
<\kappa_2(n_1)>;&<\kappa_1((-1)^s,1),\kappa_3(n_1,n_2)>:\ s+n_2\hbox{ even or odd}\\
K_1\hbox{ even}&<\kappa_3(n_1,n_2)>:\ n_2\hbox{ even or odd}&\\
\hline
(\D_{2K_1}\rl\D_{K_1};\D_{2K_2}\rl\Z_{4K_2})&<\kappa_1(\pm,1)>;\
<\kappa_2(n_1)>:\ n_1\hbox{ even or odd};&
<\kappa_1(\pm,1),\kappa_4(n_1,n_2)>:\ n_2\hbox{ even or odd};\\
K_1\hbox{ odd}&<\kappa_3(n_2)>:\ n_2\hbox{ even or odd};&
<\kappa_2(n_1),\kappa_4(n_1-K_1,n_2)>:\ n_1,n_2\hbox{ even or odd}\\
&<\kappa_4(n_1,n_2)>:\ n_2\hbox{ even or odd}&\\
\hline
(\D_{2K_1}\rl\D_{K_1};\D_{K_2}\rl\Z_{2K_2})&<\kappa_1(\pm,1)>;\
<\kappa_2(n_1,n_2)>;\ <\kappa_3(n_2)>&
<\kappa_1((-1)^s,1),\kappa_2(n_1,n_2)>:\\
K_1,K_2\hbox{ odd}&&s+n_1\hbox{ even or odd}\\
\hline
(\D_{2K_1}\rl\D_{K_1};\D_{K_2}\rl\Z_{2K_2})&<\kappa_1(\pm,1>;\
<\kappa_2(n_1,n_2)>&<\kappa_1((-1)^s,1),\kappa_2(n_1,n_2)>:\\
K_1\hbox{ even},K_2\hbox{ odd}&&s+n_1\hbox{ even or odd}\\
\hline
(\D_{3K}\rl\D_{3K};\mO\rl\mO)&<\kappa_1(+,\pm,\pm,\pm)>;&
<\kappa_1(+,(-1)^{q_1},(-1)^{q_2},(-1)^{q_3}),\kappa_3(n,0,(-1)^{q_1+q_2+1})>:\\
K\hbox{ odd} &<\kappa_2(n,r)>;\ <\kappa_3(n,r,\pm)>&
q_3+n\hbox{ even or odd}\\
\hline
\end{array}
$$
\end{sidewaystable}

\begin{sidewaystable}
$$
\begin{array}{|l|l|l|}
\hline
{\Gamma} & \Sigma & \Delta\\
\hline
(\D_{6K}\rl\D_{6K};\mO\rl\mO)&<\kappa_1(+,\pm,\pm,\pm)>;\ <\kappa_2(\pm,r)>;&
<\kappa_1(+,(-1)^{q_1},(-1)^{q_2},\pm),\kappa_5(n,0,(-1)^{q_1+q_2+1})>:\\
K\hbox{ odd} &<\kappa_3(\pm,r,\pm)>;&
n\hbox{ even or odd};\\
&<\kappa_4(n,r)>:\ n\hbox{ even or odd};&
<\kappa_2(\pm,r),\kappa_4(n,r+1)>;\ <\kappa_2(\pm,r),\kappa_5(n,r,\pm)>;\\
&<\kappa_5(n,r,\pm)>:\ n\hbox{ even or odd}&
<\kappa_3(\pm,r,\pm),\kappa_4(n,r)>:\ n\hbox{ even or odd}\\
\hline
(\D_{4K}\rl\D_{4K};\mO\rl\mO)&<\kappa_1(+,r,\pm)>;\ <\kappa_2(\pm,r,\pm)>;&
<\kappa_1(+,r,\pm),\kappa_3(n,r+1)>:\ n\hbox{ even or odd};\\
K\hbox{ odd, }K\ne3m&<\kappa_3(n,r)>:\ n\hbox{ even or odd};&
<\kappa_2(\pm,r,\pm),\kappa_3(n,r)>:\ n\hbox{ even or odd}\\
&<\kappa_4(n,r,\pm)>:\ n\hbox{ even or odd}&\\
\hline
(\D_{12K}\rl\D_{12K};\mO\rl\mO)&<\kappa_1(+,\pm,\pm,\pm)>;\ <\kappa_2(+,r,\pm)>;&
<\kappa_1(+,(-1)^{q_1},(-1)^{q_2},\pm),\kappa_5(n,0,(-1)^{q_1+q_2+1})>:\\
&<\kappa_3(\pm,r,\pm)>;&
n\hbox{ even or odd};\\
&<\kappa_4(n,r)>:\ n\hbox{ even or odd};&
<\kappa_2(+,r,\pm),\kappa_4(n,r+1)>:\ n\hbox{ even or odd};\\
&<\kappa_5(n,r,\pm)>:\ n\hbox{ even or odd}&
<\kappa_3(\pm,r,\pm),\kappa_4(n,r)>:\ n\hbox{ even or odd}\\
\hline
(\D_{3K}\rl\Z_{6K};\mO\rl\T)&
<\kappa_1(+,(-1)^{q_1},(-1)^{q_2},(-1)^{q_3})>:&
<\kappa_1(+,(-1)^{q_1},(-1)^{q_2},(-1)^{q_3}),\kappa_2(n,0,(-1)^{q_1+q_2+1})>:\\
K\hbox{ odd}& q_1+q_2+q_3\hbox{ even or odd};\ <\kappa_2(n,r,\pm)>&
q_1+n,\ q_1+q_2+q_3\hbox{ even or odd}\\
\hline
(\D_{6K}\rl\Z_{12K};\mO\rl\T)&
<\kappa_1(+,(-1)^{q_1},(-1)^{q_2},(-1)^{q_3})>:&
<\kappa_1(+,(-1)^{q_1},(-1)^{q_2},(-1)^{q_3}),\kappa_3(n,0,(-1)^{q_1+q_2+1})>:\\
K\hbox{ odd}& q_1+q_2+q_3\hbox{ even or odd};&
q_1+q_2+q_3,\ n\hbox{ even or odd};\\
&<\kappa_3(n,r,\pm)>:\ n\hbox{ even or odd}&
<\kappa_2(\pm,r),\kappa_3(n,r,\pm),>\ n\hbox{ even or odd}\\
\hline
(\D_{6K}\rl\Z_{12K};\mO\rl\T)&
<\kappa_1(+,(-1)^{q_1},(-1)^{q_2},(-1)^{q_3})>:&
<\kappa_1(+,(-1)^{q_1},(-1)^{q_2},(-1)^{q_3}),\kappa_3(n,0,(-1)^{q_1+q_2+1})>:\\
K\hbox{ even}&q_1+q_2+q_3\hbox{ even or odd}; <\kappa_2(\pm,r)>;&
q_1+q_2+q_3,\ n\hbox{ even or odd};\\
&<\kappa_3(n,r,\pm)>:\ n\hbox{ even or odd}&
<\kappa_2(\pm,r),\kappa_3(n,r,\pm),>\ n\hbox{ even or odd}\\
\hline
(\D_{4K}\rl\D_{2K};\mO\rl\T)&<\kappa_1(+,r,\pm)>;\ <\kappa_2(n,r)>;&
<\kappa_1(+,r,(-1)^s),\kappa_3(n,r,\pm)>:\ s+n\hbox{ even or odd};\\
K\hbox{ odd},\ K\ne3m&<\kappa_3(n,r,\pm)>&\\
\hline
(\D_{6K}\rl\D_{3K};\mO\rl\T)&
<\kappa_1(+,\pm,\pm,\pm)>;\ <\kappa_2(\pm,r,\pm)>;&
<\kappa_1(+,(-1)^{q_1},(-1)^{q_2},(-1)^{q_3}),\kappa_4(n,0,(-1)^{q_1+q_2+1})>:\\
K\hbox{ odd}&<\kappa_3(n,r)>;\ <\kappa_4(n,r,\pm)>&
q_3+n\hbox{ even or odd};\\
&&<\kappa_2(\pm,r,(-1)^s),\kappa_3(n,r)>:\ s+n\hbox{ even or odd}\\
\hline
(\D_{6K}\rl\D_{3K};\mO\rl\T)&
<\kappa_1(+,\pm,\pm,\pm)>;\ <\kappa_2(\pm,r,\pm)>;&
<\kappa_1(+,(-1)^{q_1},(-1)^{q_2},(-1)^{q_3}),\kappa_4(n,0,(-1)^{q_1+q_2+1})>:\\
K\hbox{ even},\ K\ne2(2m+1)&<\kappa_3(n,r)>;\ <\kappa_4(n,r,\pm)>&
q_3+n\hbox{ even or odd};\\
&&<\kappa_2(\pm,r,\pm),\kappa_3(n,r+1)>;\\
&&<\kappa_2(\pm,r,(-1)^s),\kappa_4(n,r,(-1)^{s+1})>\\
\hline
(\D_{12K}\rl\D_{6K};\mO\rl\T)&
<\kappa_1(+,\pm,\pm,\pm)>;\ <\kappa_2(+,r,\pm)>;&
<\kappa_1(+,(-1)^{q_1},(-1)^{q_2},(-1)^{q_3}),\kappa_4(n,0,(-1)^{q_1+q_2+1})>:\\
K\hbox{ odd}&<\kappa_3(n,r)>;\ <\kappa_4(n,r,\pm)>&
q_3+n\hbox{ even or odd};\\
&&<\kappa_2(\pm,r,(-1)^s),\kappa_4(n,r,\pm)>:\ s+n\hbox{ even or odd}\\
\hline
(\D_{3K}\rl\Z_{2K};\mO\rl\V)&
<\kappa_1(+,\pm,\pm,\pm)>;\ <\kappa_2(n,\pm)>&
<\kappa_1(+,(-1)^{q_1},(-1)^{q_2},(-1)^{q_3}),\kappa_2(n,(-1)^{q_1+q_2+1})>:\\
K\hbox{ odd}&&q_2+q_3+n\hbox{ even or odd}\\
\hline
\end{array}
$$
Continuation of Annex \ref{conjugacyclasses}.
\end{sidewaystable}

\begin{sidewaystable}
$$
\begin{array}{|l|l|l|}
\hline
{\Gamma} & \Sigma & \Delta\\
\hline
(\D_{6K}\rl\Z_{4K};\mO\rl\V)&
<\kappa_1(+,\pm,\pm,\pm)>;\ <\kappa_2(\pm,r)>;&
<\kappa_1(+,(-1)^{q_1},(-1)^{q_2},(-1)^{q_3}),\kappa_2(n,r,(-1)^{q_1+q_2+1})>:\\
K\hbox{ odd}&<\kappa_3(n,\pm)>:\ n\hbox{ even or odd}&n\hbox{ even or odd};\\
&&<\kappa_2((-1)^s,0),\kappa_3(n,\pm)>:\ s+n\hbox{ even or odd}\\
\hline
(\D_{6K}\rl\Z_{4K};\mO\rl\V)&
<\kappa_1(+,\pm,\pm,\pm)>;\ <\kappa_2(\pm,r)>;&
<\kappa_1(+,(-1)^{q_1},(-1)^{q_2},(-1)^{q_3}),\kappa_3(n,r,(-1)^{q_1+q_2+1})>:\\
K\hbox{ odd}&<\kappa_3(n,\pm)>:\ n\hbox{ even or odd}&n\hbox{ even or odd};\\
&&<\kappa_2((-1)^s,0),\kappa_3(n,\pm)>:\ n\hbox{ even or odd}\\
\hline
(\D_{3K}\rl\D_{3K};\I\rl\I) &
<\kappa_1>;\ <\kappa_2>&
<\kappa_1(+,(-1)^{q_1},(-1)^{q_2},(-1)^{q_3}),
\kappa_2'(n,0,(-1)^{q_1+q_2+1},(-1)^{q_1+q_3})>:\\
K\hbox{ odd}, K\ne5m&&q_1+n\hbox{ even or odd};\\
&&<\kappa_1'(\pm,r,(-1)^{s_1},(-1)^{s_2}),\kappa_2(n,r)>:\
n+s_1+s_2\hbox{ even or odd}\\
\hline
(\D_{6K}\rl\D_{6K};\I\rl\I) &
<\kappa_1>;\ <\kappa_2>:\ n\hbox{ even or odd};&
<\kappa_1(+,(-1)^{q_1},(-1)^{q_2},(-1)^{q_3}),
\kappa_2'(n,0,(-1)^{q_1+q_2+1},(-1)^{q_1+q_3+1})>:\\
K\hbox{ odd}, K\ne5m&&n\hbox{ even or odd};\\
&&<\kappa_1'(+,r,\pm,\pm),\kappa_2(n,r)>:\ n\hbox{ even or odd};\\
&&<\kappa_2(n,r),\kappa_3((-1)^s,r+1)>:\ n+s\hbox{ even or odd};\\
&&<\kappa_2'(n,(-1)^{q_1},(-1)^{q_2}),
\kappa_3'((-1)^s,r+1,(-1)^{q_1+1},(-1)^{q_1+q_2+1})>:\\
&& n+s\hbox{ even or odd}\\
\hline
(\D_{6K}\rl\D_{6K};\I\rl\I) &
<\kappa_1>;\ <\kappa_2>:\ n\hbox{ even or odd};&
<\kappa_1(+,(-1)^{q_1},(-1)^{q_2},(-1)^{q_3}),
\kappa_2'(n,0,(-1)^{q_1+q_2+1},(-1)^{q_1+q_3})>:\\
K\hbox{ even}, K\ne5m&&n\hbox{ even or odd};\\
&&<\kappa_1'(+,r,\pm),\kappa_2(n,r)>:\ n\hbox{ even or odd};\\
&&<\kappa_2(n,r),\kappa_3(\pm,r+1)>:\ n\hbox{ even or odd};\\
&&<\kappa_2'(n,(-1)^{q_1},(-1)^{q_2}),
\kappa_3'(\pm,r+1,(-1)^{q_1+1},(-1)^{q_1+q_2})>:\ n\hbox{ even or odd}\\
\hline
(\D_{5K}\rl\D_{5K};\I\rl\I) &
<\kappa_1(+,r,\pm,\pm)>;\ <\kappa_2>&
<\kappa_1(+,r,(-1)^{q_1},(-1)^{q_2}),\kappa_2(n,r)>:\
n+q_1+q_2\hbox{ even or odd};\\
K\hbox{ odd}, K\ne3m&
&<\kappa_1(+,r,\pm,\pm),\kappa_2'(n,r+s,\pm,\pm)>,\ s=1,2:\ n+s\hbox{ even or odd}\\
\hline
(\D_{10K}\rl\D_{10K};\I\rl\I) &
<\kappa_1(+,r,\pm,\pm)>;\ <\kappa_2>;&
<\kappa_1(+,r,\pm,\pm),\kappa_2(n,r)>:\ n\hbox{ even or odd};\\
K\hbox{ odd}, K\ne3m&<\kappa_3>
&<\kappa_1(+,r,\pm,\pm),\kappa_2'(n,r+s,\pm,\pm)>,\ s=1,2:\ n\hbox{ even or odd};\\
&&<\kappa_2(n,r),\kappa_3((-1)^s,r+1)>:\ n+s\hbox{ even or odd};\\
&&<\kappa_2'(n,(-1)^{q_1},(-1)^{q_2}),
\kappa_3'((-1)^s,r+1,(-1)^{q_1+1},(-1)^{q_1+q_2+1})>:\\
&&n+s\hbox{ even or odd}\\
\hline
(\D_{10K}\rl\D_{10K};\I\rl\I) &
<\kappa_1(+,r,\pm,\pm)>;\ <\kappa_2>;&
<\kappa_1(+,r,\pm,\pm),\kappa_2(n,r)>:\ n\hbox{ even or odd};\\
K\hbox{ even}, K\ne3m&<\kappa_3>
&<\kappa_1(+,r,\pm,\pm),\kappa_2'(n,r+s,\pm,\pm)>,\ s=1,2:\ n\hbox{ even or odd};\\
&&<\kappa_2(n,r),\kappa_3((-1)^s,r+1)>:\ n\hbox{ even or odd};\\
&&<\kappa_2'(n,(-1)^{q_1},(-1)^{q_2}),
\kappa_3'((-1)^s,r+1,(-1)^{q_1+1},(-1)^{q_1+q_2+1})>:\\
&&n\hbox{ even or odd}\\
\hline
\end{array}
$$
Continuation of Annex \ref{conjugacyclasses}.
\end{sidewaystable}

\begin{sidewaystable}
$$
\begin{array}{|l|l|l|}
\hline
{\Gamma} & \Sigma & \Delta\\
\hline
(\D_{15K}\rl\D_{15K};\I\rl\I) &
<\kappa_1>;\ <\kappa_2(\pm,r,\pm)>;&
<\kappa_1(+,(-1)^{q_1},(-1)^{q_2},(-1)^{q_3}),
\kappa_3'(n,0,(-1)^{q_1+q_2+1}))>:\\
&&q_q+n\hbox{ even or odd};\\
K\hbox{ odd}& <\kappa_3>&
<\kappa_1'(+,(-1)^{s_1},(-1)^{s_2}),\kappa_3(n,0)>:\
s_1+s_2+n\hbox{ even or odd};\\
&&<\kappa_2(+,r,(-1)^{s_1},(-1)^{s_2}),\kappa_3(n,r))>:\
s_1+s_2+n\hbox{ even or odd};\\
&&<\kappa_2(+,(-1)^{s_1},(-1)^{s_2}),\kappa_3'(n,0,(-1)^{s_1+s_2+1}))>:\
s_1+n\hbox{ even or odd}\\
\hline
(\D_{30K}\rl\D_{30K};\I\rl\I) &
<\kappa_1>;\ <\kappa_2(\pm,r,\pm)>;&
<\kappa_1(+,(-1)^{s_1},(-1)^{s_2},(-1)^{s_3}),
\kappa_3'(n,0,(-1)^{s_1+s_2+1}))>:\ n\hbox{ even or odd};\\
K\hbox{ odd}& <\kappa_3>:\ n\hbox{ even or odd};\ <\kappa_4>&
<\kappa_1'(+,(-1)^{s_1},(-1)^{s_2}),\kappa_3(n,0)>:\
n\hbox{ even or odd};\\
&&<\kappa_2((-1)^s,r,\pm),\kappa_3(n,r)>:\ n\hbox{ even or odd}\\
&&<\kappa_2((-1)^s,r,\pm),\kappa_3'(n,r+1,\mp,\pm)>:\ n\hbox{ even or odd}\\
&&<\kappa_3(n,r),\kappa_4((-1)^s,r+1)>:\ n+s\hbox{ even or odd}\\
&&<\kappa'_3(n,r,\mp,\pm),\kappa_4'((-1)^s,r+1,\pm,\pm)>:\ n+s\hbox{ even or odd}\\
\hline
(\D_{30K}\rl\D_{30K};\I\rl\I) &
<\kappa_1>;\ <\kappa_2(\pm,r,\pm)>;&
<\kappa_1(+,(-1)^{s_1},(-1)^{s_2},(-1)^{s_3}),
\kappa_3'(n,0,(-1)^{s_1+s_2+1}))>:\ n\hbox{ even or odd};\\
K\hbox{ even}& <\kappa_3>:\ n\hbox{ even or odd};\ <\kappa_4>&
<\kappa_1'(+,(-1)^{s_1},(-1)^{s_2}),\kappa_3(n,0)>:\
n\hbox{ even or odd};\\
&&<\kappa_2((-1)^s,r,\pm),\kappa_3(n,r)>:\ n\hbox{ even or odd}\\
&&<\kappa_2((-1)^s,r,\pm),\kappa_3'(n,r+1,\mp,\pm)>:\ n\hbox{ even or odd}\\
&&<\kappa_3(n,r),\kappa_4((-1)^s,r+1)>:\ n\hbox{ even or odd}\\
&&<\kappa'_3(n,r,\mp,\pm),\kappa_4'((-1)^s,r+1,\pm,\pm)>:\ n\hbox{ even or odd}\\
\hline
(\D_{2rK_1}/\Z_{K_1};\D_{2rK_2}/\Z_{K_2})&
<\kappa_1(1)>,\ <\kappa_2(n_1,n_2,n_3)>,&
<\kappa_1(1),\kappa_2(n_1,n_2,n_3)>\\
&<\kappa_3(n_1,n_2,n_3)>&\\
\hline
\end{array}
$$
Continuation of Annex B.
\end{sidewaystable}
}

\newpage
\section{Isotropy subgroups $\Sigma_j$ and $\Delta_j$ and the element $\gamma$
satisfying conditions {\bf C1}-{\bf C6} of lemma \ref{lem1},
where $\Sigma_1\cong\Z_k$ with $k\ge3$. For the groups
$(\D_{3K}\rl\Z_{6K};\mO\rl\T)$ and $(\D_{15K}\rl\D_{15K};\I\rl\I)$ with odd $K$ we
list subgroups satisfying conditions of lemma \ref{lem11}.}

\begin{sidewaystable}
$$
\begin{array}{|l|l|}
\hline
{\Gamma} & \Sigma_j, \Delta_j\mbox{ and }\gamma \\
\hline
(\D_{2K_1}\rl\D_{2K_1};\D_{2K_2}\rl\D_{2K_2})&
\Sigma_1=<\kappa_1(+,1)>,\ \Sigma_2=<\kappa_4(0,0)>,\
\Sigma_3=<\kappa_2(K_1)>,\ \Sigma_4=<\kappa_4(0,1)>\\
&\Delta_1=<\kappa_4(0,1),\kappa_1(+,1)>,\
\Delta_2=<\kappa_1(+,1),\kappa_4(0,0)>,\\
&\Delta_3=<\kappa_4(0,0),\kappa_2(K_1)>,\
\Delta_4=<\kappa_2(K_1),\kappa_4(0,1)>,\ \gamma=e\\
\hline
(\D_{2K_1r}\rl\Z_{2K_1};\D_{2K_2r}\rl\Z_{2K_2})_s&
\Sigma_1=<\kappa_1(+,1)>,\ \Sigma_2=<\kappa_2(0,0,0)>,\
\Sigma_3=<\kappa_1(-,1)>,\ \Sigma_4=<\kappa_2(1,0,0)>\\
&\Delta_1=<\kappa_2(1,0,0),\kappa_1(+,1)>,\
\Delta_2=<\kappa_1(+,1),\kappa_2(0,0,0)>,\\
&\Delta_3=<\kappa_2(0,0,0),\kappa_1(-,1)>,\
\Delta_4=<\kappa_2(-,1),\kappa_2(1,0,0)>,\ \gamma=e\\
\hline
(\D_{2K_1}\rl\D_{K_1};\D_{2K_2}\rl\D_{K_2})&
\Sigma_1=<\kappa_1(+,1)>,\ \Sigma_2=<\kappa_5(0,0)>\\
&\Delta_1=<\kappa_5(1,0),\kappa_1(+,1)>,\
\Delta_2=<\kappa_1(+,1),\kappa_5(0,0)>,\\
&\gamma=((\cos\theta_1,0,0,-\sin\theta_1);(0,\cos\theta_2,\sin\theta_2,0))\\
\hline
(\D_{2K_1}\rl\D_{K_1};\D_{2K_2}\rl\Z_{4K_2})&
\Sigma_1=<\kappa_1(+,1)>,\ \Sigma_2=<\kappa_3(0,0)>\\
K_1\hbox{ even}&\Delta_1=<\kappa_3(0,1),\kappa_1(+,1)>,\
\Delta_2=<\kappa_1(+,1),\kappa_3(0,0)>,\\
&\gamma=((0,\cos\theta_1,\sin\theta_1,0);(\cos\theta_2,0,0,-\sin\theta_2))\\
\hline
(\D_{2K_1}\rl\D_{K_1};\D_{2K_2}\rl\Z_{4K_2})&
\Sigma_1=<\kappa_1(+,1)>,\ \Sigma_2=<\kappa_4(0,0)>,\
\Sigma_3=<\kappa_2(K_1)>,\ \Sigma_4=<\kappa_4(0,1)>\\
K_1\hbox{ odd}&\Delta_1=<\kappa_4(0,1),\kappa_1(+,1)>,\
\Delta_2=<\kappa_1(+,1),\kappa_4(0,0)>,\\
&\Delta_3=<\kappa_4(0,0),\kappa_2(K_1)>,\
\Delta_4=<\kappa_2(K_1),\kappa_4(0,1)>,\ \gamma=e\\
\hline
(\D_{2K_1}\rl\D_{K_1};\D_{K_2}\rl\Z_{2K_2})&
\Sigma_1=<\kappa_1(+,1)>,\ \Sigma_2=<\kappa_2(0,0)>\\
K_2\hbox{ odd}&\Delta_1=<\kappa_2(1,0),\kappa_1(+,1)>,\
\Delta_2=<\kappa_1(+,1),\kappa_2(0,0)>,\\
&\gamma=((\cos\theta_1,0,0,-\sin\theta_1);(0,1,0,0))\\
\hline
(\D_{3K}\rl\D_{3K};\mO\rl\mO)&
\Sigma_1=<\kappa_1(+,+,+,+)>,\ \Sigma_2=<\kappa_3(0,0,-)>\\
K\hbox{ odd}&\Delta_1=<\kappa_3(3K,0,-),\kappa_1(+,+,+,+)>,\
\Delta_2=<\kappa_1(+,+,+,+),\kappa_3(0,0,-)>,\\
&\gamma=((1,0,0,0);(0,0,0,1))\\
\hline
(\D_{6K}\rl\D_{6K};\mO\rl\mO)&
\Sigma_1=<\kappa_1(+,+,+,+)>,\ \Sigma_2=<\kappa_5(0,0,-)>,\
\Sigma_3=<\kappa_5(3K,0,+)>\\
K\hbox{ odd}&\Delta_1=<\kappa_5(3K,0,-),\kappa_1(+,+,+,+)>,\
\Delta_2=<\kappa_1(+,+,+,+),\kappa_5(0,0,-)>,\\
&\Delta_3=<\kappa_5(0,0,-),\kappa_5(3K,0,+)>,\ \gamma=((1,0,0,0);(0,1,0,0))\\
\hline
(\D_{4K}\rl\D_{4K};\mO\rl\mO)&
\Sigma_1=<\kappa_1(+,0,+)>,\ \Sigma_2=<\kappa_3(0,1)>,\
\Sigma_3=<\kappa_2(+,1,+)>,\ \Sigma_4=<\kappa_3(1,1)>\\
&\Delta_1=<\kappa_3(1,1),\kappa_1(+,0,+)>,\
\Delta_2=<\kappa_1(+,0,+),\kappa_3(0,1)>,\\
&\Delta_3=<\kappa_3(0,1),\kappa_2(+,1,+)>,\
\Delta_4=<\kappa_2(+,1,+),\kappa_1(+,0,+)>,\ \gamma=e\\
\hline
(\D_{12K}\rl\D_{12K};\mO\rl\mO)&
\Sigma_1=<\kappa_2(+,0,+)>,\ \Sigma_2=<\kappa_4(0,1)>,\
\Sigma_3=<\kappa_3(+,1,+)>,\ \Sigma_4=<\kappa_4(1,1)>\\
&\Delta_1=<\kappa_4(1,1),\kappa_2(+,0,+)>,\
\Delta_2=<\kappa_2(+,0,+),\kappa_4(0,1)>,\\
&\Delta_3=<\kappa_4(0,1),\kappa_3(+,1,+)>,\
\Delta_4=<\kappa_3(+,1,+),\kappa_2(+,0,+)>,\ \gamma=e\\
\hline
(\D_{3K}\rl\Z_{6K};\mO\rl\T)&
\Sigma_1=<\kappa_1(+,+,+,+)>,\ \Sigma_2=<\kappa_2(0,0,-)>,\
\Sigma_3=<\kappa_1(+,+,+,-)>,\ \Sigma_4=<\kappa_2(1,0,-)>,\\
&\Delta_1=<\kappa_2(1,0,-),\kappa_1(+,+,+,+)>,\
\Delta_2=<\kappa_1(+,+,+,+),\kappa_2(0,0,-)>,\\
&\Delta_3=<\kappa_2(0,0,-),\kappa_1(+,+,+,-)>,\
\Delta_4=<\kappa_1(+,+,+,-),\kappa_2(-,0,-)>,\ \gamma=e\\
\hline
\end{array}
$$
\end{sidewaystable}

\begin{sidewaystable}
$$
\begin{array}{|l|l|}
\hline
{\Gamma} & \Sigma_j, \Delta_j\mbox{ and }\gamma \\
\hline
(\D_{6K}\rl\Z_{12K};\mO\rl\T)&
\Sigma_1=<\kappa_1(+,+,+,+)>,\ \Sigma_2=<\kappa_3(0,0,-)>,\
\Sigma_3=<\kappa_1(+,+,+,-)>,\ \Sigma_4=<\kappa_3(1,0,-)>\\
&\Delta_1=<\kappa_3(1,0,-),\kappa_1(+,+,+,+)>,\
\Delta_2=<\kappa_1(+,+,+,+),\kappa_3(0,0,-)>,\\
&\Delta_3=<\kappa_3(0,0,-),\kappa_1(+,+,+,-)>,\
\Delta_4=<\kappa_2(+,+,+,-),\kappa_3(1,0,-)>,\ \gamma=e\\
\hline
(\D_{4K}\rl\D_{2K};\mO\rl\T)&
\Sigma_1=<\kappa_1(+,0,+)>,\ \Sigma_2=<\kappa_3(0,0,+)>,\\
K\ne3m\hbox{ odd}&\Delta_1=<\kappa_3(1,0,+),\kappa_1(+,0,+)>,\
\Delta_2=<\kappa_1(+,0,+),\kappa_3(0,0,+)>,\\
&\gamma=((\cos\theta_1,0,0,-\sin\theta_1);(0,1,1,0)/\sqrt{2})\\
\hline
(\D_{6K}\rl\D_{3K};\mO\rl\T)&
\Sigma_1=<\kappa_1(+,+,+,+)>,\ \Sigma_2=<\kappa_4(0,0,-)>,\\
&\Delta_1=<\kappa_4(1,0,-),\kappa_1(+,+,+,+)>,\
\Delta_2=<\kappa_1(+,+,+,+),\kappa_4(0,0,-)>,\\
&\gamma=((\cos\theta_1,0,0,-\sin\theta_1);(0,1,-1,0)/\sqrt{2})\\
\hline
(\D_{3K}\rl\Z_{2K};\mO\rl\V)&
\Sigma_1=<\kappa_1(+,+,+,+)>,\ \Sigma_2=<\kappa_2(0,-)>,\\
K\hbox{ odd}&\Delta_1=<\kappa_2(3K,-),\kappa_1(+,+,+,+)>,\
\Delta_2=<\kappa_1(+,+,+,+),\kappa_2(0,-)>,\\
& \gamma=((1,0,0,0);(0,0,0,1))\\
\hline
(\D_{6K}\rl\Z_{4K};\mO\rl\V)&
\Sigma_1=<\kappa_1(+,+,+,+)>,\ \Sigma_2=<\kappa_3(0,-)>,\
\Sigma_3=<\kappa_2(+,0)>,\ \Sigma_4=<\kappa_3(1,-)>\\
&\Delta_1=<\kappa_3(1,-),\kappa_1(+,+,+,+)>,\
\Delta_2=<\kappa_1(+,+,+,+),\kappa_3(0,-)>,\\
&\Delta_3=<\kappa_3(0,-),\kappa_2(+,0)>,\
\Delta_4=<\kappa_2(+,0),\kappa_3(1,-)>,\ \gamma=e\\
\hline
(\D_{3K}\rl\D_{3K};\I\rl\I) &
\Sigma_1=<\kappa_1'(+,0,+,+)>,\ \Sigma_2=<\kappa_2(0,0)>,\\
K\hbox{ odd}&\Delta_1=<\kappa_2(3K,0),\kappa_1'(+,0,+,+)>,\
\Delta_2=<\kappa_1'(+,0,+,+),\kappa_2(0,0)>,\ \gamma=((1,0,0,0);(0,0,0,1))\\
\hline
(\D_{6K}\rl\D_{6K};\I\rl\I) &
\Sigma_1=<\kappa_1'(+,0,+,+)>,\ \Sigma_2=<\kappa_2(0,0)>,\
\Sigma_3=<\kappa_3(+,1)>,\ \Sigma_4=<\kappa_2(1,0)>,\\
&\Delta_1=<\kappa_2(1,0),\kappa_1'(+,0,+,+)>,\
\Delta_2=<\kappa_1'(+,0,+,+),\kappa_2(0,0)>,\\
&\Delta_3=<\kappa_2(0,0),\kappa_3(+,1)>,\
\Delta_4=<\kappa_3(+,1),\kappa_2(1,0)>,\ \gamma=e\\
\hline
(\D_{5K}\rl\D_{5K};\I\rl\I) &
\Sigma_1=<\kappa_1(+,0,+,+)>,\ \Sigma_2=<\kappa_2(0,0)>,\\
K\hbox{ odd}&\Delta_1=<\kappa_2(5K,0),\kappa_1(+,0,+,+)>,\
\Delta_2=<\kappa_1(+,0,+,+),\kappa_2(0,0)>,\ \gamma=((1,0,0,0);(0,0,0,1))\\
\hline
(\D_{10K}\rl\D_{10K};\I\rl\I) &
\Sigma_1=<\kappa_1(+,0,+,+)>,\ \Sigma_2=<\kappa_2(0,0)>,\
\Sigma_3=<\kappa_3(+,1)>,\ \Sigma_4=<\kappa_2(1,0)>,\\
&\Delta_1=<\kappa_2(1,0),\kappa_1(+,0,+,+)>,\
\Delta_2=<\kappa_1(+,0,+,+),\kappa_2(0,0)>,\\
&\Delta_3=<\kappa_2(0,0),\kappa_3(+,1)>,\
\Delta_4=<\kappa_3(+,1),\kappa_2(1,0)>,\ \gamma=e\\
\hline
(\D_{15K}\rl\D_{15K};\I\rl\I) &
\Sigma_1=<\kappa_1'(+,0,+,+)>,\ \Sigma_2=<\kappa_3(0,0)>,\
\Sigma_3=<\kappa_2(+,0,+,+)>,\ \Sigma_4=<\kappa_3(1,0)>,\\
&\Delta_1=<\kappa_3(1,0),\kappa_1'(+,0,+,+)>,\
\Delta_2=<\kappa_1'(+,0,+,+),\kappa_3(0,0)>,\\
&\Delta_3=<\kappa_3(0,0),\kappa_2(+,0,+,+)>,\
\Delta_4=<\kappa_3(1,0),\kappa_2(+,0,+,+)>,\ \gamma=e\\
\hline
(\D_{30K}\rl\D_{30K};\I\rl\I) &
\Sigma_1=<\kappa_1'(+,0,+,+)>,\ \Sigma_2=<\kappa_3(0,0)>,\
\Sigma_3=<\kappa_2(+,0,+,+)>,\ \Sigma_4=<\kappa_3(1,0)>,\\
&\Delta_1=<\kappa_3(1,0),\kappa_1'(+,0,+,+)>,\
\Delta_2=<\kappa_1'(+,0,+,+),\kappa_3(0,0)>,\\
&\Delta_3=<\kappa_3(0,0),\kappa_2(+,0,+,+)>,\
\Delta_4=<\kappa_3(1,0),\kappa_2(+,0,+,+)>,\ \gamma=e\\
\hline
(\D_{2rK_1}\rl\Z_{K_1};\D_{2rK_2}\rl\Z_{K_2})&
\Sigma_1=<\kappa_1(1)>,\ \Sigma_2=<\kappa_2(0,0,0)>,\\
&\Delta_1=\Delta_2=<\kappa_1(1),\kappa_2(0,0,0)>,\ \gamma=e\\
\hline
\end{array}
$$
Continuation of Annex C
\end{sidewaystable}

\end{document}